\documentclass[twocolumn,times,trackchanges]{aastex63}
\usepackage[utf8]{inputenc}
\usepackage{placeins, xspace}

\newcommand{\kms}[1]{\ensuremath{\mathrm{km\,s^{-1}}}\xspace}
\newcommand{\ditto}{''}

\begin{document}
\title{An Expanding Accretion Disk and a Warm Disk Wind As Seen In the Spectral Evolution of HBC 722}
\author{Adolfo Carvalho}
\affiliation{Department of Astronomy; California Institute of Technology; Pasadena, CA 91125, USA}
\author{Lynne Hillenbrand}
\affiliation{Department of Astronomy; California Institute of Technology; Pasadena, CA 91125, USA}
\author{Jerome Seebeck}
\affiliation{Department of Astronomy; California Institute of Technology; Pasadena, CA 91125, USA}
\affiliation{Department of Astronomy; University of Maryland; College Park, MD 20742, USA}
\author{Kevin Covey}
\affiliation{Department of Physics \& Astronomy, Western Washington University, MS-9164, 516 High St., Bellingham, WA 98225, USA}

\begin{abstract}
We present a comprehensive analysis of the post-outburst evolution of the FU Ori object HBC 722 in optical/near-infrared (NIR) photometry and spectroscopy. Using a modified viscous accretion disk model, we fit the outburst epoch SED to determine the physical parameters of the disk, including $\dot{M}_\mathrm{acc} = 10^{-4.0} \ M_\odot$ yr$^{-1}$, $R_\mathrm{inner} = 3.65 \ R_\odot$, $i = 79^\circ$, and a maximum disk temperature of $T_\mathrm{max} = 5700$ K. We then use a decade of optical/NIR spectra to demonstrate a changing accretion rate drives the visible-range photometric variation, while the NIR shows the outer radius of the active accretion disk expands outward as the outburst progresses. We also identify the major components of the disk system: a plane-parallel disk atmosphere in Keplerian rotation and a 2-part warm disk wind that is collimated near the star and wide-angle at larger radii. The wind is traced by classic wind lines, and appears as a narrow, low-velocity, deep absorption component in several atomic lines spanning the visible spectrum and in the CO 2.29$\mu$m band. We compare the wind lines to those computed from wind models for other FU Ori systems and rapidly accreting young stellar disks and find a 4000-6000 K wind can explain the observed line profiles. Fitting the progenitor spectrum, we find $M_* = 0.2 \ M_\odot$ and $\dot{M}_\mathrm{progenitor} = 7.8 \times 10^{-8} \ M_\odot \ \mathrm{yr}^{-1}$. Finally, we discuss HBC 722 relative to V960 Mon, another FU Ori object we have previously studied in detail. 
\end{abstract}


\section{Introduction}
FU Ori-type sources are a class of Young Stellar Object (YSO) that have undergone extreme ($\Delta V \sim 4-6$ mag) outbursts attributed to rapid increases in their circumstellar-to-star disk accretion rates. While $\sim 30$ of these objects have been identified since the eruption of FU Ori itself in 1936 \citep{Herbig_FUOri_interpretation_1966VA}, the mechanisms triggering and driving the outbursts remain elusive. 

The recent outbursts of the FU Ori objects HBC 722 \citep[][]{Semkov_HBC722Detection_2010ATel.2801} and V960 Mon \citep[][]{Maehara_2014ATel.6770} have provided unique opportunities to study outbursts as they evolve, in unprecedented detail. These outbursts happened well into the era of deep all-sky surveys, enabling multi-wavelength characterization that was not possible for the ``classic" FU Ori objects like FU Ori itself, V1057 Cyg, and V1515 Cyg. 
Both outbursts were both promptly followed-up, with the outburst peaks captured by several observatories with high precision photometry, intermediate resolution spectrophotometry, and high-resolution spectroscopy \citep{miller_evidence_2011, hillenbrand_optical_2015}. 

In the years following these two outbursts, the targets were revisited by many of these same facilities, enabling detailed study of the spectral evolution of the objects \citep[e.g.][]{Lee_HBC722_2015ApJ, park_high-resolution_2020, Carvalho_V960MonPhotometry_2023ApJ, Carvalho_V960MonSpectra_2023ApJ}. Continued observations of these targets by the American Association of Variable Star Observers (AAVSO) and the ease of access to their publicly available data has also provided crucial multi-band photometry to contextualize the spectroscopic observations. 

Following a similar approach to the work in \citet{Carvalho_V960MonPhotometry_2023ApJ} and \citet{Carvalho_V960MonSpectra_2023ApJ} for V960 Mon, here we use several Keck/HIRES, IRTF/SpeX, and Palomar/TripleSpec spectra to supplement the existing photometry of HBC 722 and better understand the many components in the accretion/outflow system. 

\begin{figure*}[!htb]
    \centering
    \includegraphics[width=0.98\linewidth]{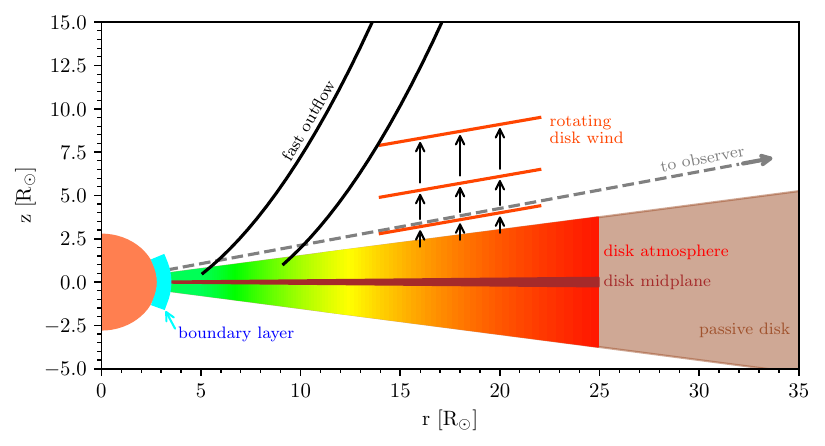}
    \caption{A simplified diagram of the HBC 722 system near the outburst peak. The color gradient represents the radial temperature profile in the disk atmosphere, which is assumed plane-parallel and peaks at $T_\mathrm{max} = 5700$ K with $L_\mathrm{acc} = 71 \ L_\odot$.  The depicted height of the disk atmosphere above the midplane is \textbf{representative of} the $\tau = 1$ surface in the \citet{Zhu_outburst_FUOri_2020MNRAS} simulation of the FU Ori system. The midplane (in maroon) is marked for reference and given a height similar to the average scale height in the \citet{Zhu_outburst_FUOri_2020MNRAS} simulation. The outer, passive disk represents that modeled in Section \ref{sec:spex}, which is reprocessing the bright inner disk emission. The fast outflow lines represent a magnetocentrifugal wind following the BP82 solutions, which appear consistent with what we observe in the the wind-sensitive lines of the spectrum \textbf{(Section \ref{sec:windLines})}.}
    \label{fig:cartoon}
\end{figure*}

Our picture of the system near the outburst peak is shown in Figure \ref{fig:cartoon}, 
illustrating the various absorption and emission components that we detect in the spectra of the object. Our interpretation of the structure of the inner disk region is consistent with past studies of FU Ori objects \citep{Calvet_FUOriModel_1993, hartmann_fu_1996, Zhu_outburst_FUOri_2020MNRAS}. The presence of each component is identified empirically and is agnostic to the potential outburst trigger or active instability in this system. In our data, we are able to clearly identify absorption from a rotating, plane-parallel disk atmosphere, a high-velocity collimated disk wind, a low-velocity disk wind with a wide opening angle, and emission from the winds. The disk atmosphere is generally consistent with a modified viscously-heated accretion disk \citep{Shakura_sunyaev_alpha_1973A&A, Kenyon_FUOri_disks_1988ApJ, Rodriguez_model_2022}, while the disk winds are similar to the \citet[][hereafter BP82]{BlandfordPayneWind_1982MNRAS} magnetocentrifugal wind model.


Over the course of this article, we will present our evidence for the existence of each of these components and how they varied as HBC 722 evolved through the 3 main stages of its post-outburst lightcurve. 

In Section \ref{sec:data_spec}, we describe the different datasets we used and the data reduction steps for the spectra. We then explain, in Section \ref{sec:SEDFit}, the SED fits with which we predict the time-evolution of the outer boundary of the active region of the disk and its atmosphere. In Section \ref{sec:highResolutionModels} we present our high resolution models of the system. In Section \ref{sec:excess} the wind components of the system are identified in the high resolution spectra. We then provide an interpretation and summary of our analysis in Section \ref{sec:interp}.

\section{Data} \label{sec:data_spec}
To understand the parameters of the HBC 722 system both at outburst and as they evolve in time, we combine several years of photometry, flux-calibrated medium-resolution spectra, and high resolution spectra. We mostly rely on previously unpublished data, though we include some published data in our analysis. The spectra sample the lightcurve well, especially in the 3 main epochs: the outburst peak, the subsequent fade (which we refer to as the ``dip"), and the second outburst into the plateau phase. An observing log for the spectra is given in Table \ref{tab:obs}.

\subsection{Photometry}
\begin{figure*}[!htb]
    \centering
    \includegraphics[width=0.98\linewidth]{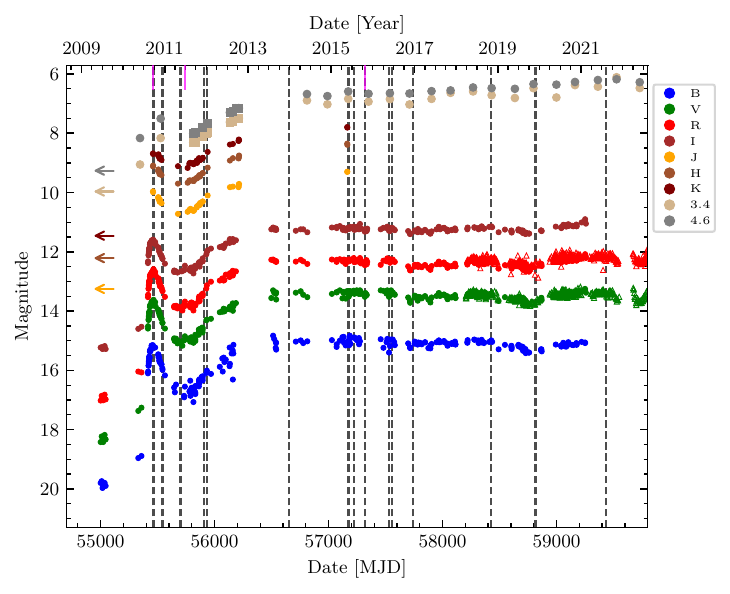}
    \caption{The multiband lightcurve of HBC 722, illustrating the rapid rise during outburst, brief post-outburst fade (``dip") and subsequent rise and ``plateau". Filters are as given in the legend; see text for details. Green and red triangles show ZTF photometry shifted by $g - 0.9$ and $r-0.4$ to overlap with the $V$ and $R$ photometry. Square points in the 3.4 and 4.6 $\mu$m bands mark the Spitzer/IRAC photometry from \citet{kospal_hbc722_2016A&A}, distinguishing it from the WISE and NEOWISE measurements. Left-pointing arrows indicate the progenitor photometry taken prior to 2009 \citep{miller_evidence_2011}. Epochs of our optical HIRES spectra are shown as vertical black dashed lines,  
    while epochs of infrared SpeX and TripleSpec spectra are marked with short vertical magenta lines.}
    \label{fig:AAVSO_lc}
\end{figure*}

We use photometry from \citet{miller_evidence_2011}, \citet{semkov_photometric_2013}, and \citet{Semkov_HBC722Photometry_2021Symm}, as shown in Figure \ref{fig:AAVSO_lc}. 

To briefly summarize the photometric evolution of the system: HBC 722 began a slow rise in 2009 ($\Delta V/\Delta t \sim -1 \ \mathrm{mag \ yr}^{-1}$), which culminated in a strong ($\Delta V \sim -4 \ \mathrm{mag}$) outburst in late 2010. The outburst maximum brightness lasted only a few days before the object began to fade. In the following months, the object faded at a rate of $\Delta V/\Delta t \sim 2 \ \mathrm{mag \ yr}^{-1}$, reaching a post-outburst minimum brightness in May 2011 before steadily brightening again at a rate of $\Delta V/\Delta t \sim - 0.75 \ \mathrm{mag \ yr}^{-1}$ until late 2013. Since 2014, the system brightness has remained relatively constant, with occasional $|\Delta V| \sim 0.1$ mag variations over several months.

\subsection{Medium Resolution Spectrophotometry}
We make use of several medium resolution flux-calibrated spectra to study the photometric evolution of the spectrum simultaneously with the changing molecular absorption. 

Near the outburst epoch,
the visible range medium resolution spectrum we use to constrain the model is from September 16, 2010, taken using the Kast spectrograph \citep{MillerStone_KastSpec_1993} on the Lick observatory 3 m Shane Telescope. The near-infrared (NIR) medium resolution spectrum was taken on September 25, 2010, with the TripleSpec spectrograph \citep{Herter_TripleSpecInstrument_2008SPIE} on the Palomar 5 m Hale telescope. Details of the data reduction for both spectra can be found in \citet{miller_evidence_2011}. We use these spectra without further manipulation or reduction beyond that described in \citet{miller_evidence_2011}.

For the ``dip" and ``plateau" epochs,
NIR medium resolution spectra were taken with the SpeX spectrograph \citep{Rayner_spex_2003PASP} on the 3 m NASA Infrared Telescope Facility (IRTF) on June 26, 2011, and October 27, 2015. The spectrum from 2015 has been previously published as part of the \citet{connelley_near-infrared_2018} survey of FU Ori objects and the data reduction is described there. 
The previously unpublished (2011) SpeX observations were obtained using the cross-dispersed echelle gratings between 0.8 and 2.45 $\mu$m,  using the 0.5$^{\prime \prime}$ wide slit. Seeing was 0.5-0.7$^{\prime \prime}$, with light to heavy cirrus. Spectral extraction, wavelength calibration, telluric correction, and flux calibration were done using the Spextool reduction package \citep{Vacca_telluricMethod_2003PASP, Cushing_spextool_2004PASP}. Observations of the A0V standard HD 205314 obtained immediately prior and at similar airmass were used for telluric correction and spectral slope calibration. Contemporaneous $K$ band photometry was used to correct for slit losses by fixing the flux level of the spectrum at 2.2 $\mu$m.

\subsection{High Resolution spectra}
\subsubsection{Keck/HIRES}

We obtained visible range high-dispersion spectra from the Keck Observatory's HIgh Resolution Echelle Spectrograph \citep[HIRES;][]{Vogt1994}. The spectra were taken in slightly different spectrograph settings over time, but broadly overlap in their wavelength coverage. Table \ref{tab:obs} gives the epochs, coverage, and resolution for the spectra. 

The spectra with 3600-7500 \AA\ coverage were obtained upon request by the California Planet Search collaboration and
processed using their standard reduction scripts \citep{Howard_CPSI_2010ApJ}; the first such spectrum was published previously by us in \cite{miller_evidence_2011}.
The spectrum for 14 November 2010 was obtained from the Keck Observatory Archive (PI: G. Herbig) and was processed by their automated pipeline, which uses the MAKEE pipeline reduction package written by Tom Barlow\footnote{ {\url{https://sites.astro.caltech.edu/~tb/makee/}}}. All other spectra were processed with the 2008 version of MAKEE.

We normalize the spectra by fitting the continuum using a regularized Asymmetric Least-Squares-based algorithm \citep{eilers2005baseline}. The technique relies on non-parameteric least-squares fitting with a regularization term to smoothly reproduce major features in the data. The error function the algorithm aims to minimize is $S = \Sigma_i d_i^2 + \lambda \Sigma_i (\Delta^2 z_i)^2 $, where $d_i$ is the residual between $y_i$, the input, and $z_i$, the model. The $\Delta^2 z_i$ term is the second derivative of the model and ensures the smoothness of the fit. The regularization parameter, $\lambda$, enables us to tune the smoothness of the continuum fits to avoid removing broad molecular absorption features or blended atomic features \citep[see][Section III.A for a clear description of the algorithm]{oller2014adaptive}. 

The technique is also more robust to the edges of the spectrum than polynomial fitting. Orders with emission lines (e.g., H$\alpha$, weak forbidden emission lines, and the Ca II Infrared Triplet) need special treatment. We mask emission lines in the spectrum and use the linear interpolation from the redward continuum point to the blueward continuum point on either side of the lines as the continuum under those lines. The normalized spectra are shown in Section \ref{sec:HIRESModelsComp}.

We measure a heliocentric system velocity of $-10.0$ km s$^{-1}$ for the system by centering the disk profiles of isolated disk lines (e.g., Ca I 6439, Ca I 6449, and Li I 6707).

\subsubsection{Keck/NIRSPEC}
We obtained a NIR spectrum spanning 1.0-2.5 $\mu$m on 30 October 2023 using the Keck Observatory Near InfraRed SPECtrograph \citep[NIRSPEC,][]{McLean_nirspecDesign_1998SPIE, Martin_NIRSPECupgrade_2018SPIE10702E, Lopez_NIRSPECUpgradePerformance_2020SPIE11447E}. The spectrum was taken in good weather and seeing, yielding a signal-to-noise ratio of $\sim 300$. Four sets of spectrograph settings were required to cover the wavelength range, one for each of Y, J, H and K bands (the coverage and resolution are given in Table \ref{tab:obs}). We acquired the spectra in ABBA nod pairs to facilitate efficient background subtraction, nodding the telescope $\sim 6^{\prime\prime}$ along the slit between integrations. The exposure times differed in each band due to the target SED and the different filter throughput values. The exposure numbers and times for each band were as follows: 8$\times$30 s ($Y$), 8$\times$30 s ($J$), 4$\times$60 s ($H$), 8$\times$10 s ($K$).

We reduced the spectrum using the newly developed NIRSPEC pipeline \citep{CarvalhoDoppmann_NIRSPEC_2024ascl} in the $\mathtt{PypeIt}$ package \citep{prochaska_pypeit_2020JOSS}. The pipeline uses the lamp flats to identify the edges of the orders of the spectra, which it then traces to construct slit masks. The slits are then extracted and rectified. We use Ar, Xe, Kr, Ne arc lamp spectra to wavelength-calibrate the spectra. $\mathtt{PypeIt}$ fits the emission lines with Gaussian profiles to identify their centroids in pixel space. We then manually identify the emission lines using the references produced by Greg Doppmann and published on the NIRSPEC instrumentation webpage\footnote{\url{https://www2.keck.hawaii.edu/inst/nirspec/cals.html}}. The A/B nod pairs are then subtracted from one another and the target traces extracted using an optical extraction algorithm. The A0V telluric standards and the lamp flats are then used to estimate the sensitivity and blaze functions prior to coadding the individual spectra. 

We telluric corrected the spectra by first continuum normalizing the telluric standard spectrum and fitting an A0V PHOENIX \citep{Husser_Phoenix_2013A&A} stellar model spectrum to it. We then continuum normalize the target spectrum and divide it by the telluric spectrum. The continuum normalization was done using the same procedure we used for the HIRES spectra.  

\subsubsection{Gemini/IGRINS}
We also use a spectrum from the Immersion GRating INfrared Spectrograph \citep[IGRINS,][]{Park_IGRINS_2014SPIE} to study the observed NIR variability at high resolution. The spectrum was taken on November 20, 2014 when IGRINS was installed on the Harlan J. Smith Telescope at the McDonald Observatory and was previously published by \citet{Lee_HBC722_2015ApJ}. The spectrum was reduced using the IGRINS Pipeline Package \citep{Lee_IGRINS_Pipeline_2017zndo} and is available on the Raw and Reduced IGRINS Spectral Archive\footnote{\url{https://igrinscontact.github.io}}. We continuum normalize the IGRINS spectrum using the same procedure as we used for the HIRES and NIRSPEC spectra.

\begin{deluxetable}{cccc}[!htb]
\caption{Spectroscopic Observations Log}\label{tab:obs}
	\tablehead{\colhead{Date} 	     & \colhead{Instrument} &  \colhead{Band ($\mu$m)} & \colhead{$\lambda/\Delta\lambda$}}

\startdata
\hline
 	 2010-09-25  & Keck/HIRES & $0.36-0.79$ & 30,000 \\
 	 2010-11-14  & $\ditto$ & $0.43-0.86$ & $\ditto$ \\
 	 2010-12-13  & $\ditto$ & $0.43-0.86$ & $\ditto$ \\
 	 2011-05-20  & $\ditto$ & $0.36-0.79$ & $\ditto$ \\
 	 2011-12-10  & $\ditto$ & $0.47-0.92$ & $\ditto$ \\
 	 2012-01-06  & $\ditto$ & $0.44-0.86$ & $\ditto$ \\
 	 2013-12-27  & $\ditto$ & $0.47-0.92$ & $\ditto$ \\
 	 2015-06-02  & $\ditto$ & $0.47-0.92$ & $\ditto$ \\
 	 2015-07-24  & $\ditto$ & $0.47-0.92$ & $\ditto$ \\
 	 2015-10-27  & $\ditto$ & $0.47-0.92$ & $\ditto$ \\
 	 2016-05-20  & $\ditto$ & $0.47-0.92$ & $\ditto$ \\
 	 2016-06-15  & $\ditto$ & $0.47-0.92$ & $\ditto$ \\
 	 2016-12-22  & $\ditto$ & $0.47-0.92$ & $\ditto$ \\
 	 2018-11-03  & $\ditto$ & $0.47-0.92$ & $\ditto$ \\
 	 2019-11-29  & $\ditto$ & $0.47-0.92$ & $\ditto$ \\
 	 2021-08-07  & $\ditto$ & $0.36-0.79$ & $\ditto$  \\
\hline
        2010-09-16 & Lick/Kast & $0.35-1.00$ & 6000  \\
        2010-09-23 & Palomar/TripleSpec & $0.90-2.50$ & 2700  \\
        2011-06-26 & IRTF/SpeX & $0.80-2.50$ & 2000  \\
        2015-10-27 & $\ditto$  & $0.70-5.00$ &$\ditto$\\
\hline
        2014-11-20 & Gemini/IGRINS & $1.40-2.50$ & 45,000 \\
        2023-10-30 & Keck/NIRSPEC & $1.00-2.50$ & 15,000 \\
\enddata
\end{deluxetable}

\section{SED Fits} \label{sec:SEDFit}
We adopt a viscous accretion disk model that follows the \citet{Shakura_sunyaev_alpha_1973A&A} temperature profile, 
\begin{equation}
T_\mathrm{eff}(r)^4 = \frac{3 G M_* \dot{M}_\mathrm{acc}}{8 \pi \sigma_{SB} r^3} \left[1 - \sqrt{\frac{R_\mathrm{inner}}{r}}  \right],    
\end{equation}
where $R_\mathrm{inner}$ is the innermost radius of the disk, $M_*$ is the stellar mass, $\dot{M}_\mathrm{acc}$ is the mass accretion rate (hereafter we will simply use $\dot{M}$ for the disk-to-star accretion rate during the FU Ori outburst), $\sigma_{SB}$ is the Stefan-Boltzmann constant, and $G$ is the gravitational constant. Following \citet{Kenyon_FUOri_disks_1988ApJ}, we set $T_\mathrm{eff}(r < \frac{49}{36} R_\mathrm{inner}) = T_\mathrm{eff}(\frac{49}{36} R_\mathrm{inner}) = T_\mathrm{max}$. Kinematically, the gas is assumed to be in Keplerian orbit around the star and thus follows the velocity profile: $v_\mathrm{Kep}(r) = \sqrt{GM_*/r}$. Our model is described in detail in \citet{Carvalho_V960MonPhotometry_2023ApJ}.

We fit the model to our spectrophotometry of HBC 722 to constrain the physical parameters of the disk and how they evolve post-outburst. We first fit the outburst epoch, allowing several system parameters to vary. We then fix most of the parameters to their outburst epoch best-fit values and demonstrate that the post-outburst evolution can be modeled by changing just $\dot{M}$ and $R_\mathrm{outer}$. We detail these two main modeling stages below. 

\subsection{Fitting the Outburst Epoch SED} \label{sec:outburstSEDFit}
We constrain the HBC 722 system outburst epoch parameters by fitting an SED constructed from 2 medium-resolution spectra taken at the peak of the outburst and spanning 0.4-2.5 $\mu$m: the 16 September 2010 Lick/Kast spectrum and the 23 September 2010 Palomar/TripleSpec spectrum \citep{miller_evidence_2011}. Since our focus in the SED fit is to match the continuum and broad molecular emission in the target, we resample the spectra to a grid with $\lambda / \Delta \lambda = 100$. This also accelerates the model construction (and therefore fitting) significantly. 

Our SED fitting technique follows a similar procedure to that outlined in \citet{Carvalho_V960MonPhotometry_2023ApJ}, combining information from existing literature and the high resolution spectra to constrain an MCMC \citep[$\mathtt{emcee}$,][]{FM_emcee_2013PASP} fit. The parameters we will need to compute our models are: $M_*$, $\dot{M}$, $R_\mathrm{inner}$, $A_V$, $R_\mathrm{outer}$, $d$ (the distance to the target), and $i$ (the disk inclination). We will also demonstrate in Section \ref{sec:spex} that we will need to also vary $T_\mathrm{min}$, the coolest temperature atmosphere model used in the accretion disk model, where for $T(r)<T_\mathrm{min}$, $F_\lambda(r) = B_\lambda(T(r))$. For the fits described below, we fix $T_\mathrm{min} = 2400$ K.

The existence of pre-outburst observations of the target \citep{Fang_NorthAmerica_2020ApJ, miller_evidence_2011} enables us to use the progenitor $M_*$ and $R_*$ to guide our modeling of the post-outburst state. The pre-outburst visible range spectrum indicates the progenitor had a $T_\mathrm{eff} = 3100$ K, while the pre-outburst SED \citep{miller_evidence_2011} indicates an $R_* = 2.8 \ R_\odot$, corresponding to $M_* = 0.2 \ M_\odot$ in the $10^5$ yr PARSEC \citep{Nguyen_parsec_2022A&A} isochrone \citep[see Appendix \ref{app:progenitor} and ][]{Fang_NorthAmerica_2020ApJ}. We therefore fix $M_* = 0.2 \ M_\odot$ for our disk model. We can then restrict the $R_\mathrm{inner} \geq R_* = 2.8 \ R_\odot$. We also adopt the Gaia-derived 745 pc as the distance to the target \citep{Kuhn_NorthAmericaDistance_2020RNAAS}. 

The 25 Sept 2010 HIRES spectrum shows clear disk-like absorption line profiles that are consistent with a maximum disk velocity, $v_\mathrm{max} \sim 90$ km s$^{-1}$ (see samples from the spectra in Section \ref{sec:HIRESModelsComp} and Section \ref{sec:ccfs}). The line profiles are distinctly non-Gaussian and are characterized by having flat ``cores" and are well-matched by Keplerian rotation profiles. We use this to impose a prior on the $R_\mathrm{inner}$ and inclination ($i$) in the MCMC fit by requiring that $v_\mathrm{max}$ should be drawn from a Gaussian distrubution centered at 90 km s$^{-1}$ with an uncertainty of 10 km s$^{-1}$.

Since we have $d=745$ pc and $M_*=0.2 \ M_\odot$, we only need to fit $\dot{M}$, $R_\mathrm{inner}$, $i$, and $A_V$. The very weak K band emission in this epoch implies an "active" region much smaller than the $R_\mathrm{outer} \sim 100 \ R_\odot$ typically assumed for FU Ori objects \citep[e.g., ][]{Rodriguez_model_2022}. For this epoch we fix $R_\mathrm{outer} = 25 \ R_\odot$, which matches the data well (and larger values overestimate the $K$ band continuum flux). We discuss the $R_\mathrm{outer}$ value for this and other epochs of HBC 722 spectra in Section \ref{sec:spex}.

\begin{figure}[!htb]
    \centering
    \includegraphics[width = \linewidth]{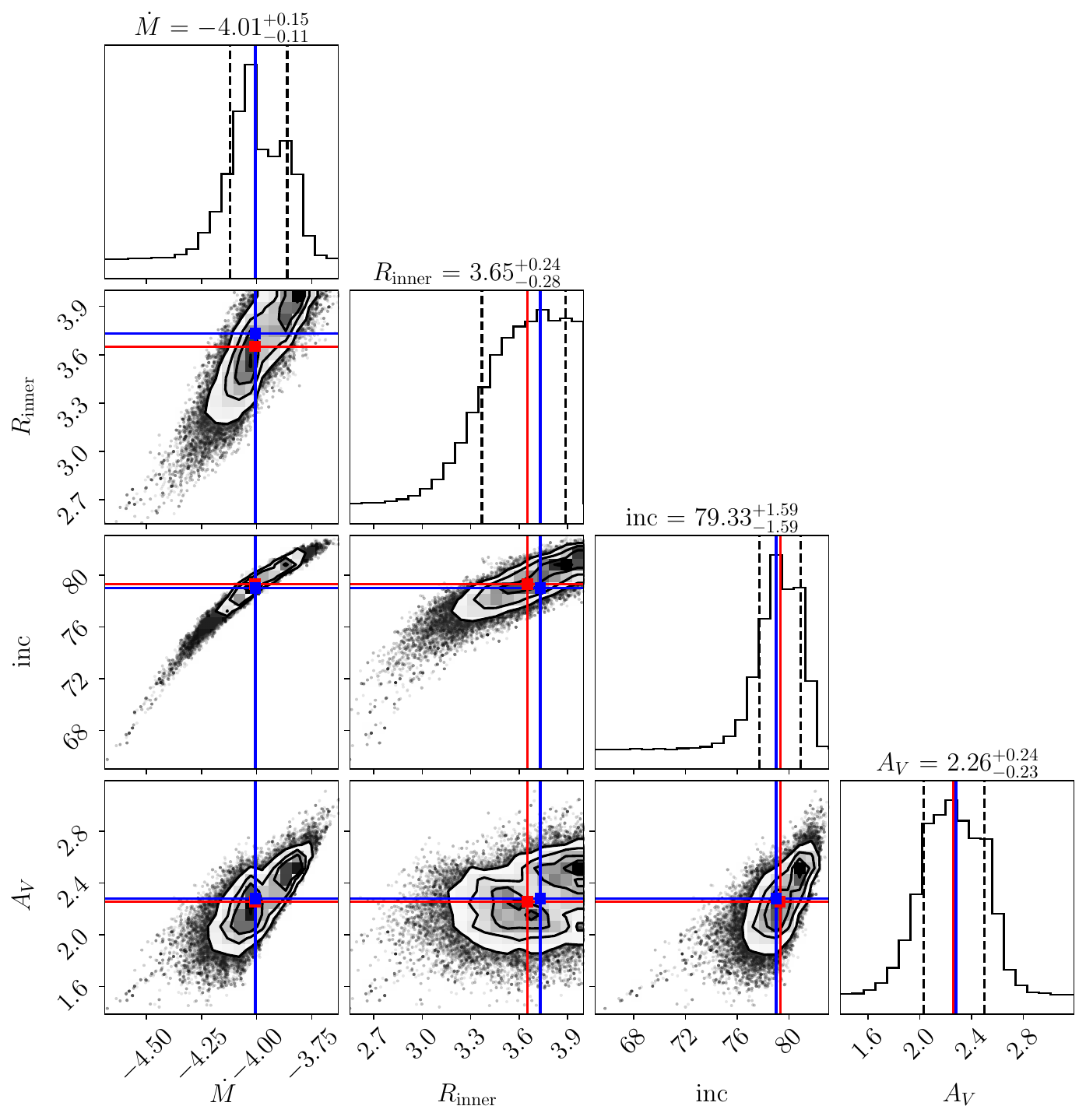}
    \caption{The $\mathtt{corner}$ plot for the MCMC fit of the outburst epoch SED showing the posterior distributions for
    the $\dot{M}$, $R_\mathrm{inner}$, $i$, and $A_V$ model parameters. 
    The median (mode) parameter values in the posterior distributions are marked by the red (blue) horizontal and vertical lines.}
    \label{fig:MCMCCorner}
\end{figure}

The $\mathtt{corner}$ \citep{corner_FM_2016} plot with the results of the fit is shown in Figure \ref{fig:MCMCCorner}. The posterior distributions are well-behaved, with good agreement between the modal and median values for each of the 4 parameters, so we take as our best-fit model parameters: $\dot{M} = 10^{-4.0} \ M_\odot$ yr$^{-1}$, $R_\mathrm{inner} = 3.65 \ R_\odot$, $i = 79^\circ$, and $A_V = 2.3$ mag. The best-fit $A_V = 2.3$ mag is slightly smaller than our estimated $A_V = 2.5$ mag to the progenitor, but is still formally consistent with the value. The best-fit model is shown in Figure \ref{fig:spex_fits} as the outburst epoch model. The $R_\mathrm{inner}$ and $i$ values are reflected in Figure \ref{fig:cartoon}, while the $\dot{M}$ sets the maximum temperature represented by the rainbow colorscale in the Figure.


\subsection{Determining the Evolution of Disk Parameters} \label{sec:spex}
Following the procedure described in Section \ref{sec:outburstSEDFit}, we are able to successfully fit the outburst epoch Lick/Shane + Palomar/TripleSpec spectrophotometry well. We then sought to fit the other two NIR (IRTF/SpeX) spectra to better understand the system evolution in the dip (26 Jun 2011) and the plateau (27 Oct 2015). 

We find that the color-magnitude evolution of the target as it faded into the dip and brightened into the plateau is inconsistent with the model for the post-outburst fade of V960 Mon presented in \citet{Carvalho_V960MonPhotometry_2023ApJ}, in which 
$R_\mathrm{inner} \propto \dot{M}^{-2/7}$ and $T_\mathrm{max} \propto \dot{M}^{13/28}$. 
Therefore, we do not assume the same behavior for this system and instead choose to fix $R_\mathrm{inner}$ in our models. The relatively constant-color evolution favors a fixed $A_V$ mag over time, 
so we also fix that parameter. For these two fixed parameters, we adopt the best-fit values from Section \ref{sec:SEDFit}: $R_\mathrm{inner} = 3.65 \ R_\odot$ and $A_V = 2.3$ mag. Of the parameters that we varied in the MCMC fit described in Section \ref{sec:outburstSEDFit}, we only varied $\dot{M}$ to match the other two NIR spectrophotometric epochs. For the outburst epoch, we use the best-fit parameters described above, including $\dot{M} = 10^{-4.0} \ M_\odot$ yr$^{-1}$. For the dip, we find that $\dot{M} = 10^{-4.4} \ M_\odot$ yr$^{-1}$ best-matches the spectrum, whereas in the plateau we find a value of $\dot{M} = 10^{-3.9} \ M_\odot$ yr$^{-1}$.

Since the NIR data show a rapid increase in the strength of the $H_2 O$ absorption features between 1.3-1.4 $\mu$m and 1.8-1.9 $\mu$m, we found it necessary to also increase $R_\mathrm{outer}$ and decrease $T_\mathrm{min}$ for the 2011 (dip) and 2015 (plateau) epochs. This is further corroborated by the steady brightening of the 3-5 $\mu$m continuum flux (see Figure \ref{fig:AAVSO_lc}), especially after HBC 722 recovered from the dip. The brightening 3-5 $\mu$m continuum indicates that an increasing area of the disk is hotter than predicted by the passive disk model and is better matched by the viscously heated disk component. 

The $T_\mathrm{min}$ for the outburst epoch is 2400 K, whereas matching the dip epoch requires a minimum of 1500 K and matching the plateau spectrum requires a minimum of 1100 K. We also find that between the outburst and the dip the $R_\mathrm{outer} = 25 \ R_\odot$ grows to $R_\mathrm{outer} = 35 \ R_\odot$, but then must increase by another 65 $R_\odot$ to match the plateau epoch spectrum. The implied outward motion of the outer boundary of the active region of the disk is 65 $R_\odot$ over the course of 1950 days. This translates to an outward velocity of 0.27 km s$^{-1}$.

\begin{figure*}[!htb]
    \centering
    \includegraphics[width = 0.48\linewidth]{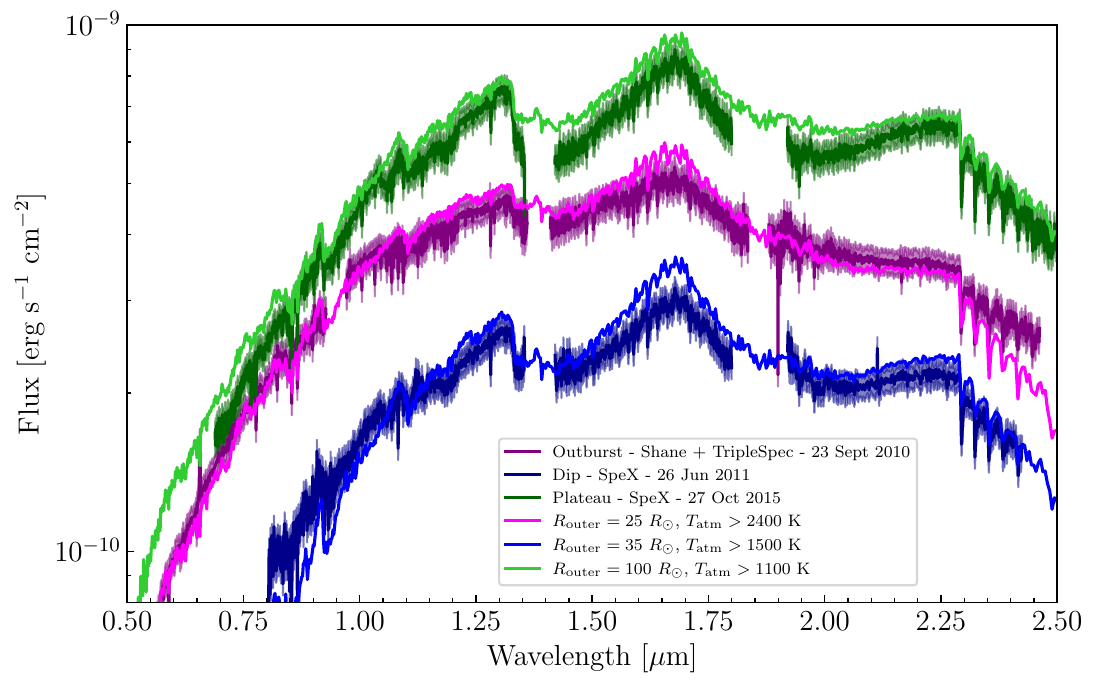}
    \includegraphics[width = 0.48\linewidth]{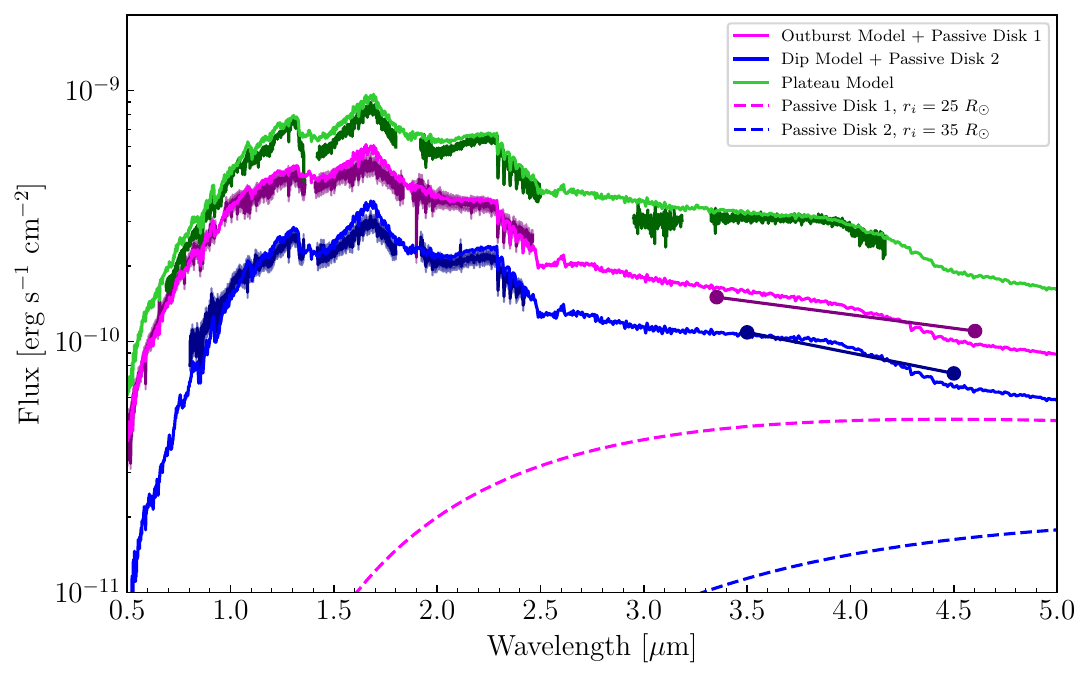}
    \caption{The 3 epochs of NIR spectra that trace the major parts of the lightcurve - the outburst peak (purple), the dip (blue), and the plateau (green). From outburst to the dip to the plateau, the evolution of the spectra is consistent with a general increase in $R_\mathrm{outer}$ and a decrease in the minimum $T_\mathrm{eff}$ of the model grid used to generate the disk model spectrum. Right panel includes extended wavelength coverage, and in order to match the photometry, a model for the passive disk in the outburst and dip epochs. For the plateau epoch, the active disk is large enough that a passive disk component is not necessary to model the spectrum in the $3-5$ $\mu$m range. }
    \label{fig:spex_fits}
\end{figure*}

The models are shown in Figure \ref{fig:spex_fits}. The $\dot{M}$ progression described above enables us to match the overall flux level and $I$ to $J$ band slope well, while the $R_\mathrm{outer}$ and $T_\mathrm{min}$ progression produce an excellent match to the molecular absorption in $J$ and $H$ band and a good fit to the $K$ band continuum. The larger $R_\mathrm{outer}$ values in the 2011 and 2015 epochs are critical to producing the appropriate continuum shape in $H$ and $K$ bands. 

We also model the contribution of a passive disk for the 2010 (peak) and 2011 (dip) epochs to better match the WISE and Spitzer photometry at $3-5$ $\mu$m. The model we adopt is that of a passive disk irradiated by the inner actively accreting region of the disk, as described in \citet{Carvalho_V960MonPhotometry_2023ApJ}. We use three $T_\mathrm{eff}$ components in the model for the incident radiation, based on $T_\mathrm{max}$, $T (1.5 \ R_\mathrm{inner})$ and $T(2.0 \ R_\mathrm{inner})$ of the disk in each epoch. We assume a mildly flaring disk, with $\frac{h}{r} = 0.2 \left( \frac{r}{r_i}  \right)^{1.0}$, where $h$ is the height of the disk atmosphere, $r$ is the radial coordinate, and $r_i$ is the inner radius of the passive disk. The flaring index of 1.0 is similar to that found for the outer disk of the other outbursting YSOs \citep{cieza_v883Ori_2018MNRAS}. For both epochs, we set $r_i = R_\mathrm{outer}$, so that the passive disk begins where the active disk ends (see Figure \ref{fig:cartoon} for an illustration of this in the outburst epoch). 

For the 2015 (plateau) epoch, the $3-5$ $\mu$m spectrum is well-fit by the active disk model alone, indicating that the active region of the disk dominates the flux at those bands, and that the passive component may be detected only in longer wavelength observations. The models including the passive disk components are shown in the right panel of Figure \ref{fig:spex_fits}.

\section{The Disk Model in High Resolution Spectra} \label{sec:highResolutionModels}
The HIRES and NIRSPEC spectra enable us to test the simple thin-disk viscous accretion disk model across a broad range of spectrum, with nearly continuous coverage from 0.4-2.5 $\mu$m. The broad wavelength range allows us to probe emission presumed to arise from a large fraction of the active region of the disk, from $r \sim R_\mathrm{inner}$ to $r \sim 10 \ R_\mathrm{inner} = 36 \ R_\odot$ (see Appendix \ref{app:RadLambda}). In this section, we demonstrate the strong agreement between our disk model and the data for the HIRES and NIRSPEC spectra, which we interpret as evidence of the presence of the disk atmosphere depicted in Figure \ref{fig:cartoon}.

\subsection{The Accurately-predicted Visible Range Spectra} \label{sec:HIRESModelsComp}
\begin{figure*}[!htb]
    \centering
    \includegraphics[width=0.6\linewidth]{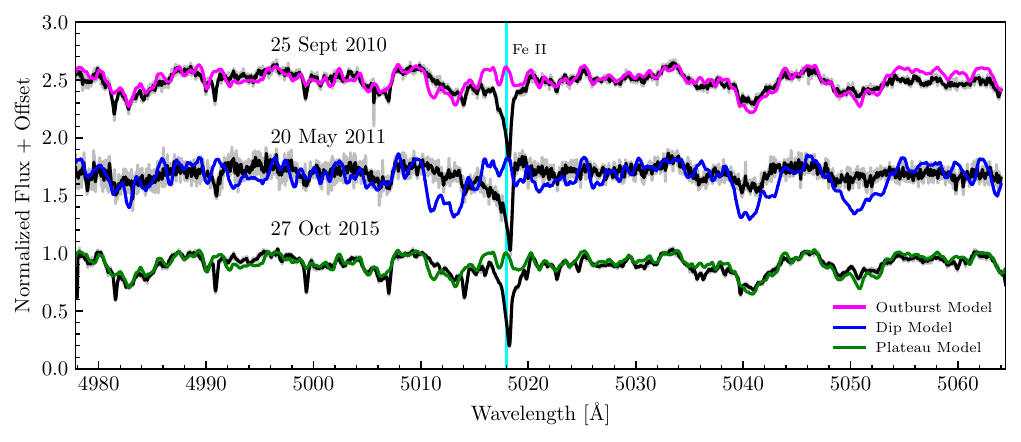}
    \includegraphics[width=0.6\linewidth]{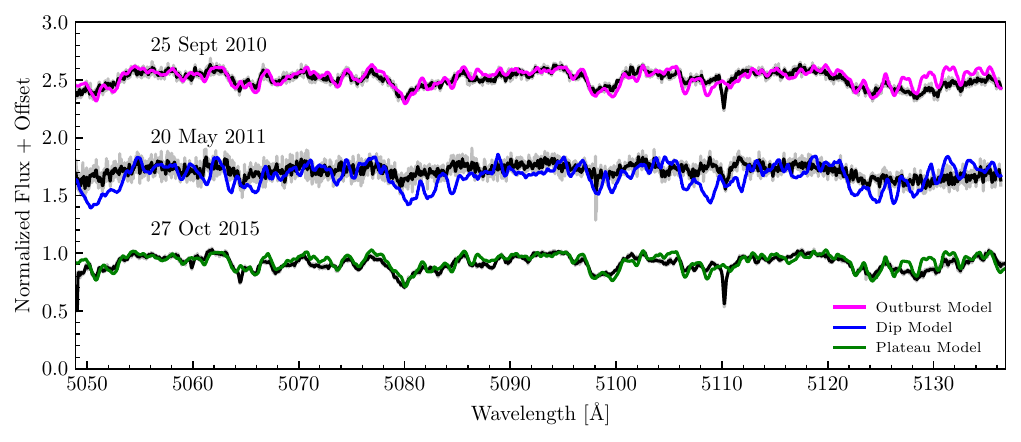}
    \includegraphics[width=0.6\linewidth]{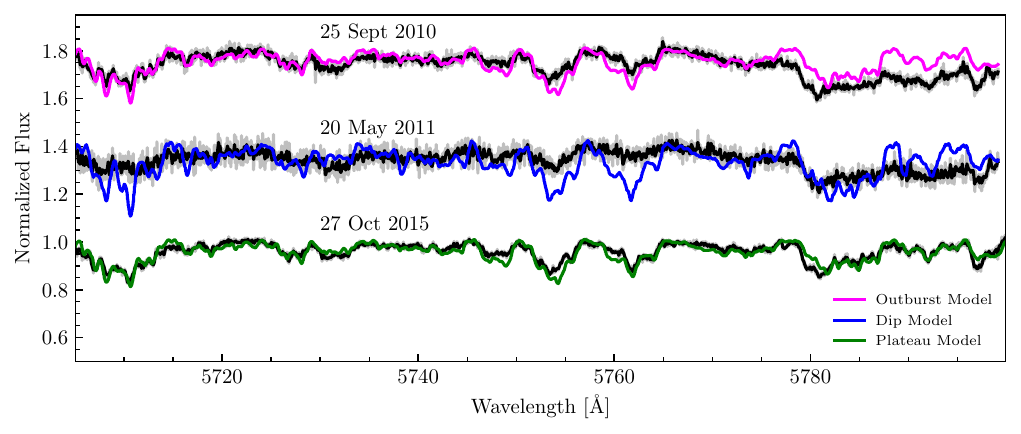}
    \includegraphics[width=0.6\linewidth]{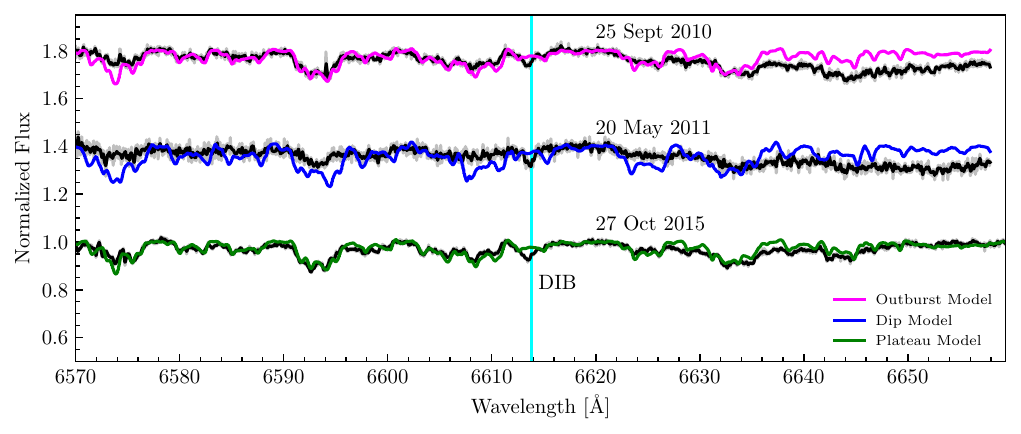}
    \includegraphics[width=0.6\linewidth]{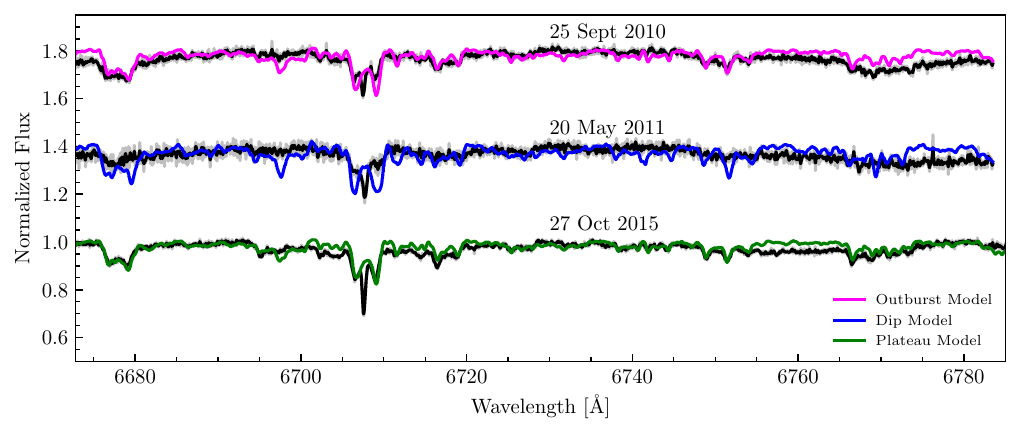}
    \caption{Selected orders from the HIRES spectra corresponding to the 3 SpeX epochs shown in Figure \ref{fig:spex_fits}. Notice the strong agreement with the disk model spectrum that was derived to fit the lower resolution infrared spectrum, across a large range of visible wavelengths that include many atomic species.  The most discrepant lines (e.g. Fe II 5018) trace wind absorption and are discussed in Section \ref{sec:windLines} while others (e.g. Li I 6707) exhibit a narrow blue-shifted absorption that grows over time and are discussed in Section \ref{sec:narrow_blue}}
    \label{fig:HIRESSpec}
\end{figure*}

Using the system parameters derived from the SED fits, we predict high resolution spectra corresponding to the three epochs for which we modeled the medium-resolution spectrophotometry. We compare these predicted models to the three HIRES spectra that were taken nearly coincident with the medium-resolution spectra. The HIRES epochs we select are: 25 Sept 2010 (outburst), 20 May 2011 (dip), and 27 Oct 2015 (plateau). The high resolution spectra show significant evolution between the outburst, dip, and plateau stages.  While we find that our models of the outburst epoch and the plateau epoch match the HIRES spectra well, the dip epoch is not very well-fit by the disk model, with the observations showing additional structure distinct from the line profiles predicted by the pure-disk model. The disk models and several of the HIRES spectral orders for the three epochs are shown in Figure \ref{fig:HIRESSpec}.
    
The outburst epoch HIRES spectrum agrees with the $T_\mathrm{max} = 5700$ K best-fit model for the outburst SED, especially toward the redder orders. The bluer orders, especially blueward of 4500 \AA, are dominated by wind absorption that is not included in our model so the fits are worse there than in the red. 

During the dip, the HIRES spectrum no longer looks like a disk spectrum. This seems surprising given the similarity between the dip epoch model and the 26 Jun 2011 SpeX spectrum. The inconsistency between the optical and near-infrared spectra may indicate that although the temperature profile still follows $T(r) \propto r^{-3/4}$, the disk atmosphere may no longer be rotating as a Keplerian disk would be. We discuss this possibility further in Section \ref{sec:interp}.

The plateau HIRES spectrum is also well-matched by the $T_\mathrm{max} = 6000$ K plateau SED model. The absorption lines in the spectrum have also recovered their disk profiles, indicating the disk atmosphere has recovered its Keplerian velocity profile. 

As was observed in the \citet{Carvalho_V960MonSpectra_2023ApJ} models of V960 Mon, the TiO bandhead at 8860 \AA\ is predicted to be quite strong in all of our models but is totally absent in the data. This mismatch between the modeled and observed spectra of FU Ori objects was noted in \citet{herbig_high-resolution_2003, PetrovHerbig_FUOriLineStructure_2008AJ} as a flaw with accretion disk models of these objects. \citet{Zhu_FUOriDifferentialRotation_2009ApJ} demonstrated that the excessively deep TiO absorption can be revolved by lowering the assumed $T_\mathrm{max}$ in the model so that the flatter part of the temperature distribution constributes relatively more in the visible range. The excess TiO absorption in our model of HBC 722 is surprising because the target is much cooler than V960 Mon, and, following the logic of \citet{Zhu_FUOriDifferentialRotation_2009ApJ}, would not have such strong predicted TiO. This consistent TiO overprediction highlights a need to better understand the location in the disk where these features are formed to explain how the conditions differ from those assumed in our model. 


\subsection{The Accurately-predicted NIR Spectra} \label{sec:NIRSPECModelsComp}
The disk model also predicts the line profiles in the 30 Oct 2023 NIRSPEC spectrum, which spans 1.0-2.5 $\mu$m. Though the spectrum was taken later than the plateau HIRES and SpeX epochs, the line profiles in the 1.0-2.5 $\mu$m wavelength range are relatively insensitive to the small difference in accretion rate between the two epochs.

\begin{figure*}[!htb]
    \centering
    \includegraphics[width=0.48\linewidth]{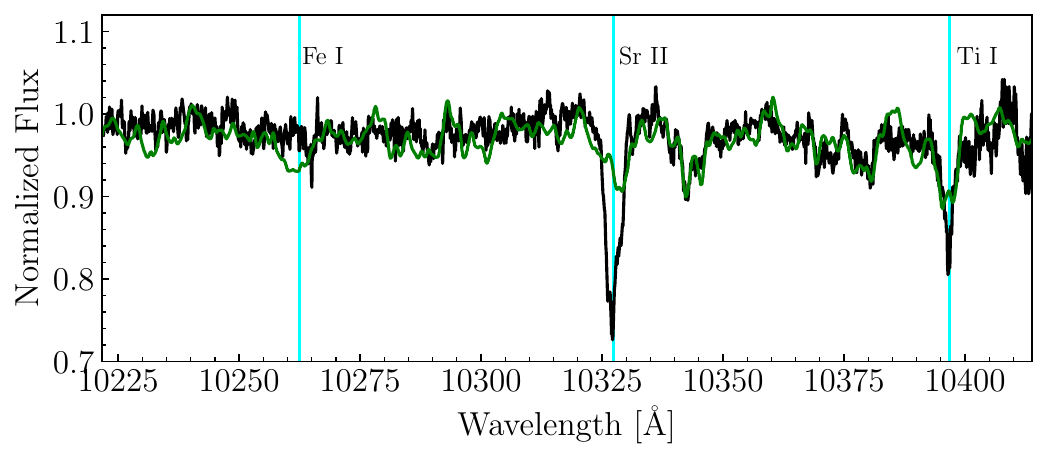}
    \includegraphics[width=0.48\linewidth]{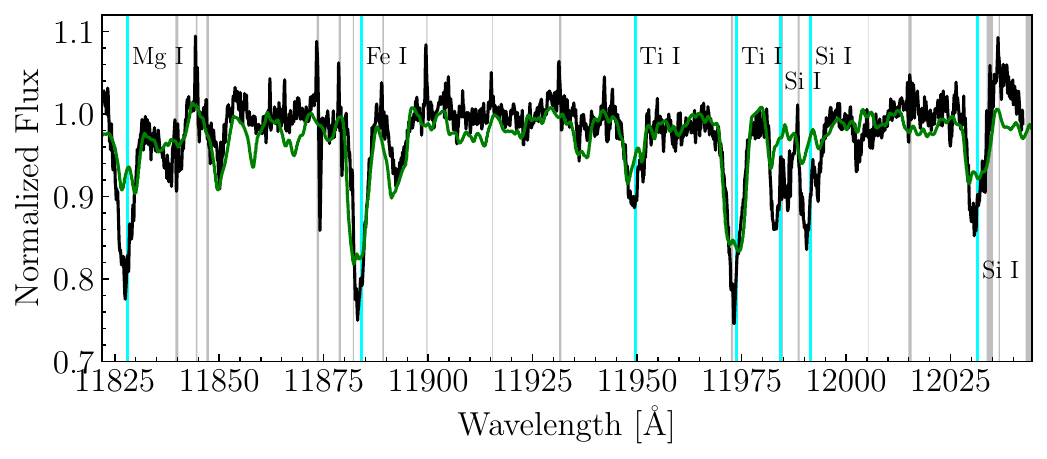}
    \includegraphics[width=0.48\linewidth]{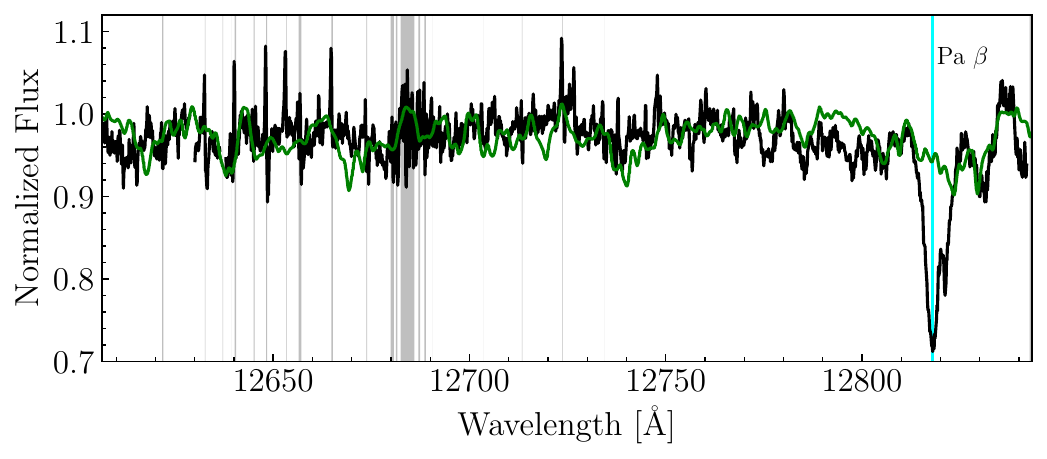}
    \includegraphics[width=0.48\linewidth]{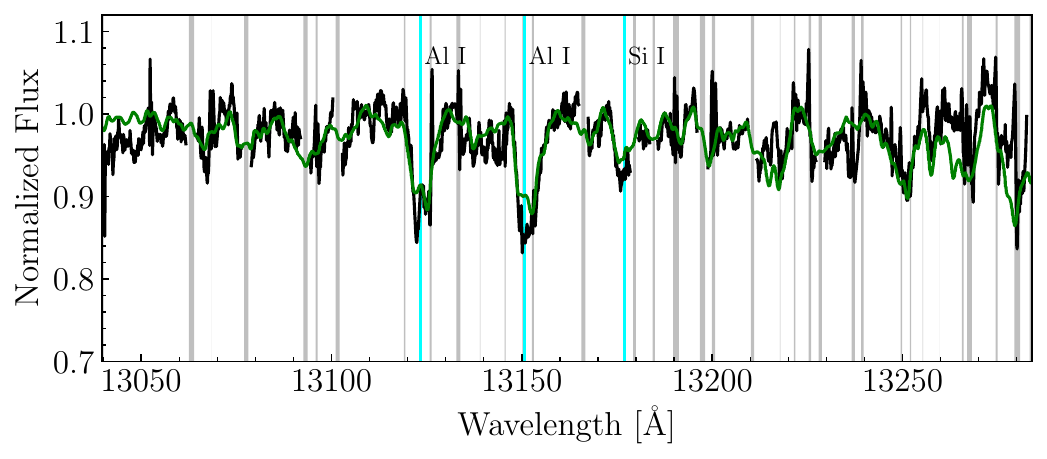}
    \includegraphics[width=0.48\linewidth]{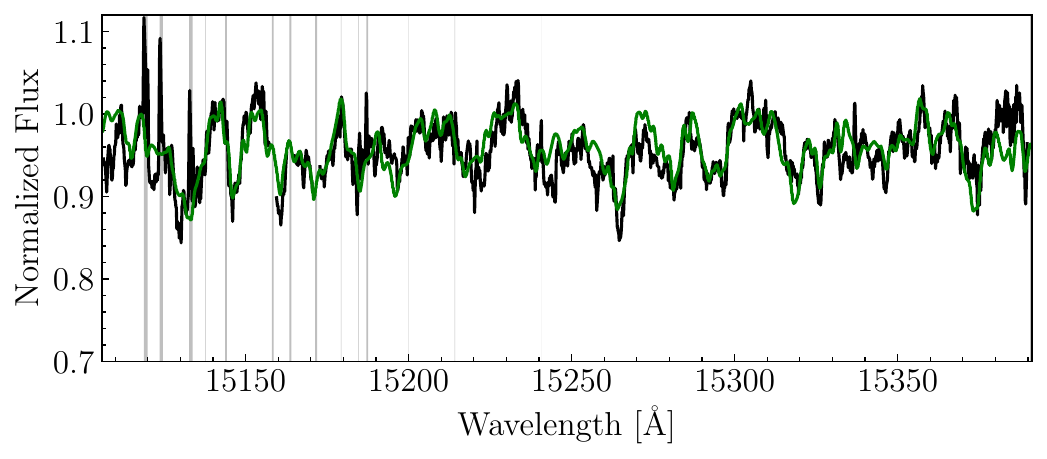}
    \includegraphics[width=0.48\linewidth]{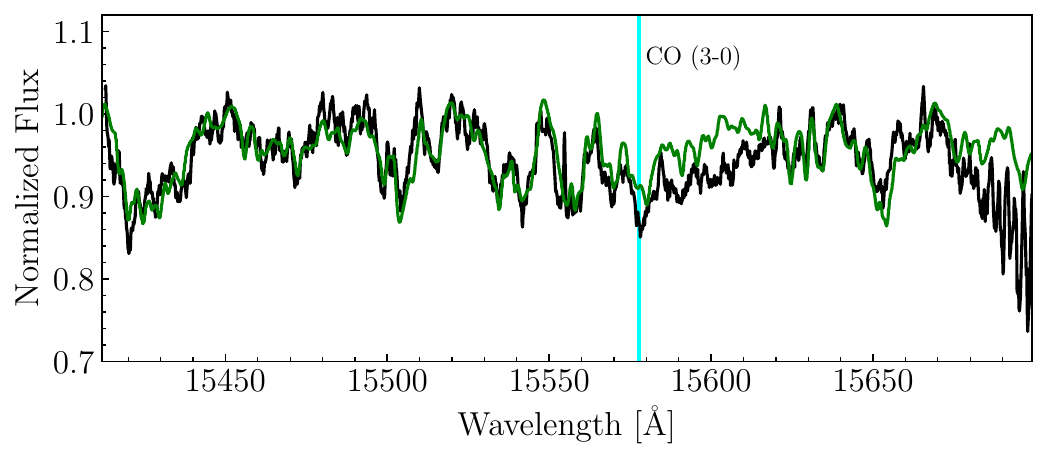}
    \includegraphics[width=0.48\linewidth]{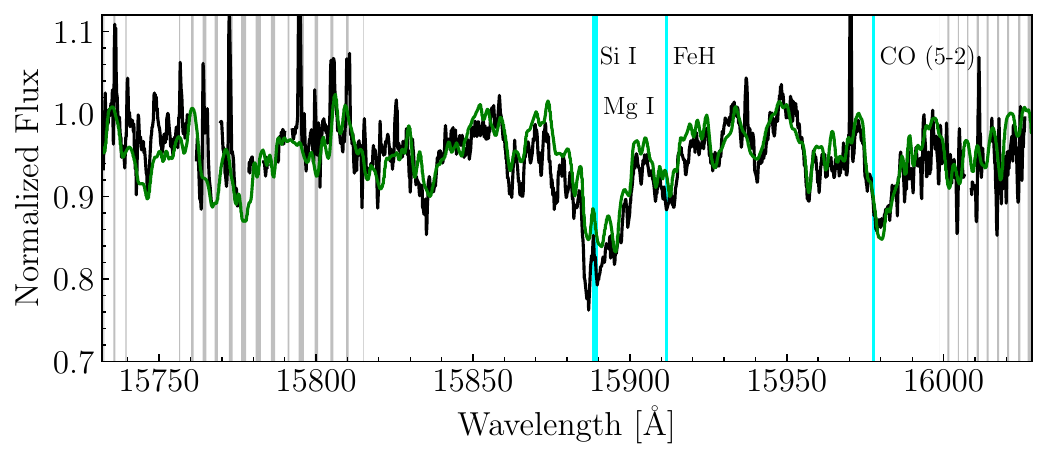}
    \includegraphics[width=0.48\linewidth]{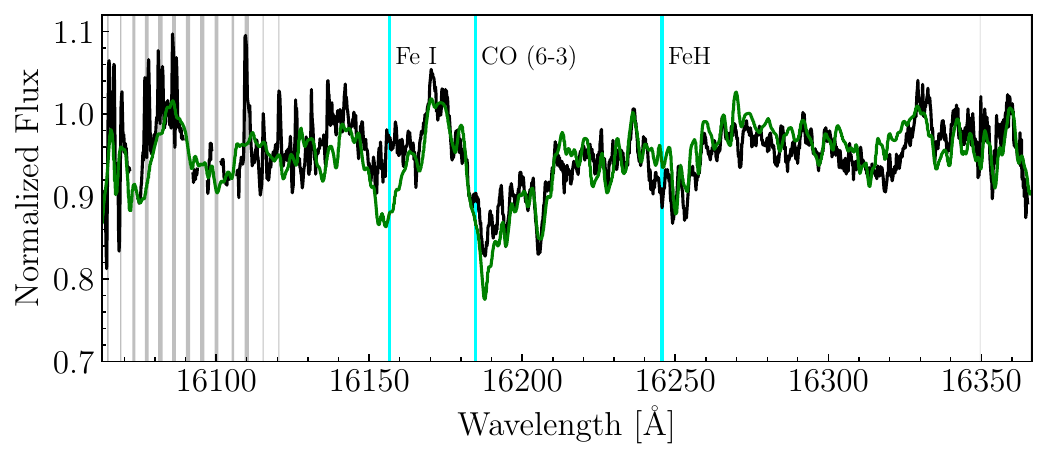}
    \includegraphics[width=0.48\linewidth]{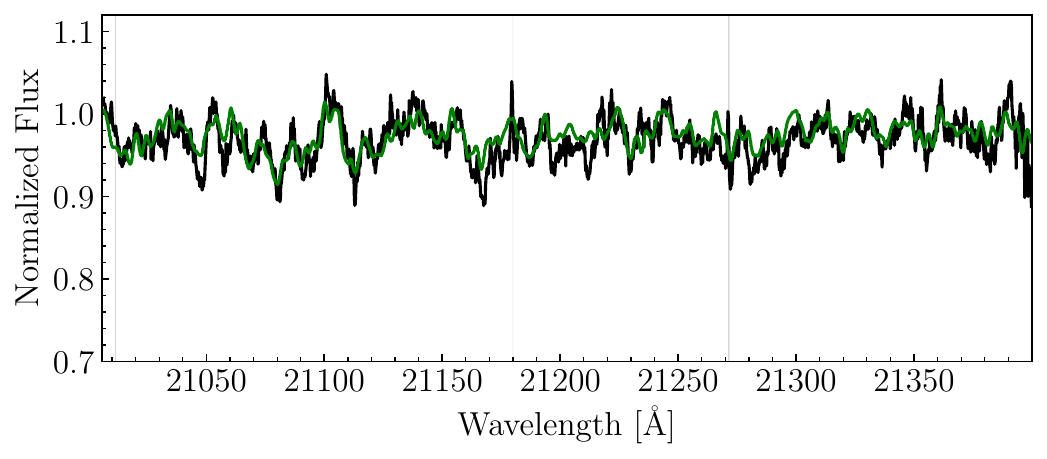}
    \includegraphics[width=0.48\linewidth]{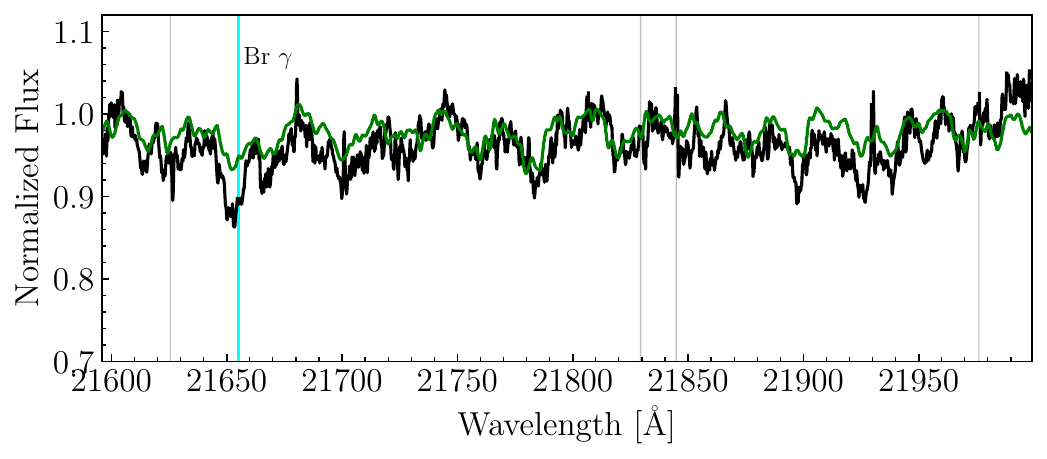}

    \caption{Selected orders from the NIRSPEC spectrum (in black), spanning 1.0-2.2 $\mu$m.  Grey vertical lines mark locations of strong telluric absorption. The plateau epoch model is shown in green. The lines tracing the wind outflow, such as Pa$\beta$ and Br$\gamma$, as well as several disk atmosphere lines including Ti I, Mg I, and Fe I, are marked with cyan vertical lines.}
    \label{fig:NIRSPECOrders}
\end{figure*}

Representative orders of the NIRSPEC spectrum are shown in Figure \ref{fig:NIRSPECOrders}. For most of the atomic and molecular features across the entire NIRSPEC range, the disk model matches the data very closely. Major exceptions to this are wind-sensitive features \citep[e.g. He I 10830 and the H absorption lines,][]{Dupree_HE10830HotWind_2005ApJ}.
or low excitation potential (EP) lines tracing a disk wind (e.g., \ion{Ti}{1} 10396 \AA\ and \ion{Fe}{1} 11879 \AA, see Section \ref{sec:narrow_blue}). The excess absorption in Sr II 10326 relative to the disk model is an indicator of the very low gravity in the disk \citep{Rayner_IRTF_2009} and a result of the strong departure in the line from the local thermodynamic equilibrium (LTE) assumptions in the model spectra \citep{Korotin_SrII_NLTE_2020MNRAS}. 

The lines that trace the disk absorption in the NIR tend to have higher EP than the disk tracing lines in the visible range (typically $>$ 2 eV). It is notable that across the entire NIR range shown in Figure \ref{fig:NIRSPECOrders}, the line profiles are matched in both depth and width by our disk model. This, combined with the good model-data agreement in HIRES, indicates that the plane-parallel disk atmosphere with Keplerian rotation is present in the system for the entire region of the disk that is probed by the 0.4-2.5 $\mu$m continuum. 

The wind features in the NIR do not show the same dramatic high-velocity absorption we see in the HIRES spectra (discussed in Section \ref{sec:windLines}). The absorption in both Pa$\beta$ and Br$\gamma$ are similar to the most recent plateau HIRES spectra, as might be expected. Both lines show blue-shifted absorption out to $-200$ km s$^{-1}$, consistent with the absorption velocities seen in the later H$\beta$ profiles, which indicates they are tracing the same massive wind (see Sections \ref{sec:highVwind} and \ref{sec:diskWind}). The He I 10830 \AA\ line shows a symmetric disk profile, indicating the line is tracing mostly the disk atmosphere and potentially the launching point of the rotating wind. 

The CO absorption in the NIR spectra is particularly illuminating. Although the modeled CO ($\Delta \nu=3$) bands are consistent with those observed in the NIRSPEC spectrum (see the analysis in Section \ref{sec:COAbs}), there is a significant discrepancy between the modeled CO ($\Delta \nu = 2$) bands and the data, where the observed CO ($\Delta \nu = 2$) bands show deep, narrow, blue-shifted absorption that is not predicted by the model. We attribute this absorption to a low-velocity disk wind, that is increasing in density over time. We show the CO (2-0) and (3-1) bands and contrast them to the disk model in Section \ref{sec:COAbs} and we discuss their connection to the disk wind in Section \ref{sec:diskWind}.

\begin{figure*}[!htb]
    \centering
    \includegraphics[width = 0.32\linewidth]{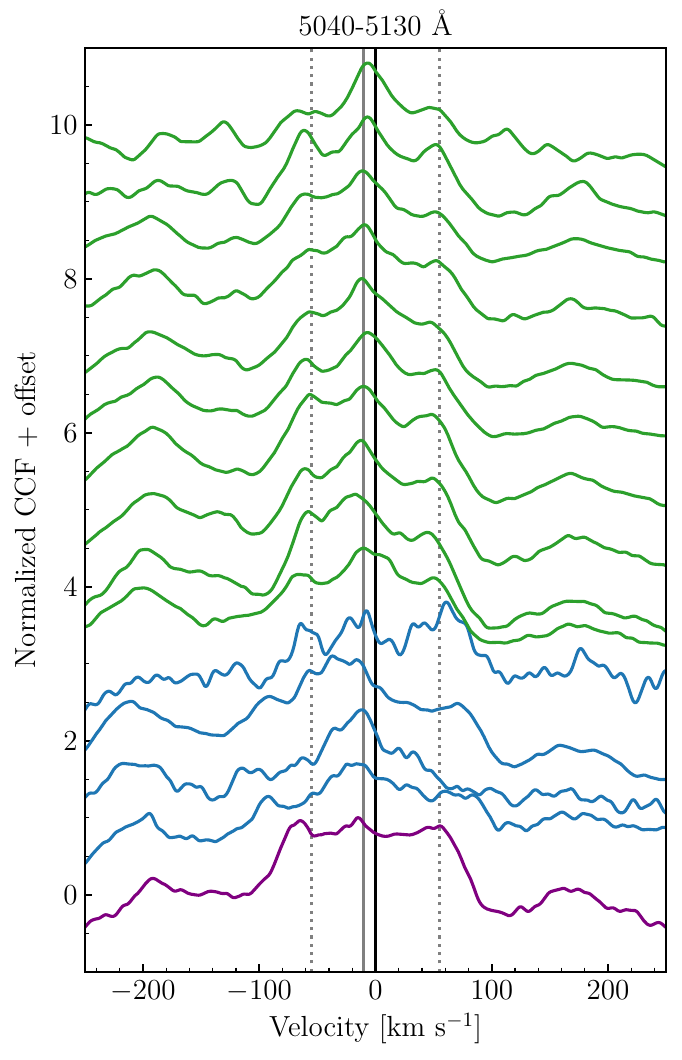}
    \includegraphics[width = 0.32\linewidth]{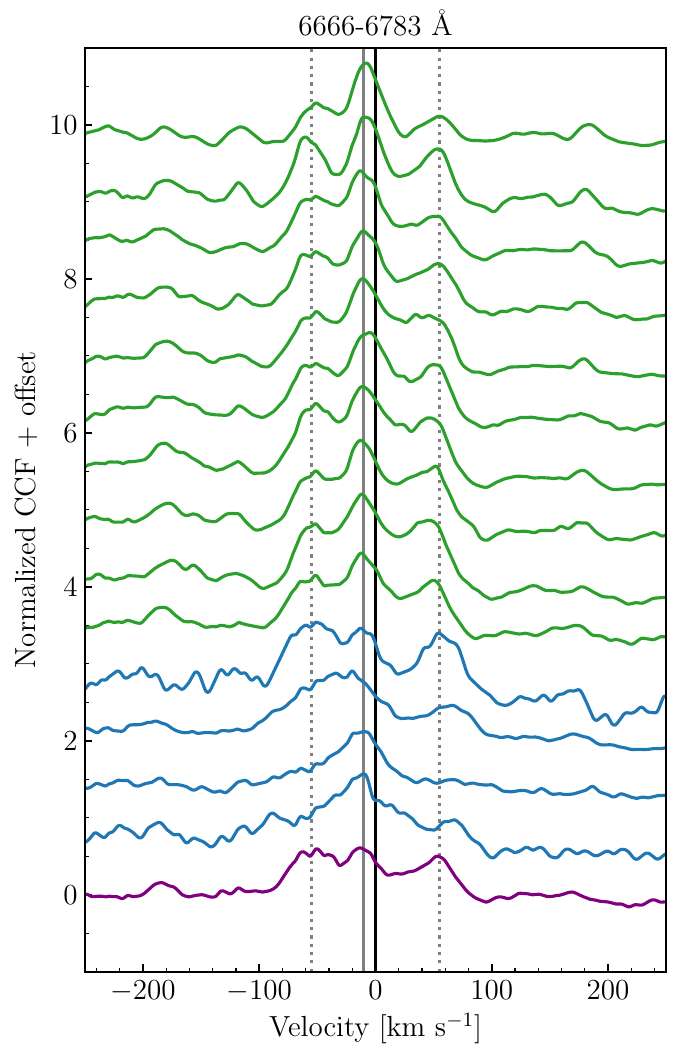}
    \includegraphics[width = 0.32\linewidth]{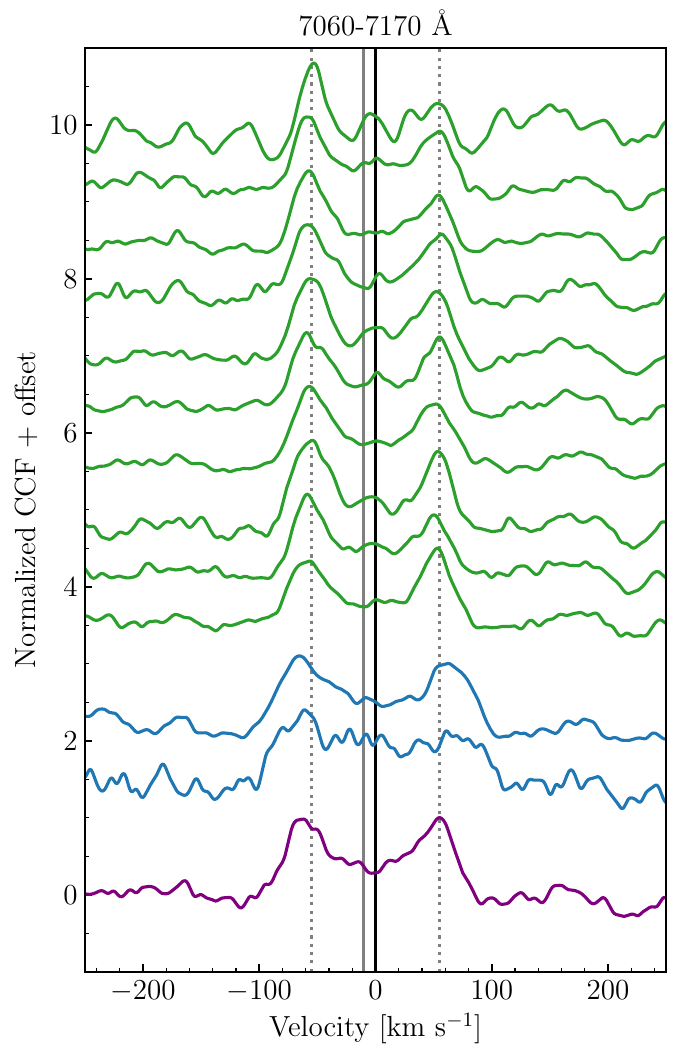}
    \caption{CCFs computed for 3 HIRES orders that are covered in 15/16 HIRES epochs, showing the evolution of the disk absorption over time. The purple profile indicates the 25 Sept 2010 outburst epoch, the blue profiles are those spanning the dip (Dec 2010 - Jan 2013) and the green profiles show the epochs in the plateau (2013-2022). The dotted grey lines show the approximate velocities of the peak absorption ($\pm 55$ km s$^{-1}$), whereas the solid grey line marks the $-10$ km s$^{-1}$ at which we see a narrow absorption feature growing over time (see Section \ref{sec:narrow_blue}). For most of the epochs and across the HIRES wavelength range, the line profiles are a consistent width, indicating an unchanging $v_\mathrm{max}$ over time.}
    \label{fig:CCFs_HIRES}
\end{figure*}

\subsection{Cross Correlation Analysis of the HIRES Spectra} \label{sec:ccfs}
To highlight the profile evolution of atomic lines in the HIRES spectra, we computed the cross-correlation function (CCF) between different spectral orders and a $T_\mathrm{eff} = 4500$ K, log$g=1.5$ PHOENIX model stellar spectrum\footnote{$T_\mathrm{eff} = 4500$ K is representative of the expected region in the disk from which the 5000-8000 \AA\ continuum arises (see Appendix \ref{app:RadLambda}). The log$g$ = 1.5 is similarly a typical expected gravity in this region. }. We chose 3 orders that appear in almost all HIRES epochs and represent a large fraction of the HIRES wavelength coverage: 5040-5130 \AA, 6666-6783 \AA, and 7060-7170 \AA. The CCFs, shown in Figure \ref{fig:CCFs_HIRES}, have clear disk line profiles in the outburst and plateau epochs. During the dip, as mentioned above, the profiles differ from disk absorption profiles and take on a more centrally absorbed structure. 

As has been observed for many of the classical FU Ori objects \citep{herbig_high-resolution_2003}, there is little wavelength-dependence in the width of the line profiles across the visible range. This may in part be due to the phenomenon described in \citet[][see also Appendix \ref{app:RadLambda}]{Carvalho_V960MonSpectra_2023ApJ} whereby atomic lines that probe different temperature components do not simply follow the expected continuum wavelength dependence. There are several lines in the redder visible orders (e.g., \ion{Ca}{2} 8927 \AA) that probe hotter regions of the disk than some of the lines in the bluer orders (e.g., \ion{Fe}{1} 5328 \AA). 

Our detailed disk modeling, in addition to previous work \citep{welty_FUOriV1057CygDiskModelAndWinds_1992ApJ, Zhu_FUOriDifferentialRotation_2009ApJ}, has shown that the rotational broadening of atomic lines in the optical wavelength range has a very weak wavelength dependence and that NIR observations are necessary to observe the phenomenon.

In the outburst and plateau epochs, the line profiles remain strongly disk-like and have peaks at $\pm 55$ km s$^{-1}$.
At later times in the plateau, we begin see a strong, narrow, blue-shifted absorption component that grows in strength against the disk absorption, which we discuss in Section \ref{sec:narrow_blue}. The centrally absorbed lines seen in the 13 Dec 2010 and 20 May 2011 CCFs are indicative of a departure from Keplerian rotation in these epochs and potentially increased turbulence in the disk atmosphere (see Section \ref{sec:thinDisk}).

\section{Lines with Excess Absorption or Emission Relative to the Disk Model} \label{sec:excess}
As in V960 Mon, the high resolution spectra of HBC 722 contain several lines that show absorption in excess of that predicted by the disk model. In this section we highlight several of these lines and describe their time-evolution. 

We divide the lines into 2 main groups: the wind-tracing lines, many of which show P-Cygni-like profiles and show both high- and low-velocity blue-shifted excess absorption, in addition to some emission, and a series of atomic lines with low-velocity and narrow blue-shifted absorption. We then also identify the molecular absorption that appears suddenly in the 2021 HIRES spectrum and the similar narrow CO absorption in the NIR spectra.

\subsection{The Variable Profiles of Wind-tracing Lines} \label{sec:windLines}
The lines typically known to trace outflows, H$\beta$, H$\alpha$, \ion{Fe}{2} 5018 \AA, \ion{Mg}{1} 5183 \AA, the Na I D doublet, and the Ca II H \& K and Infrared Triplet (IRT) all show significant post-outburst evolution. Although in other YSO disk systems these features are also known to trace accretion, in FU Ori objects they are only known to trace outflows \citep{Lima_HalphaLineEmissionWinds_2010A&A}. We consider the line profiles of HBC 722 within the three main stages of post-outburst evolution discussed above: outburst (the 25 Sept 2010 epoch), dip (13 Dec 2010 - 06 Jan 2012), and plateau (27 Dec 2013 - 07 Aug 2021). 

The outburst line profiles are generally similar to the plateau profiles, so we will discuss them together, though there remain some subtle differences between them that we will address. The spectrum of HBC 722 during the dip differs strongly from the outburst and plateau epochs, so we will address the line profile behavior during the dip separately. We will also then highlight the persistant low-velocity absorption present in all of the wind-tracing lines and the emission component of the H$\alpha$ and Ca II IRT lines.

Regarding the Ca II lines, there are two important notes. First, in FU Ori objects, the strong blue-shifted wind absorption in the H$\epsilon$ line at 3970 \AA\ blends into the red side of the Ca II H line profile. For our discussion we will focus on the Ca II K line since it is isolated. Second, of the lines in the Ca II IRT, we will focus on 8498 \AA, because it is the only one covered in almost all of the HIRES spectra.

\subsubsection{The Outburst and Plateau Profiles} \label{sec:wind_outburst_plateau}

\begin{figure*}[!htb]
    \centering
    \hfill
    \includegraphics[width = 0.32\linewidth]{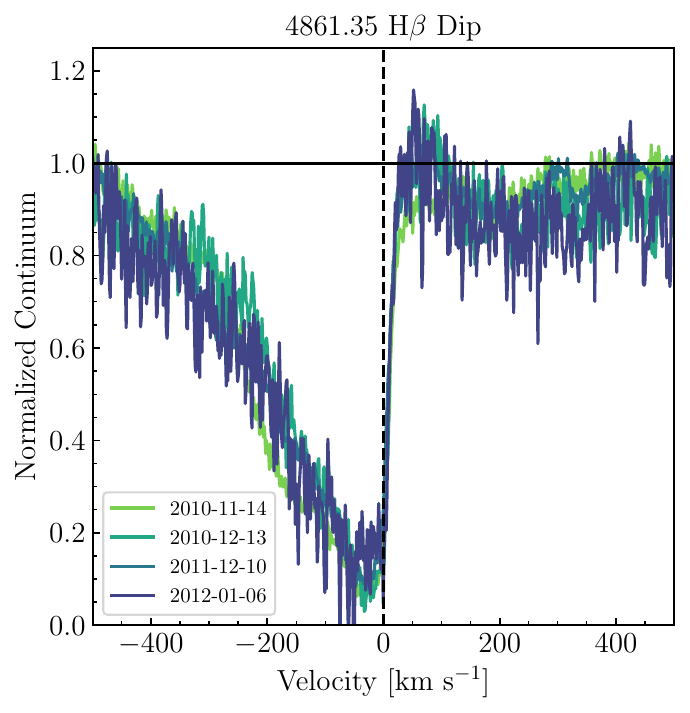}
    \includegraphics[width = 0.32\linewidth]{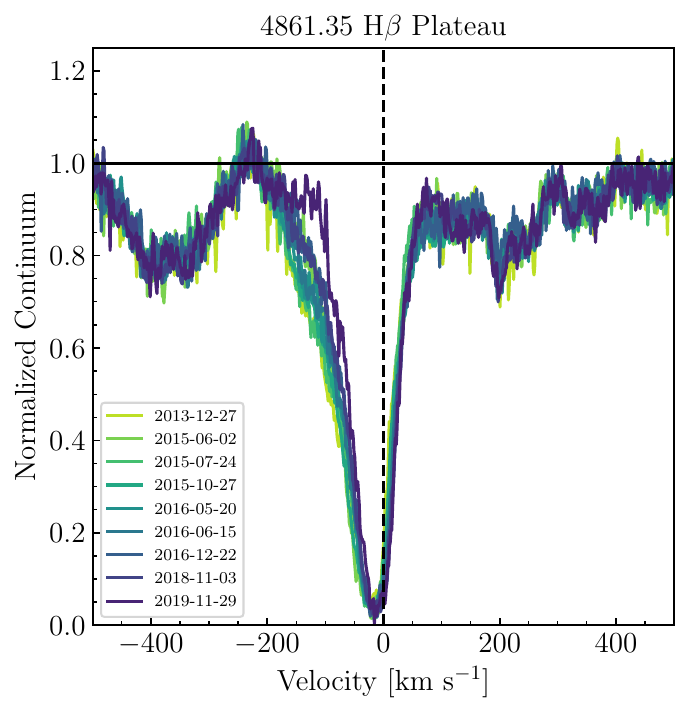}
    \newline
    \includegraphics[width = 0.32\linewidth]{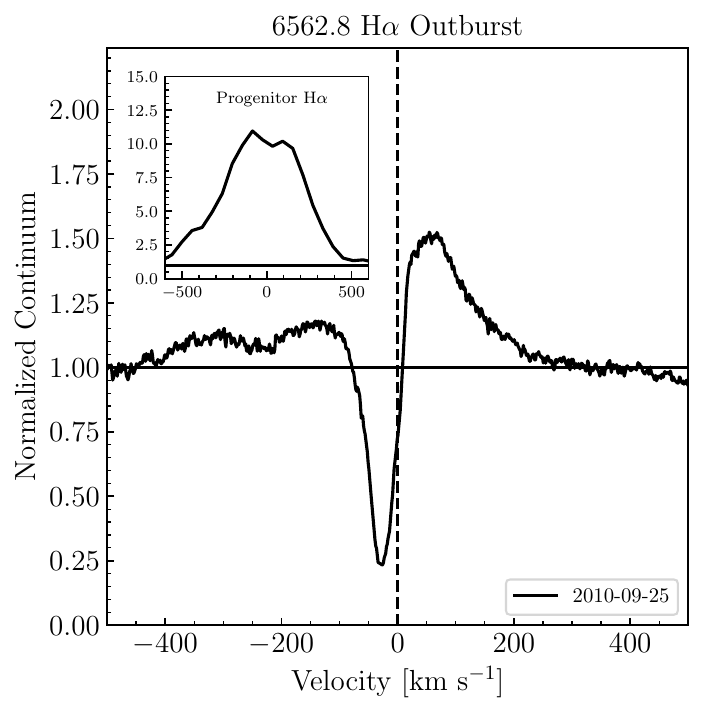}
    \includegraphics[width = 0.32\linewidth]{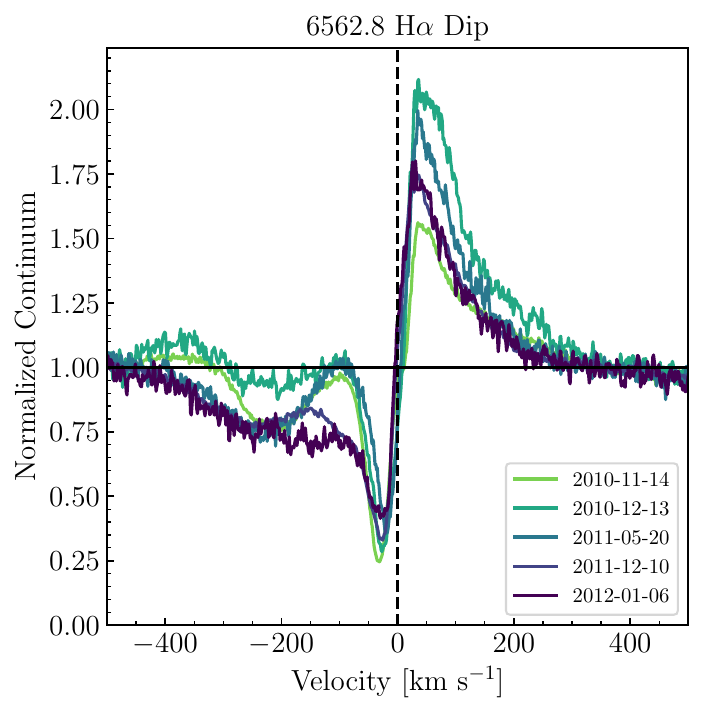}
    \includegraphics[width = 0.32\linewidth]{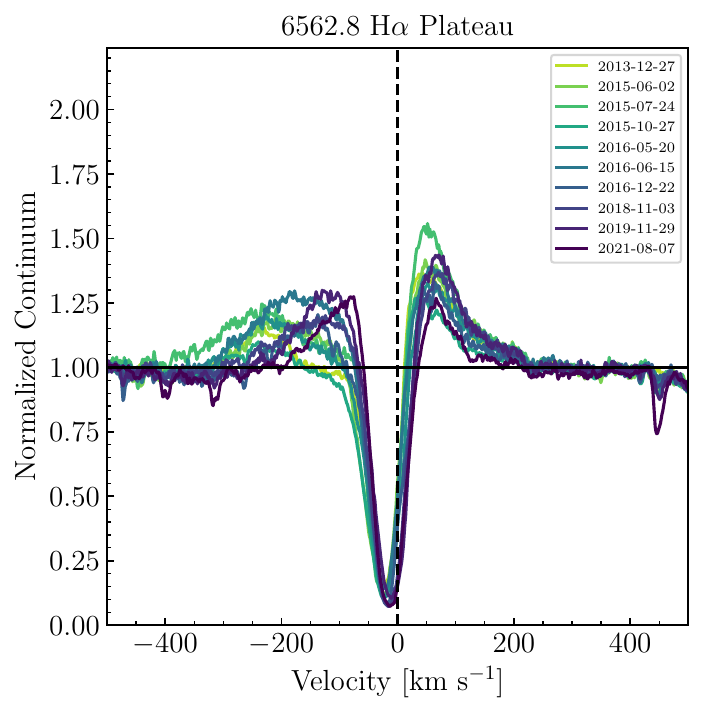}
    \caption{H$\beta$ (upper row) and H$\alpha$ (lower row) lines in the HBC 722 HIRES spectra. The H$\beta$ line is not covered in the outburst epoch, so only the dip (left panel) and plateau (right panel) epochs are shown. The outburst H$\alpha$ profile is shown in the inset in the bottom left panel. The lines are highly variable as the target evolves but all consistently trace a massive outflow with $v = -10$ to $-30$ km s$^{-1}$.}
    \label{fig:BalmerLines}
\end{figure*}

\begin{figure*}[!htb]
    \centering
    \includegraphics[width = 0.32\linewidth]{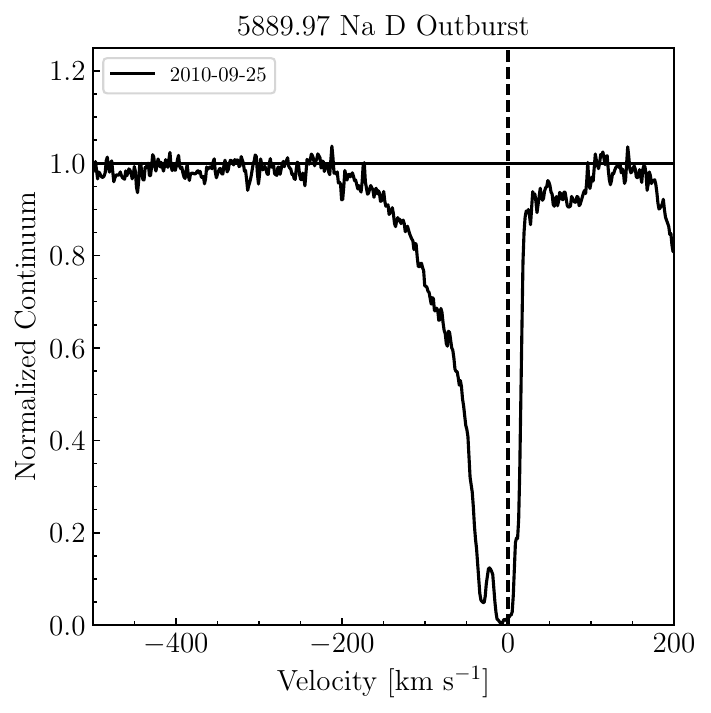}
    \includegraphics[width = 0.32\linewidth]{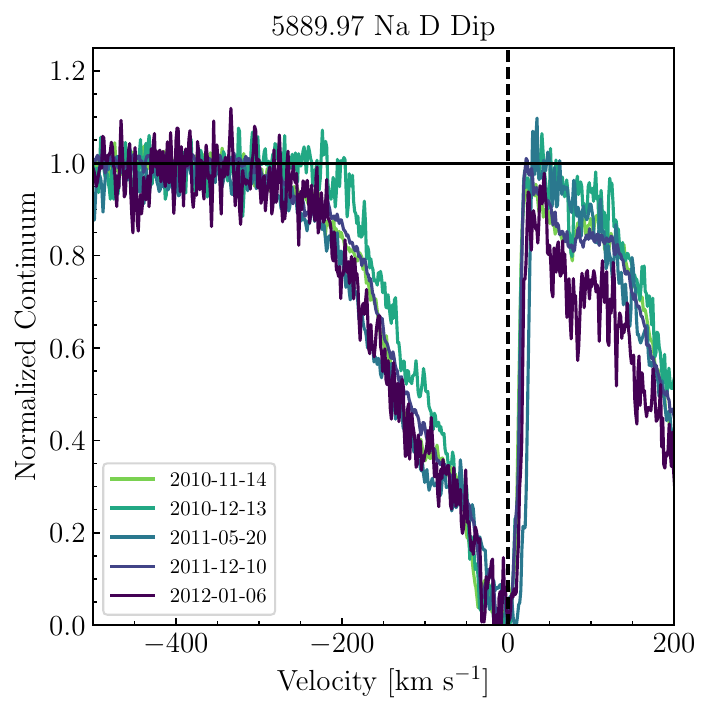}
    \includegraphics[width = 0.32\linewidth]{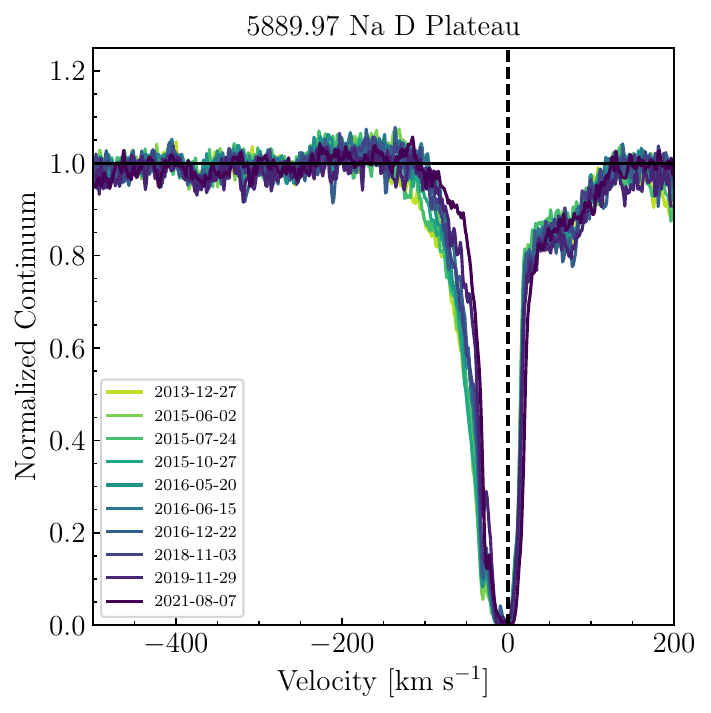}
    \caption{The evolution \ion{Na}{1} D 5889 line in the HIRES spectra shown for the 3 main stages of the lightcurve: the outburst (left), the dip (center), and the plateau (right). Notice the line profile during the outburst epoch looks very similar to the profiles during the plateau. However, during the dip, the profiles are dominated by wind absorption and show almost no disk broadening in the red wing. }
    \label{fig:NaD}
\end{figure*}

As mentioned above, all of the wind line profiles look remarkably similar between the outburst epoch the plateau epochs. The lines all have multi-component blue-shifted absorption profiles and, in general, the red-shifted absorption is contained within the velocity range of the disk profile (see the absorption in the H$\alpha$ line in Figure \ref{fig:BalmerLines} compared with the disk profile CCFs in Figure \ref{fig:CCFs_HIRES}), indicating the absorbing material is moving slower than the $v_\mathrm{max}$ of the disk model. We detail below the components of each of the wind lines and how they evolve during the outburst and plateau epochs. The evolution of the profiles during the dip epochs is discussed in Section \ref{sec:wind_dip}. 

The H$\alpha$ profiles are the most variable and complex of the wind lines, as can be seen in Figure \ref{fig:BalmerLines}. The blue-shifted H$\alpha$ absorption has two primary components in the outburst epoch: one at high velocity, absorbing from $-150$ km s$^{-1}$ to $-300$ km s$^{-1}$ and one at lower velocity, absorbing from $0$ km s$^{-1}$ to $-100$ km s$^{-1}$. The high velocity component is not apparent in the plateau profiles, though in the profiles prior to 2016 there is a weak intermediate velocity absorption component ranging from $-100$ km s$^{-1}$ to $-250$ km s$^{-1}$, which disappears after 2016. The low-velocity component persists almost unchanged and is especially deep, with a maximum absorption depth of 0.75 at $-30$ km s$^{-1}$ that deepens during the plateau to almost 0.9 and slows to $-15$ km s$^{-1}$.  

The spectrograph settings in the outburst epoch did not cover H$\beta$; during the plateau, however, the line profiles show only the slow wind component at $-15$ km s$^{-1}$ that is persistent in H$\alpha$. With an absorption depth of 0.95, the slow component in H$\beta$ is even deeper than in H$\alpha$. 

The Na I D lines (shown in Figure \ref{fig:NaD}) also primarily trace this deep, slow absorption, and are almost totally absorbed from $0$ km s$^{-1}$ out to $-30$ km s$^{-1}$. At outburst, the Na I D lines have a higher velocity component from $-100$ km s$^{-1}$ to $-200$ km s$^{-1}$. In the plateau spectra, the faster component has disappeared and only the slow component remains. On the red side of the line, both outburst and plateau spectra show a disk-like absorption profile, though the absorption is around $2\times$ deeper in the plateau than at outburst. 

\begin{figure*}[!htb]
    \centering
    \includegraphics[width = 0.32\linewidth]{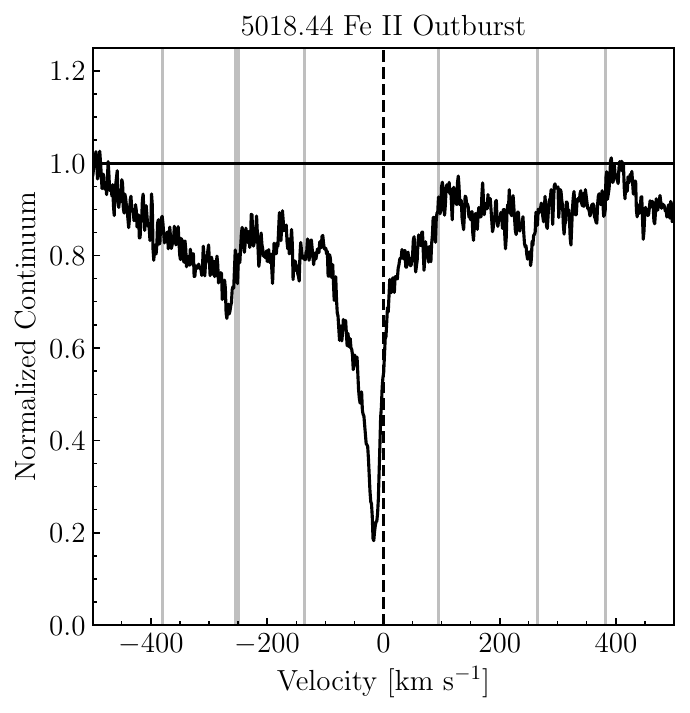}
    \includegraphics[width = 0.32\linewidth]{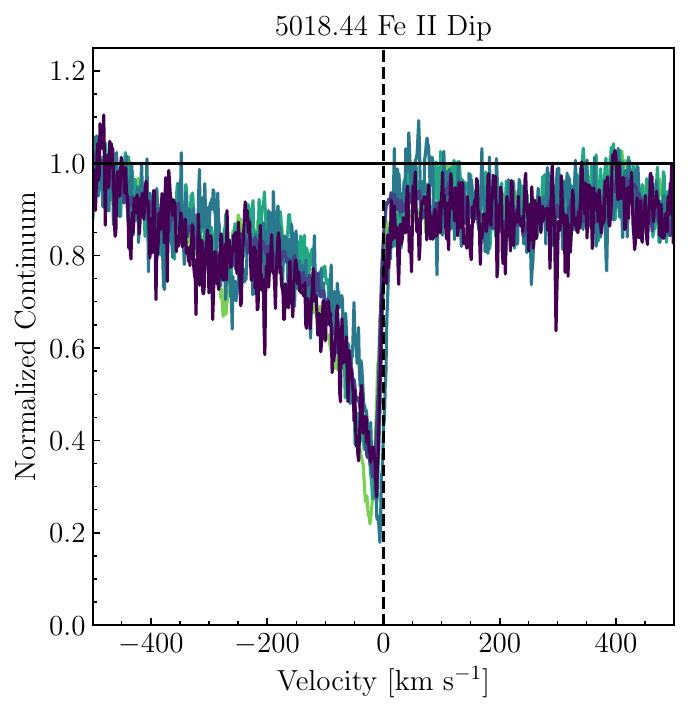}
    \includegraphics[width = 0.32\linewidth]{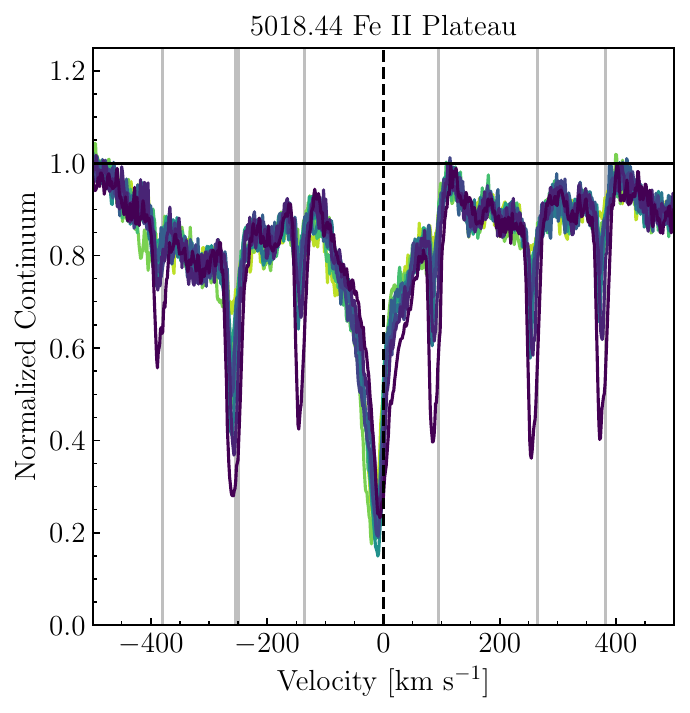}
    \includegraphics[width = 0.32\linewidth]{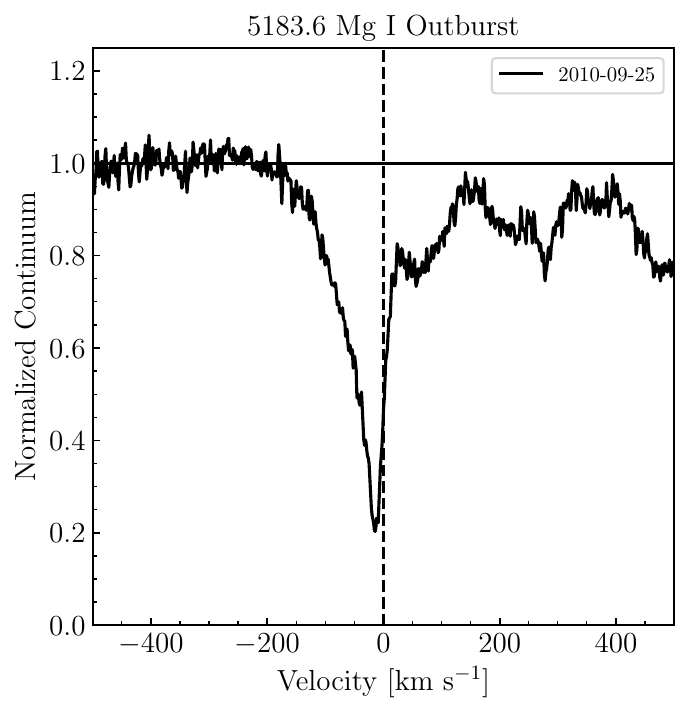}
    \includegraphics[width = 0.32\linewidth]{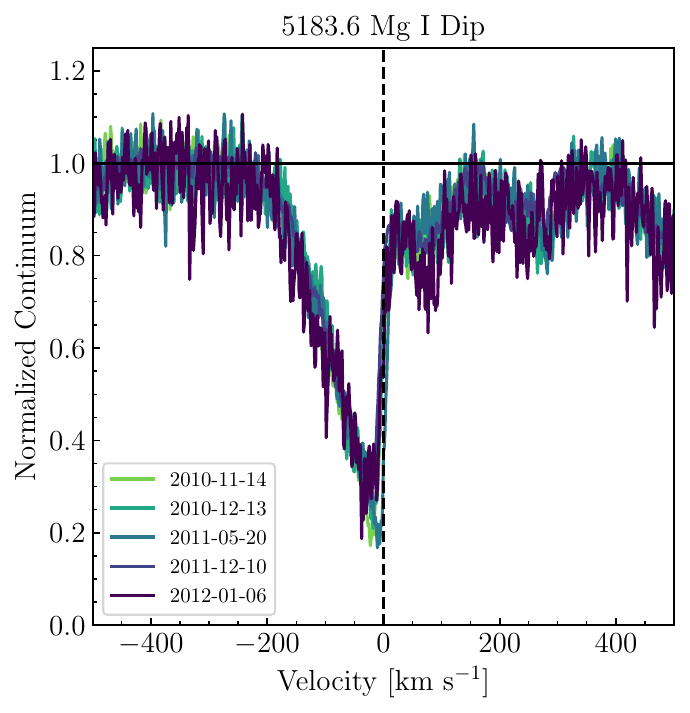}
    \includegraphics[width = 0.32\linewidth]{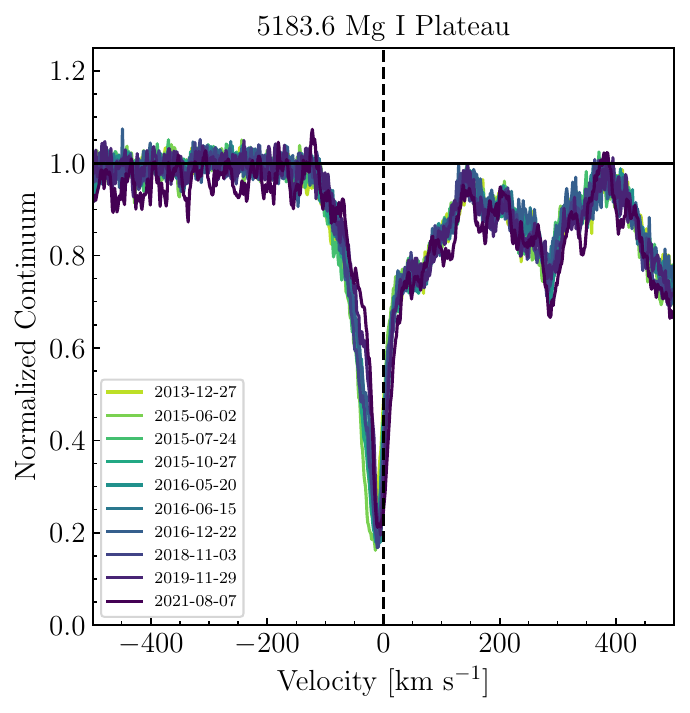}
    \includegraphics[width = 0.32\linewidth]{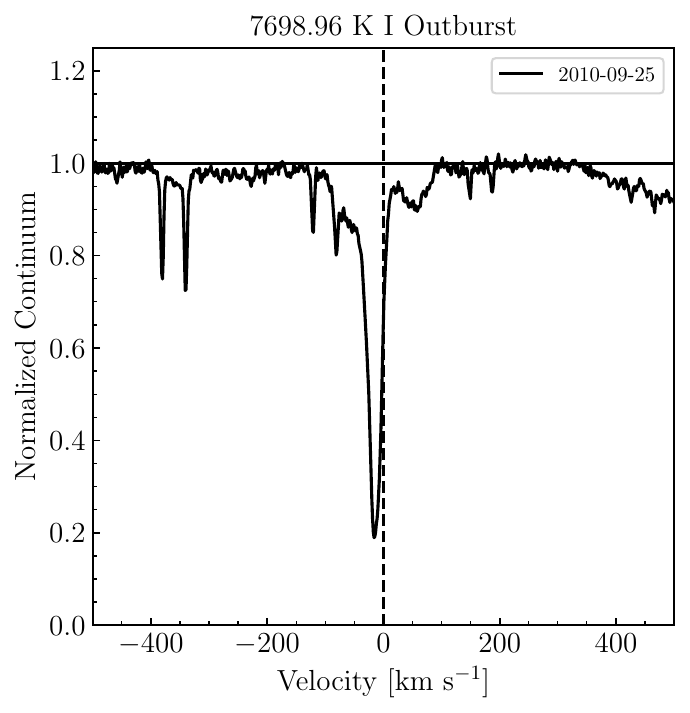}
    \includegraphics[width = 0.32\linewidth]{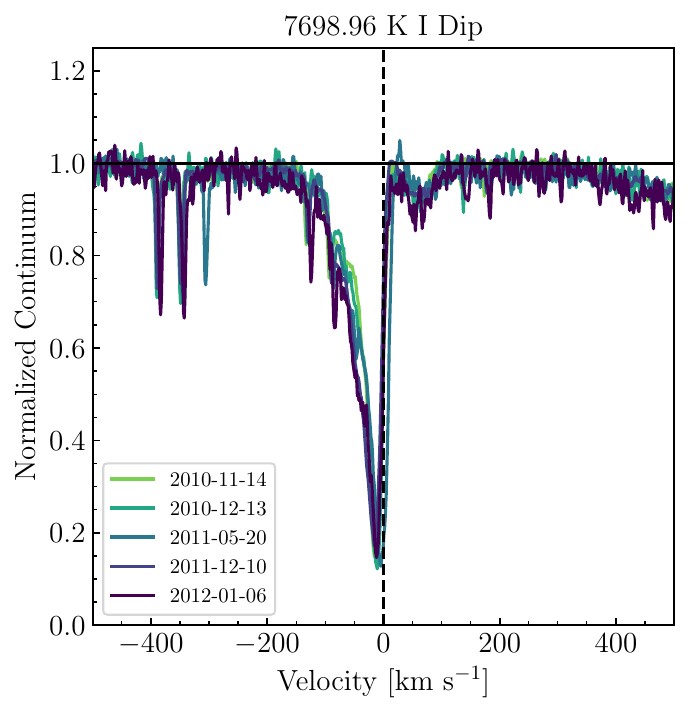}
    \includegraphics[width = 0.32\linewidth]{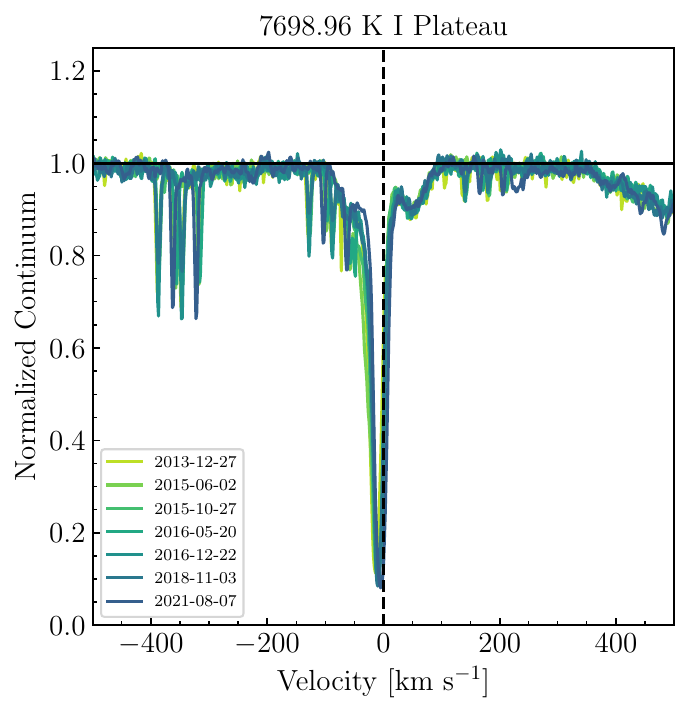}
    \caption{The evolution \ion{Fe}{2} 5018 (upper row) and Mg I 5183 (middle row) and \ion{K}{1} 7699 (lower row) lines in the HIRES spectra shown for the 3 main stages of the lightcurve: the outburst (left), the dip (center), and the plateau (right). Notice the line profiles during the outburst epoch look very similar to the profiles during the plateau. However, during the dip, the profiles are dominated by wind absorption and show almost no disk broadening in the red wing. The narrow, blue-shifted absorption traced by several Ti I lines (see Section \ref{sec:narrow_blue}) can be seen growing stronger in the plateau epochs of the Fe II 5018 profiles and their rest-wavelengths are marked with grey vertical lines. The epochs of the Fe II 5018 line are the same as the Mg I 5183 line, so we only print them once. In the \ion{K}{1} panels, the narrow absorption at $-80$, $-130$, $-350$, and $-400$ km s$^{-1}$ is due to telluric features and should be disregarded.
    }
    \label{fig:FeIILine}
\end{figure*}

The Fe II 5018, Mg I 5183, and K I 7899 (Figure \ref{fig:FeIILine}) lines evolve similarly in our spectra. All three lines have a relatively unchanging $-15$ km s$^{-1}$ absorption component with an absorption depth of 0.8. The Fe II and K I lines lack the $v < -100$ km s$^{-1}$ absorption present in the Mg II line at outburst (though it is also not in the Mg II plateau profiles). The red wing of the lines has a disk-like profile in both epochs, though its "boxiness" is more pronounced in the outburst epoch. The most dramatic change in both line profiles is the emergence of the narrow Ti I and Fe I absorption, which appears only after the dip and grows dramatically over time in the plateau stage. This absorption is discussed in more detail in Section \ref{sec:narrow_blue}. 

\begin{figure*}[!htb]
    \centering
    \includegraphics[width = 0.32\linewidth]{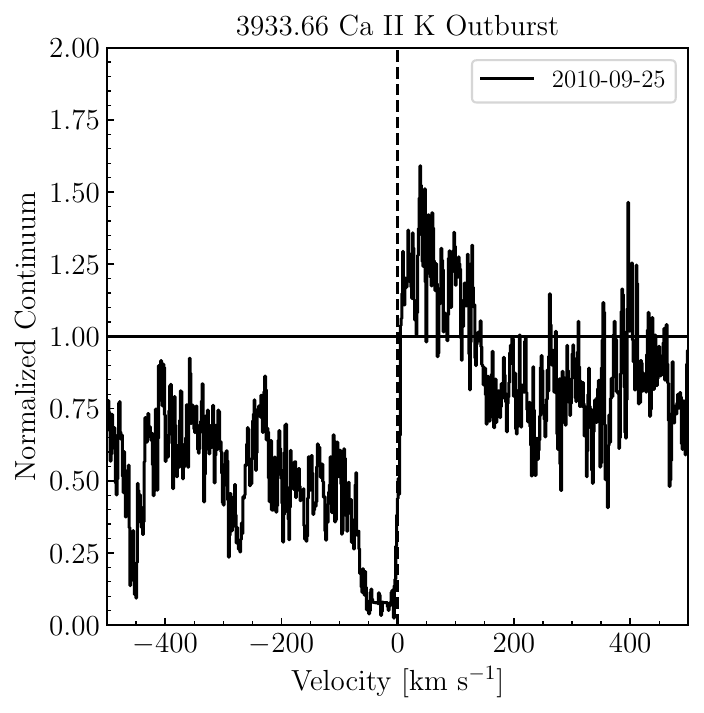}
    \includegraphics[width = 0.32\linewidth]{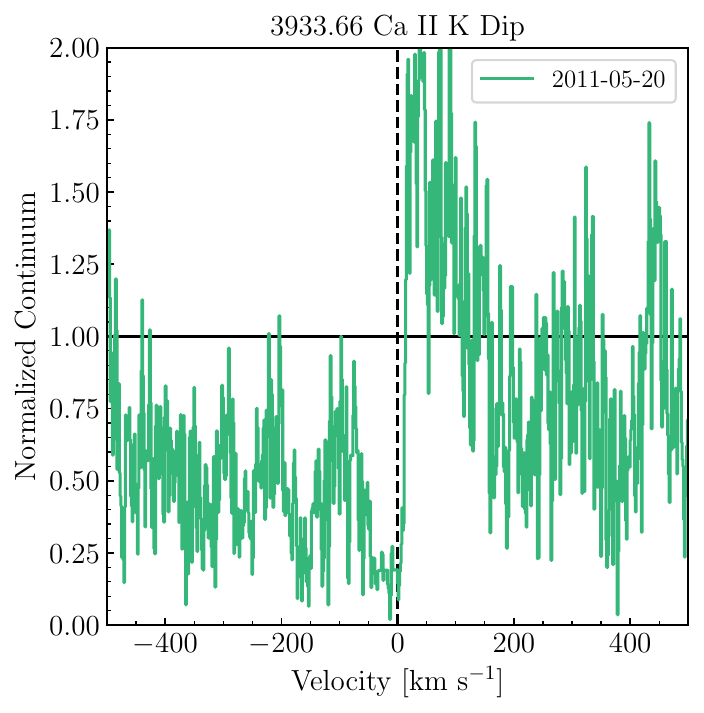} 
    \includegraphics[width = 0.32\linewidth]{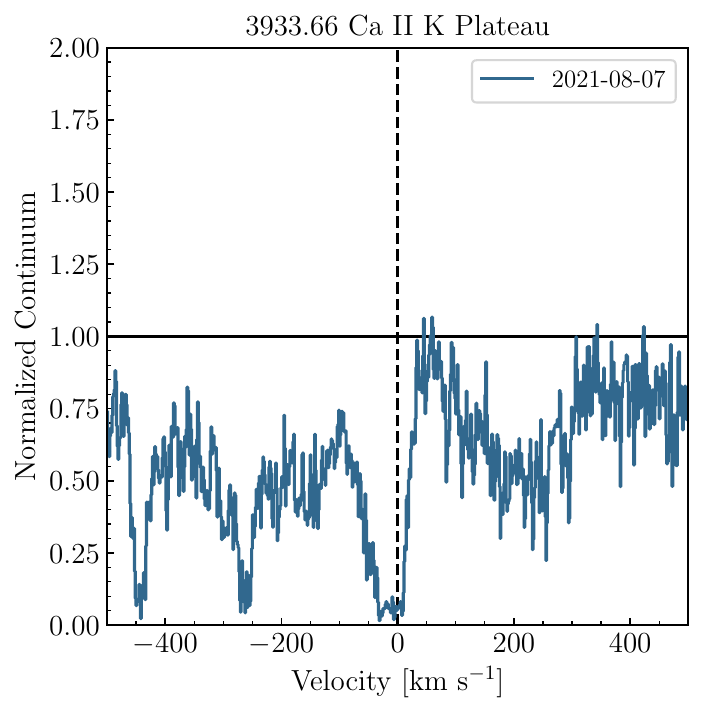} \\
    \hfill
    \includegraphics[width = 0.32\linewidth]{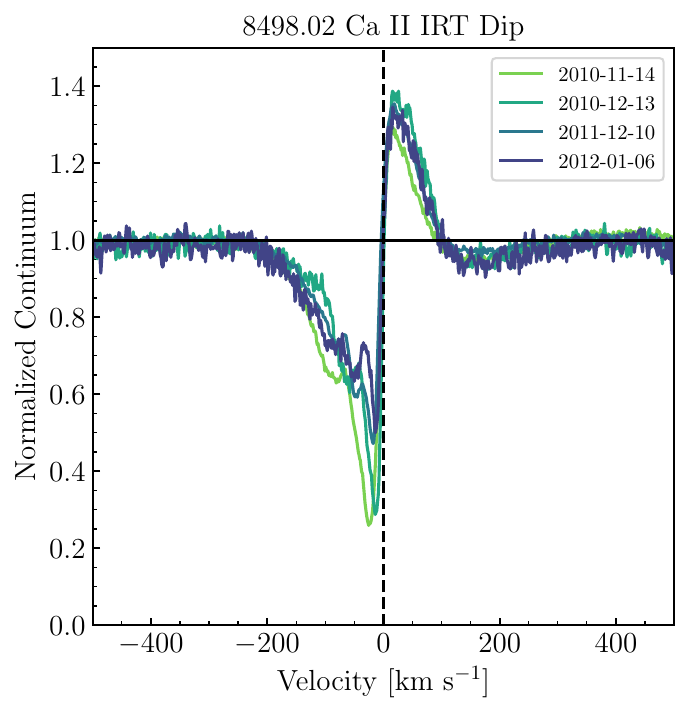}
    \includegraphics[width = 0.32\linewidth]{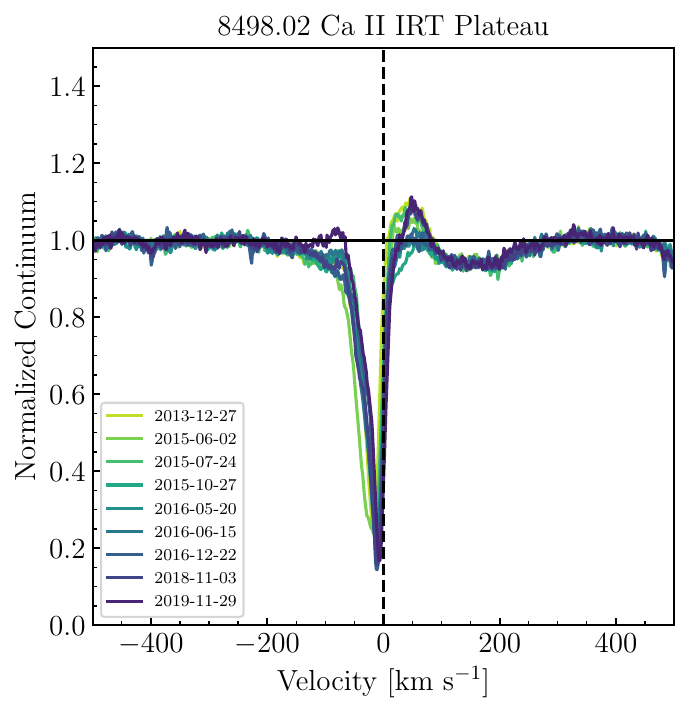}
    \caption{\ion{Ca}{2} lines in the HBC 722 HIRES spectra. The upper row shows the Ca II K line at 3933 \AA\ for the spectra that cover it, including the outburst epoch (left), dip epoch (center), and the most recent plateau spectrum (right). The outburst epoch spectrum did not cover the Ca II IRT 8498 \AA\ line so we only show the dip (left) and plateau epochs (right) that do cover it. Both sets of lines show similar profile behavior, particularly in the red-shifted emission. The wind absorption in the Ca II K line is much stronger than in the Ca II IRT. The blue-shifted absorption in the Ca II K profiles blends into two Fe I features at $-450$ km s$^{-1}$ and $-275$ km s$^{-1}$ (3927.919 \AA\ and 3930.296 \AA, respectively).}
    \label{fig:CaIRT}
\end{figure*}

The Ca II K line absorption (shown in Figure \ref{fig:CaIRT}) is totally saturated in the outburst and plateau epochs. The profile is absorbed to $\sim 0$ from 0 to $-60$ km s$^{-1}$ at outburst and from 0 to $-40$ km s$^{-1}$ in the plateau. It is difficult to determine whether the line has any high velocity absorption at outburst due to blending with several lines to the blue of Ca II K. The Ca II IRT (also in Figure \ref{fig:CaIRT}) does not appear in the outburst spectrum, but we can discuss the line profiles in the plateau epochs. The lines show relatively little evolution during the plateau and their low-velocity, deep absorption is similar to that of the Fe II 5018 and Mg I 5183 lines. The blue-shifted absorption reaches a maximum depth of 0.8. The core of the absorption is at $-15$ km s$^{-1}$ and the blue wing of the low-velocity component ends at $-80$ km s$^{-1}$. There is some hint of a higher velocity component from $-80$ km s$^{-1}$ to $-150$ km s$^{-1}$ in some epochs but not all. 

\subsubsection{The Dip Profiles} \label{sec:wind_dip}

The wind absorption profiles during the dip show several features that do not appear in the outburst or plateau wind line profiles. The first notable difference is that all of the wind lines show a strong high-velocity absorption component with maximum velocities of $-200$ km s$^{-1}$ (in H$\beta$, Fe II 5018, Na I D, Mg I 5183 and the Ca II IRT) up to $-400$ km s$^{-1}$ (in H$\alpha$). The second is that the red-shifted disk absorption only appears in Mg I 5183. 

Beginning again with the H$\alpha$ line profiles, we see the high velocity blue-shifted absorption extending out to $-400$ km s$^{-1}$. The wind is absorbing against the blue wing of the H$\alpha$ emission, which can be seen centered at $-80$ km s$^{-1}$. Taking into consideration the blue emission wing of the line, it becomes clear that the high velocity wind component blends into the low-velocity component. The low velocity component has weakened to a depth of 0.7 and slowed to a velocity of $- 20$ km s$^{-1}$. 

The high velocity absorption in the H$\beta$ line appears to extend to similarly high velocities but there is a \ion{Fe}{1} feature at 4854 \AA\ that may be responsible for the absorption blueward of $-300$ km s$^{-1}$. However, there is certainly absorption from $-200$ km s$^{-1}$ to $-300$ km s$^{-1}$ that is not present in the plateau epochs. As in the outburst and plateau spectra, the H$\beta$ absorption during the dip is deeper than the H$\alpha$ absorption. 

The distinction between the high velocity component and low velocity component is more clear in the Na I D lines, which show a clear ``kink" at $-80$ km s$^{-1}$. The Na I D high velocity component only extends to $-200$ km s$^{-1}$. The low velocity component is consistent with that seen in the other wind lines and remains as strong as in the outburst and plateau epochs.

In the Fe II 5018 and Mg I 5183 lines, the high velocity component also extends to $-200$ km s$^{-1}$ and blends into the low velocity component, though the Fe II 5018 line has a slight kink at $-80$ km s$^{-1}$ like that in the Na I D profiles. The K I line continues to only trace the lower velocity absorption, although it is deeper at $-100$ km s$^{-1}$ than in the outburst or plateau epochs. A similarity between all of the profiles features highlighted in this section is that during the dip, their profiles are all distinctly wedge-shaped, with a nearly vertical slope near 0 km s$^{-1}$ and almost no red shifted absorption. 

The Ca II K line absorption remains relatively unchanged during the dip. Despite the apparently weaker blue-shifted absorption, the line minimum may still be consistent with zero due to the low signal-to-noise in the region. The Ca II IRT has a clear 2-component blue-shifted absorption profile, with absorption minima at $-10$ km s$^{-1}$, $-90$ km s$^{-1}$, which later (by 2012) becomes a 3-component profile with a third minimum at $-50$ km s$^{-1}$. Though the slowest of the three can be attributed to the same absorbing material that has contributed to the low-velocity absorption in the other lines, the $-50$ km s$^{-1}$ component is more difficult to explain. It may not be a distinct absorption component but rather an artifact of stronger blue-shifted emission that splits the broader absorption by filling in a narrow range of velocities during this epoch, as is the case in H$\alpha$.

\subsubsection{The Persistent Slow Absorption Component} \label{sec:wind_slow}
Despite the variability in the high-velocity absorption components of the wind lines, the low-velocity component remains relatively constant throughout the outburst, dip, and plateau epochs. In the Fe II 5018, Na I D, and Ca II IRT lines, this component is centered at around $-10$ km s$^{-1}$ and has a half width at half depth (HWHD) of approximately $20-25$ km s$^{-1}$. In the Na I D lines the core is saturated, but ranges from $0$ km s$^{-1}$ to $-20$ km s$^{-1}$.

In the Balmer lines, the low-velocity component is more dynamic. The H$\beta$ profile in the dip epochs appears to contain a low-velocity component that is centered at $-40$ or $-50$ km s$^{-1}$, though it is blended with the higher velocity component. In the plateau epochs, the component slows to $-10$ km s$^{-1}$ with a HWHD of $\sim 30$ km s$^{-1}$. The line is nearly saturated at low velocities in both sets of spectra. 

The H$\alpha$ line at outburst shows a relatively faster low-velocity component, centered around $-50$ km s$^{-1}$ with a HWHD of $\sim 20$ km s$^{-1}$. In the dip, the low-velocity component is initially similar to that at outburst but, in the 10 Dec 2011 and 06 Jan 2012 epochs, the component becomes a bit shallower and broader, with a HWHD of $\sim 30$ km s$^{-1}$. In the plateau spectra, The low-velocity component tends to slow from $-30$ km s$^{-1}$ to $-20$ km s$^{-1}$. The HWHD remains $\sim 25$ km s$^{-1}$.

\subsubsection{Emission in the H$\alpha$ and Ca II Lines} \label{sec:emissionComp}
In all (or most) of the HIRES epochs, the H$\alpha$ and Ca II (K and IRT) profiles show both red- and blue-shifted emission components. The emission components are consistent with the emission expected for a hot, collimated disk wind, as we will discuss in Section \ref{sec:diskWind}. Here, we describe the characteristics of the emission in the two sets of lines and how it evolves post-outburst.

In the pre-outburst, progenitor spectrum, the H$\alpha$ line shown in Figure \ref{fig:BalmerLines} (left panel inset) is purely in emission and extremely bright, with a peak-to-continuum flux ratio of 11.0. Accounting for the $R \sim 2000$ of the spectrograph used to obtain the spectrum \citep{Fang_NorthAmerica_2020ApJ}\footnote{To account for this, we measure the FWHM of the line, then divide by $2\sqrt{2}\ln 2$ to compute the appropriate measured Gaussian width $\sigma_m$. Assuming the instrument broadening is approximately Gaussian with a width $\sigma_R$, we can estimate the underlying line width by calculating $\sigma_{H \alpha} = \sqrt{\sigma_m^2 - \sigma_R^2}  $.}, the width of the progenitor emission is $\sim 250$ km s$^{-1}$. In the post-outburst spectra, the red-shifted emission extends from $0$ km s$^{-1}$ to $200$ km s$^{-1}$, though the emission peak is strongly absorbed blue-ward of 50 km s$^{-1}$. The blue-shifted emission extends to greater velocities than the red-shift emission, reaching $-400$ km s$^{-1}$ in the outburst epoch and the 2013-2014 epochs of the plateau, potentially indicating forward-scattering of emission line by the outflow. The blue-shifted emission between 0 and $-100$ km s$^{-1}$ may explain why the H$\alpha$ absorption is shallower than the H$\beta$ absorption, as is seen in V960 Mon \citep{Carvalho_V960MonSpectra_2023ApJ} and V1331 Cyg \citep{Petrov_V1331_FacingTheWind_2014MNRAS}.

Like H$\alpha$, the Ca II K line and the IRT also show some blue- and red-shifted emission, though during the plateau it is much weaker. The red side of the Ca II H line is heavily absorbed by the blue wing of the H$\epsilon$ line, so we cannot see any potential red-shifted emission in the feature. We will also focus on the 8498 \AA\ component of the Ca II IRT, since it is the one most consistently covered by the HIRES spectra. However, the 2012-01-06 spectrum all three Ca II IRT components are visible and at their greatest peak-to-continuum ratios, enabling us to better compare the emission from all three. The ratios of the emission peaks of the three are 1:1:0.7 in increasing wavelength order. If the emission were optically thin, the expected peak flux ratios would be those of the log$g_f$ values between the lines, or 1:9:5 \citep{vald_reference_2015PhyS}, which indicates the source of the Ca II IRT optically thick, as is seen most T Tauri stars and Herbig Ae/Be stars \citep{HamannPersonn_EmissionLineStudiesI_TTSs_1992ApJS, HamannPersson_EmissionLineStudies_II_HerbigStars_1992ApJS}. 

The Ca II K and 8498 \AA\ emission components are brightest during the dip, with peak-to-continuum ratios of 2.0 and 1.35, respectively. The emission of the Ca II K line extends to almost 150 km s$^{-1}$, almost as high velocity as the H$\alpha$ emission. The maximum velocity of the Ca II IRT red-shifted emission peak is only $100$ km s$^{-1}$, much slower than the H$\alpha$ emission. The difference may be due to significant Paschen absorption to the red side of every IRT component, (e.g., line at 8502.49 \AA\ may absorb the highest velocity emission from the wing of the Ca II 8498 \AA\ line). 

For H$\alpha$ and both sets of Ca II lines, the emission peak-to-continuum ratios are significantly anti-correlated with the brightness of the $R$ band continuum, such that in the deepest part of the dip (the 13 Dec 2010 and 20 May 2011 spectra) the blue and red emission components are strongest. 

\begin{figure*}[!htb]
    \centering
    \includegraphics[width = 0.8\linewidth]{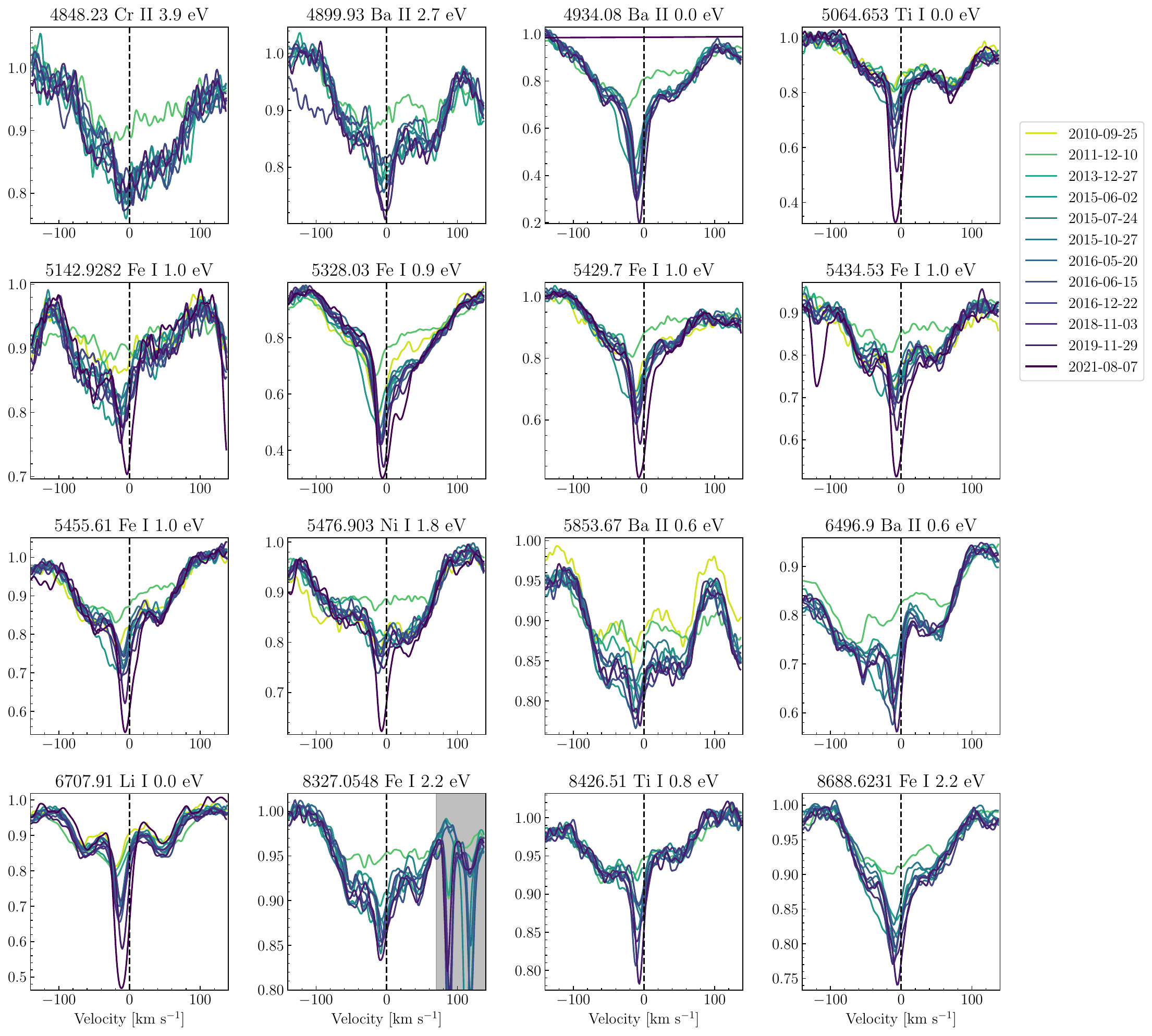}
    \caption{A selection of lines possessing the narrow absorption features in HBC 722. The features deepen toward later times, eventually almost dominating the absorption in some lines. The blue-shifted line center indicates it is tracing a low-velocity outflow.}
    \label{fig:CentralAbsGrid}
\end{figure*}

\begin{figure*}[!htb]
    \centering
    \includegraphics[width = 0.8\linewidth]{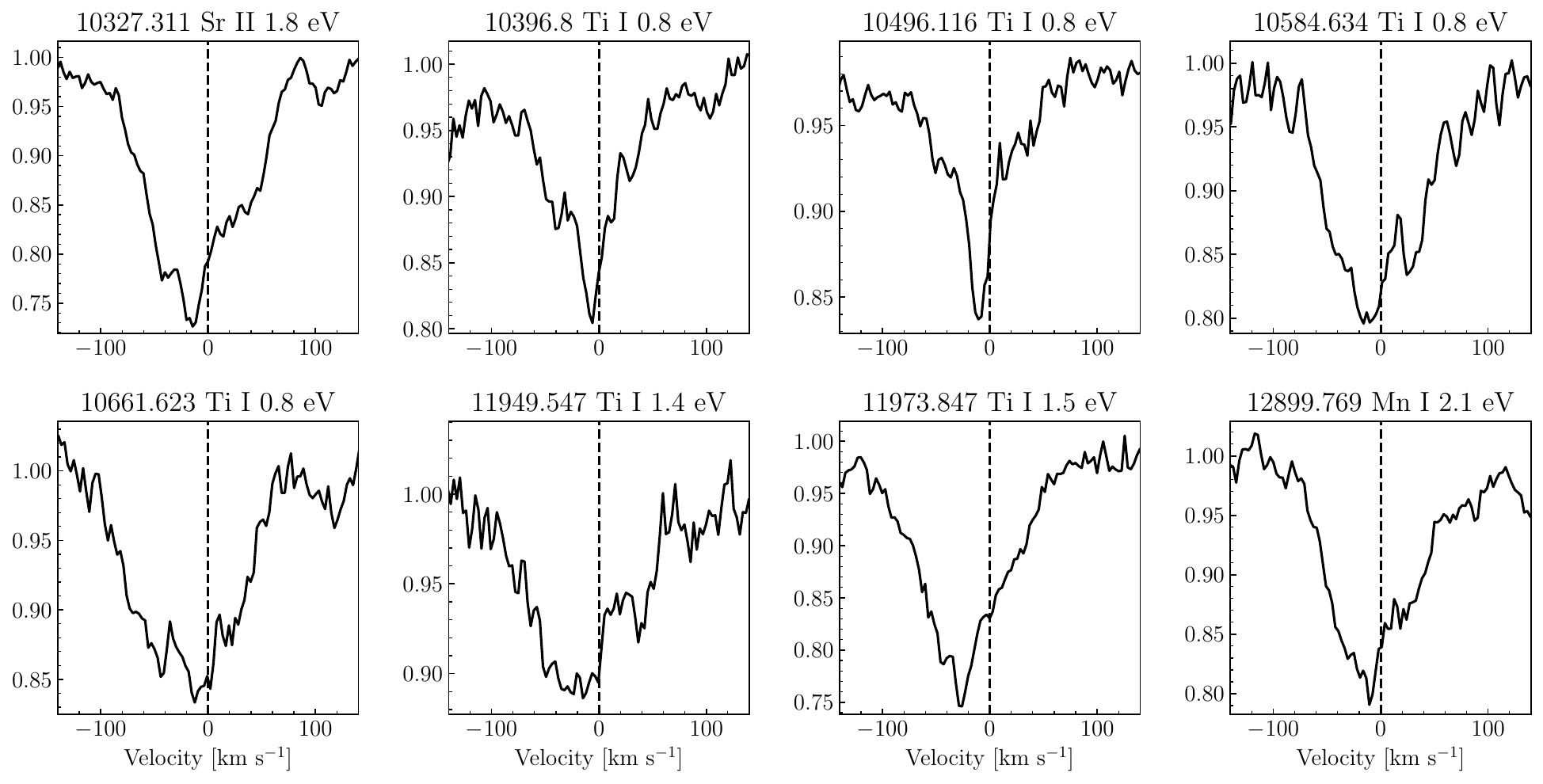}
    \caption{A selection of NIR atomic lines in the NIRSPEC spectrum spanning the Y to J band that show the same narrow absorption features we see in the HIRES spectra. The feature is at the same blue-shift and has a similar width to its visible-range counterparts, indicating it traces the same material. } 
    \label{fig:CentralAbsGridNIRSPEC}
\end{figure*}

\subsection{A Forest of Narrow Blue-shifted Absorption} \label{sec:narrow_blue}

Above, we discussed the high-velocity blue-shifted absorption visible in the wind-sensitive lines. We also identified a low-velocity component in these lines. There is another, distinct, family of lines that show a similar narrow, low-velocity blue-shifted absorption component in the HBC 722 spectra, a subset of which we show in Figure \ref{fig:CentralAbsGrid}. These lines are typically neutral species like Fe I and Ti I and almost all have low ($< 1.5$ eV) excitation potentials (EPs). We find 130 of these features in the HIRES spectra between 4800-8700 \AA, the vast majority of which are Ti I lines with EP $< 1.0$.

The narrow, blue-shifted absorption can be seen weakly in the outburst spectrum but not in the dip spectra, though it reappears as soon as the target emerges from the dip. As can be seen in Figure \ref{fig:CentralAbsGrid}, the feature appears consistently at $-10$ km s$^{-1}$ and is very narrow, with a HWHD $\sim 10$ km s$^{-1}$. We see it across the entire HIRES spectral range and, in the Aug 7 2021 spectrum it dominates over the disk absorption in several lines. Weak absorption lines tracing this feature are so ubiquitous, they appear 6 and 13 times over $\sim 800$ km s$^{-1}$ in the plateau line profile snippets in Figure \ref{fig:FeIILine} of the Fe II 5018 (upper right panel) Mg I 5183 (lower right panel) lines, respectively. 

The strong absorption in the region surrounding the Fe II 5018 line is especially notable. There are 6 weak Ti I lines within $\pm 400$ km s$^{-1}$ of the Fe II 5018 line that reappear during the plateau phase and grow to be very strong during the plateau. The features have rest wavelengths (marked in Figure \ref{fig:FeIILine}) of 5012.0676, 5014.187, 5014.276, 5016.1608, 5020.0263, 5022.868 and 5024.844 \AA, though they are all blue-shifted by 10 km s$^{-1}$, and they all have $0.8 \ \mathrm{eV}< \mathrm{EP} < 0.9 \ \mathrm{eV}$ (except 5014.187 \AA, which has $\mathrm{EP}=0$ eV). During the plateau, they reappear and grow consistently in strength, some reaching depths of 70 \% (e.g. Ba II 4934). The absorption is strongest and the growth more rapid in the lower EP features.

We also observe similarly narrow features in certain atomic lines in the Y and J bands of the NIRSPEC spectrum, shown in Figure \ref{fig:CentralAbsGridNIRSPEC}. Unfortunately, many of the features suffer from severe blending with other atomic absorption lines, so the line profiles are not as clear as those in the HIRES spectra. However, both the Ti I 10396 \AA\ and Ti I 10496 \AA\ lines have profiles that are similar to the Ti I lines seen in the 2021 epoch of the HIRES spectra. Overall, in the NIR, the lines tend to be of slightly higher EP values than those in the visible, though they are still almost all $< 1.5$ eV. 

In both the HIRES and NIRSPEC spectra, the deepest narrow absorption features appear in lines showing strong disk broadening already. We note that in the HIRES spectra the absorption depth of these narrow features grows rapidly, while the disk component of the profiles remains constant in both depth and width. This contrast demonstrates that the material carrying these lines must be absorbing against the underlying disk atmosphere. We interpret the behavior of these lines to be evidence of a low-velocity, wide angle disk wind, which we discuss in detail in Section \ref{sec:diskWind}.

\begin{figure*}[!htb]
    \centering
    \includegraphics[width = 0.65\linewidth]{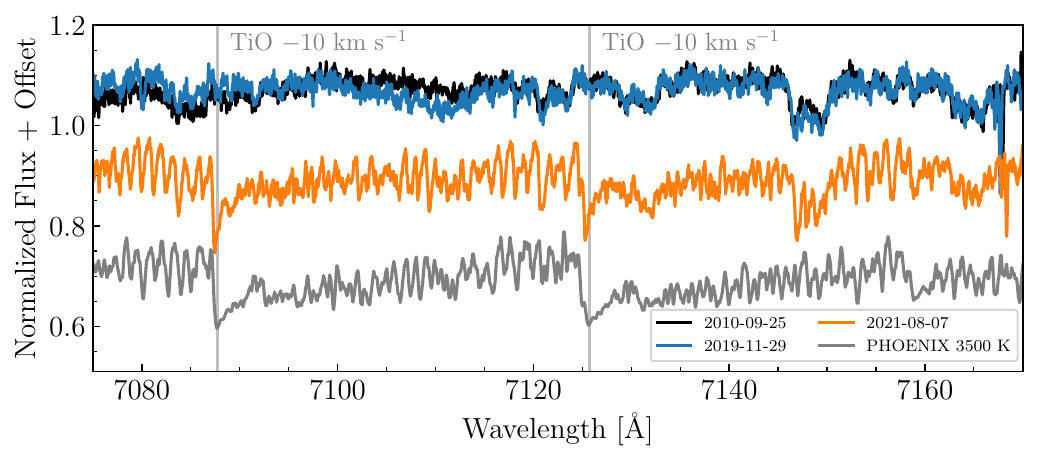}
    \includegraphics[width = 0.3\linewidth]{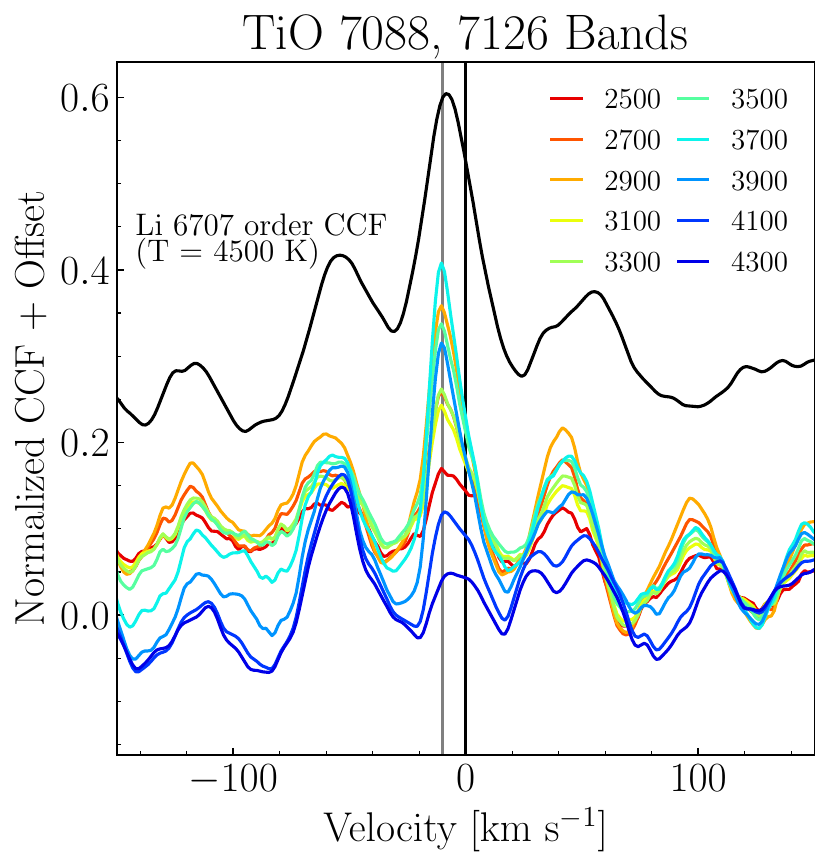}
    \caption{
    \textbf{Left:} The 7070 \AA\ - 7170 \AA\ region of the HIRES spectra showing the emergent TiO absorption in the most 2021 epoch compared to previous epochs where it did not appear. Notice that the bandheads in the red spectrum are offset -10 km s$^{-1}$ from their typical wavelengths, as highlighted by the grey vertical lines. The outburst (25 Sept 2010), penultimate (29 Nov 2019), and most recent (07 Aug 2021) spectra are shown in black, blue, and orange, respectively. A 3500 K, log$g=1.5$ PHOENIX model spectrum, rotationally broadened to 10 km s$^{-1}$, blue-shifted by 10 km s$^{-1}$, and scaled by 0.2, is shown in grey for reference. The scaled PHOENIX model has TiO absorption similar to that seen in the data. \textbf{Right:} CCFs computed with the TiO bandhead that appears in the 2021 HIRES spectrum (red in left panel). Notice that the strongest peak appears at 3500-3800 K and is blue-shifted by -10 km s$^{-1}$.}
    \label{fig:TiOEmerges}
\end{figure*}

\subsection{TiO Absorption} \label{sec:TiOAbs}

In addition to the sudden increase in strength in the narrow low-EP absorption feature in the 7 Aug 2021 HIRES spectrum we also see strong TiO absorption bands at 7088 and 7126 that did not appear in any previous epochs. The spectral range containing both bandheads is shown in Figure \ref{fig:TiOEmerges}, compared with the previous HIRES epoch (29 Nov 2019) and the outburst epoch. Notice that in the 9 years between the 2010 outburst spectrum and the 2019 spectrum, there is almost no evolution in this wavelength range. 

The bands are blue-shifted by $\sim$ 10 km s$^{-1}$ and have widths of $\sim 10$ km s$^{-1}$, consistent with the narrow blue-shifted absorption features described above. They are also relatively shallow, with bandhead absorption depths of 0.2, but they are similar in structure to those in a 3500 K, log$g=1.5$ atmosphere. A 3500 K, log$g=1.5$ PHOENIX model that has been rotationally broadened by 10 km s$^{-1}$ is shown in Figure \ref{fig:TiOEmerges} for reference. 

To better understand the line structure of the TiO absorption and where in the system it may originate, we performed a CCF analysis similar to that described in Section \ref{sec:HIRESModelsComp}. Here, we computed the CCFs between the 2021 HIRES spectrum and several different $T_\mathrm{eff}$ PHOENIX models in the wavelength range of 7060 \AA\ to 7170 \AA. The temperature range we explored spanned 2500 K to 4300 K, though we found the strongest CCF power to be between $3000$ K and $4000$ K. The CCFs are plotted in the right panel of Figure \ref{fig:TiOEmerges}. We also computed the CCF between the HIRES order containing the 6707 Li I feature and a $T_\mathrm{eff} = 4500$ K PHOENIX model, to serve as a reference for the profiles of the atomic absorption features.

The line profiles of the TiO bands are similar to the line profiles of the atomic absorption in the rest of the HIRES spectrum. The lines show the double-peak of the disk absorption profile at $\pm 60$ km s$^{-1}$ and the strong, narrow, blue-shifted absorption component that is dominant in the 2021 spectrum. The CCF peaks tracing this narrow component are also similar in width to the Li I order CCF. We therefore conclude that the TiO absorption traces the same component of the HBC 722 system as the narrow blue-shifted absorption in the low-EP atomic lines.

The CCF analysis shows that the preferred temperature for the TiO absorption feature is 3500-3700 K, though the 2900 K CCF peak is also quite strong. We discuss the implications of the TiO absorption in relation to the narrow blue-shifted atomic absorption and its preferred temperature range in Section \ref{sec:diskWind}.

\begin{figure*}[!htb]
    \includegraphics[width = 0.58\linewidth]{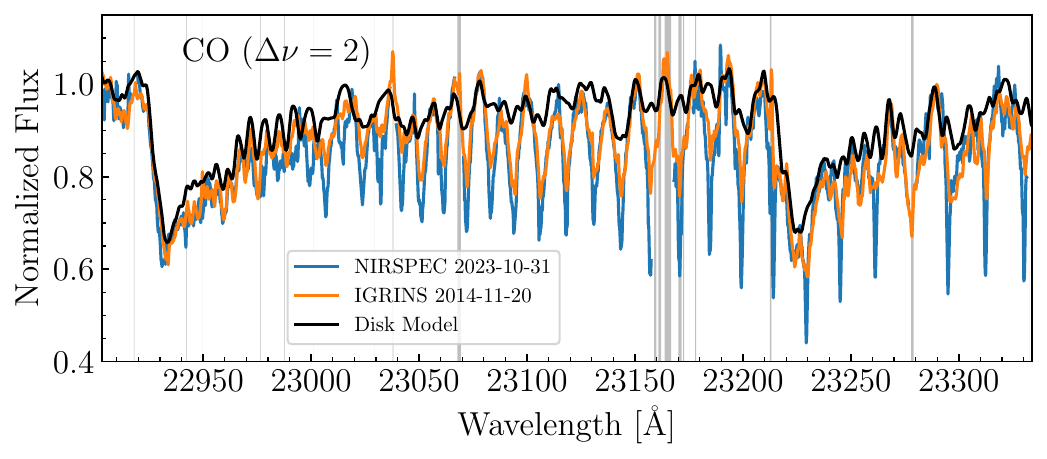}
    \includegraphics[width = 0.38\linewidth]{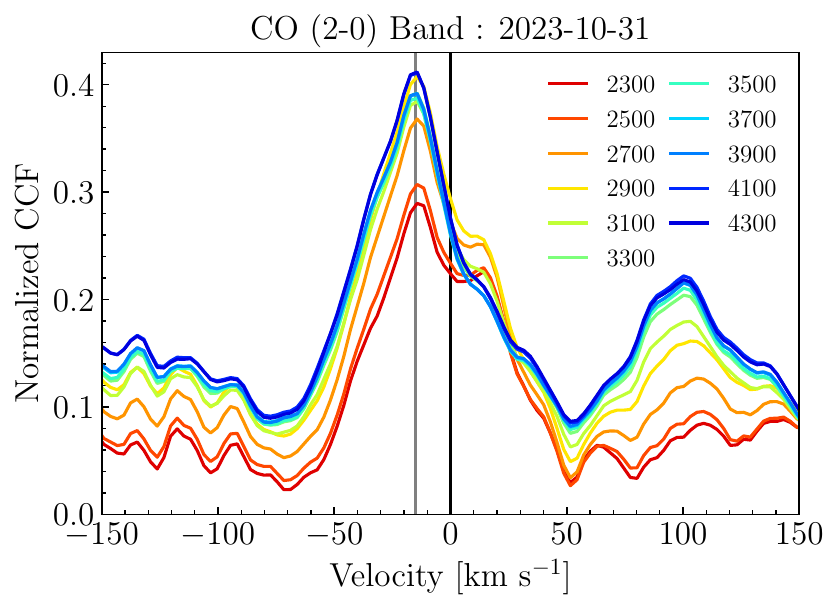}
    \caption{\textbf{Left:} The CO (2-0) and (3-1) bands from the Keck/NIRSPEC 30 Oct 2023 spectrum (blue), the HJST/IGRINS 20 Nov 2014 spectrum (orange) and the high resolution disk model spectrum (black). As in the HIRES spectra, we see a narrow excess absorption feature that grown stronger from 2014 to 2023. \textbf{Right:} The CCFs of the CO (2-0) band computed with different $T_\mathrm{eff}$ PHOENIX model spectra. At cooler temperatures, the red-shifted disk emission becomes apparent, whereas the hotter temperature profiles are dominated by the blue-shifted wind absorption.}
    \label{fig:COBandhead}
\end{figure*}

\subsection{CO Absorption} \label{sec:COAbs}
So far in this Section, we have discussed the typical visible and NIR wind features which we expect to probe a component not included in the disk model, as well as a new narrow absorption component not previously seen. Another notable discrepancy between the model and data appears in the CO (2-0) band at 2.29$\mu$m. 

The CO (2-0) band is expected to originate in the disk and has been shown to match a disk profile well in other FU Ori objects \citep[e.g., V960 Mon, ][]{park_high-resolution_2020}. However, the CO (2-0) band is dominated by narrow absorption lines much deeper than those predicted by the disk model, as can be seen in Figure \ref{fig:COBandhead}, where we show the (2-0) bands in the IGRINS and NIRSPEC spectra compared with our disk model prediction. 

Relative to the disk model spectrum, the (2-0) band shows a deep excess absorption component in both epochs. We note that the near-continuum wings of the lines in the (2-0) band are similar in width to the model lines, indicating that the absorption is growing \textit{against} the underlying disk spectrum. The absorption is reminiscent of that seen growing in the HIRES spectra and which appears in some atomic features in the Y and J bands of the NIRSPEC spectrum.

We again turn to CCF analysis to study the temperature sensitivity of the line profiles of the CO (2-0). For consistency with our analysis of the visible range TiO absorption, we use the same log$g = 1.5$ models, and explore $2300$ K $< T_\mathrm{eff} < 4300$ K. The wavelength ranges over which we computed the CCFs was $22920-23205$ \AA\ and the CCFs are shown in Figure \ref{fig:COBandhead}. We will focus on the NIRSPEC spectrum here, but we discuss the CO (2-0) band in the IGRINS spectrum and the time evolution of the narrow CO absorption in Appendix \ref{app:COBandTemps}. 

In general, the higher $T_\mathrm{eff}$ CCFs are dominated by the blue-shifted wind absorption feature and only for $T_\mathrm{eff} < 3000$ K do we see the red-shifted disk atmospheric absorption appear. Still, even the cooler model CCFs show strong wind absorption and a very weak red-shifted disk absorption profile. These facts indicate that the wind traced by CO (2-0) has a temperature range of $3000 \ \mathrm{K} < T_\mathrm{eff} < 4300 \ \mathrm{K}$, whereas the region of the disk from which the CO (2-0) atmospheric absorption arises has a temperature range of $2300 \ \mathrm{K} < T_\mathrm{eff} < 3000 \ \mathrm{K}$. The wind temperature range is consistent with the temperature range found for the TiO wind absorption discussed in Section \ref{sec:TiOAbs}. The disk temperatures are only slightly cooler than those predicted by our plateau epoch model (see Appendix \ref{app:RadLambda}).

\section{Interpretation}\label{sec:interp}

The fact that HBC 722 
had dedicated follow-up following its 2010 outburst to collect spectra that sample the entire post-outburst lightcurve to date, enables us to construct a detailed picture of the system. We find that a multi-component accretion disk system, with high and low velocity warm disk wind components and an evolving disk, are all necessary to explain the observed data. In this Section, we describe our interpretation of the spectroscopic and photometric evolution of the HBC 722 system in light of the picture presented in Figure \ref{fig:cartoon}. 

\subsection{The $\alpha$ Disk} \label{sec:thinDisk}
The primary component we observe is the accretion disk atmosphere, which at outburst has an $L_\mathrm{acc} = 71 \ L_\odot$, dominating the emission from the system. The relatively simple disk model with $T_\mathrm{max} = 5700$ K, $M_* = 0.2 \ M_\odot$, $\dot{M}_\mathrm{outburst} = 10^{-4.0} \ M_\odot \ \mathrm{yr}^{-1}$, $R_\mathrm{inner} = 3.65 \ R_\odot$, $A_V = 2.3$ mag, $i = 79^\circ$ and $R_\mathrm{outer} = 25 \ R_\odot$ reproduces the data excellently. The model matches the spectrophotometry of HBC 722 (Figure \ref{fig:spex_fits}) and is able to accurately predict the line profiles of the spectrum at high resolution spanning 0.4-2.5 $\mu$m (Figures \ref{fig:HIRESSpec} and \ref{fig:NIRSPECOrders}).

Varying only 3 parameters in the model is also able to reproduce the evolution of the system post-outburst. The 0.3-1.3 $\mu$m spectrum at outburst, in the dip, and during the plateau can be reproduced by simply varying the $\dot{M}$ of the system from $\dot{M}_\mathrm{outburst} = 10^{-4.0} \ M_\odot \ \mathrm{yr}^{-1}$ to $\dot{M}_\mathrm{dip} = 10^{-4.4} \ M_\odot \ \mathrm{yr}^{-1}$ to $\dot{M}_\mathrm{plateau} = 10^{-3.9} \ M_\odot \ \mathrm{yr}^{-1}$. The accretion luminosities and temperatures are correspondingly $L_\mathrm{acc} = 71 \ L_\odot \rightarrow 28 \ L_\odot \rightarrow 90 \ L_\odot$ and $T_\mathrm{max} = 5700 \ \mathrm{K} \rightarrow 4500 \ \mathrm{K} \rightarrow 6000 \ \mathrm{K}$. The $H_2 O$ absorption at 1.35 $\mu$m and 1.8-2.0 $\mu$m, along with the 2.2 $\mu$m continuum, help to constrain the values of $R_\mathrm{outer}$ and $T_\mathrm{min}$ in each of the three stages. 

In Section \ref{sec:spex}, we demonstrated that the spectophotometric evolution indicates an increasing $R_\mathrm{outer}$ value, from 25 $R_\odot$ at outburst to 35 $R_\odot$ in the dip and 100 $R_\odot$ in the plateau. During the earlier two epochs, a passive disk component starting at $r_i = 25 \ R_\odot$ and $r_i = 35 \ R_\odot$, respectively, is necessary to match the flux level of the $3-5$ $\mu$m photometry, whereas the 2015 spectrum is well-fitted with only a passive disk component. 
This is consistent with a model wherein the viscous accretion region of the disk propagates outward and ``activates" the passive disk.

Accompanying the increase in $R_\mathrm{outer}$, we also found that the $T_\mathrm{min}$ value of the active disk evolved from 2400 K to 1500 K to 1100 K. The decreasing $T_\mathrm{min}$ is consistent with the outer boundary of the high-viscosity, high accretion rate region of the object propagating outward. The decreasing $T_\mathrm{min}$ may also indicate cooler annuli are establishing plane-parallel disk atmospheres with significant molecular absorption, rather than simply emitting as blackbodies. The outward propagation rate of $0.27$ km s$^{-1}$ is well below the expected sound speed of the disk midplane \citet{Zhu_outburst_FUOri_2020MNRAS}. Future simulations are necessary to test the ability of different instabilities to propagate outward in the disk as we observe in this system. 

Notably, we showed in Section \ref{sec:highResolutionModels} that the HIRES (outburst and plateau epochs) and NIRSPEC spectra are closely-matched by the high resolution spectra predicted by our disk model. That spectra of FU Ori objects are consistent with disk model spectra has been shown for some of the classic objects like FU Ori and V1057 Cyg \citep{Kenyon_FUOri_disks_1988ApJ, welty_FUOriV1057CygDiskModelAndWinds_1992ApJ, Zhu_FUOriDifferentialRotation_2009ApJ} but many of these studies rely on relatively isolated regions of the spectrum, usually in visible wavelengths (e.g., 6100-6200 \AA). Here, we demonstrate agreement between model and data to within $\sim 5 \ \%$ for the entire 0.4-2.5 $\mu$m spectrum, excepting only lines known to trace massive disk winds, as discussed in Section \ref{sec:excess}. This bolsters the veracity of a high viscosity, relatively thin \citep[$h_p/r < 0.02$, where $h_p$ is the disk pressure scale height,][]{Zhu_outburst_FUOri_2020MNRAS} accretion disk as a model for the innermost regions of FU Ori accretion disks.

Although the $\alpha$ disk model successfully reproduces the intermediate resolution spectra in all epochs and the HIRES and NIRSPEC spectra during the outburst and plateau, the model is a poor fit to the HIRES spectra during the dip. This is not a difficiency in the disk model but rather illustrates that during the dip the spectra show a distinct and different average line profile. In the dip spectra, the atomic line profiles generally do not show the tidy Keplerian disk profiles seen before and after (see Section \ref{sec:highResolutionModels}). Rather, the CCFs show that the atomic lines are dominated by the low-velocity narrow absorption component, and have relatively triangular profiles. Interestingly, the CCFs of the reddest order do show much more disk-like profiles than are seen in the bluer orders. 

Altogether, this may be an indication that as the accretion rate decreased during the dip and the innermost annuli cooled, the region nearest the star became dominated by turbulence rather. The triangular line profiles can be reproduced by convolving a Keplerian disk kernel ($v_\mathrm{kep} \sim 90$ km s$^{-1}$) with a spherically rotationally broadened line ($v_\mathrm{rot} \sim 30-40$ km s$^{-1}$). This is similar to the $\sim 20$ km s$^{-1}$ turbulence observed in the V960 Mon system \citet{Carvalho_V960MonSpectra_2023ApJ} and in the \citet{Zhu_outburst_FUOri_2020MNRAS} simulation of FU Ori, though it is at a higher velocity and is only seen during the dip. Such high turbulence would imply that the \citet{Shakura_sunyaev_alpha_1973A&A} viscosity parameter $\alpha \sim 1-5$ in the innermost region of the disk during the dip. However, the turbulent motion we observe may not necessary translate to an elevated viscosity in the $\alpha$ disk model but may instead by due to the disruption of the inner disk by the decrease in accretion rate.


\subsection{The High Velocity Collimated Outflow} \label{sec:highVwind}
In the outburst spectrum, during the dip, and early into the plateau, a very high velocity $v \sim 100-400$ km s$^{-1}$ wind can be seen as blue-shifted absorption in the H$\alpha$ line profiles. Lower velocity regions of this outflow can also be observed in some of the other wind lines discussed in Section \ref{sec:windLines}: H$\beta$ (dip, $100-300$ km s$^{-1}$), Na I D (outburst and dip, $100-250$ km s$^{-1}$), Fe II 5018 and Mg I 5183 (dip, $100-200$ km s$^{-1}$), Ca II IRT (dip, $100-250$ km s$^{-1}$). 

The highest velocity wind absorption is seen especially in the dip in the H$\alpha$ line. During the dip, the absorption reaches a depth of $\sim 0.3$ at $-200$ km s$^{-1}$ to $-300$ km s$^{-1}$. The absorption then weakens at higher velocities but is still $\sim 0.1$ at $-400$ km s$^{-1}$. 

The line profiles of the Na I D, Fe II 5018, and Mg I 5183 lines are consistent with those created with the BP82 magnetocentrifugally driven disk wind model. The solid curves in our Figure \ref{fig:cartoon} are generated using the BP82 prescription for the streamlines of the outflow. The pure-absorption case for Model MHD1i of \citet{Milliner_FUOri_NaDWindModels_2019MNRAS} illustrates the BP82 model with the best-match to the profiles in HBC 722 being the 55$^\circ$ inclination case shown in their Figure 14.  

The BP82 solutions predict winds that accelerate and decrease in density as they rise. This may be observable in the different maximum velocities seen in different wind lines. When the wind is denser and nearer the disk, it is traced by all of the wind lines, so we see the $-100$ km s$^{-1}$ to $-200$ km s$^{-1}$ component in the Ca II IRT, Na I D, Fe II 5018, and Mg I 5183 lines. During the dip, we also see this component in the H$\beta$ line. However, as the wind accelerates, the only line strong enough to trace the more diffuse material is H$\alpha$. Therefore, we see the $-200$ km s$^{-1}$ to $-400$ km s$^{-1}$ absorption in H$\alpha$. 

The H$\alpha$ line profiles of HBC 722 shown in \citet{Lee_HBC722_2015ApJ} have 10 more epochs sampling the dip, which help to clarify the evolution of the profiles immediately post-outburst and as the target recovers from the dip more clearly. As can be seen in our Figure \ref{fig:BalmerLines} and Figure 1 of \citet{Lee_HBC722_2015ApJ}, the high velocity component in H$\alpha$ is strongest as the target is fading and when it reaches its minimum brightness in the dip. As the target brightens, starting in August 2012, the H$\alpha$ line dramatically changes over just 28 days from one showing absorption to $-400$ km s$^{-1}$ to showing only low velocity wind absorption. The rapid disappearance of the high velocity component must be due to a change in the wind structure, where either the wind no longer reaches such high velocities or the absorbing material is too diffuse to be detected. 

The \citet{Zhu_outburst_FUOri_2020MNRAS} magneto-hydrodynamical (MHD) simulation of the FU Ori outburst provides some potential context for interpreting our observations. They find a wind that accelerates a relatively short distance above the disk atmosphere, where the estimated optical depth is $\tau \sim 0.01$. The wind quickly reaches velocities along the rotation axis of $v_z > 50$ km s$^{-1}$, velocities along the disk radial axis of $v_r > 100$ km s$^{-1}$, and rotational velocities of $v_\phi > 100$ km s$^{-1}$. The high inclination of HBC 722 implies the projected velocities from the outflow in our direction will be dominated by $v_r$ and $v_\phi$. Given the rapid wind acceleration in the $r$ direction, the $v_r$ component may be more responsible for the high-velocity blue-shifted wind absorption than the $v_z$. The absorption would be present only in the strongest wind lines due to the lower optical depth in the region where the wind reaches the velocities we observe.

\subsection{The Low Velocity Disk Wind} \label{sec:diskWind}

The low-velocity absorption seen consistently in the wind-tracing lines (Section \ref{sec:wind_slow}) can be attributed to a wide opening-angle, rotating disk wind, like that modeled in \citet{Calvet_FUOriModel_1993}. The massive wind modeled by \citet{Calvet_FUOriModel_1993} was demonstrated to match the line behavior of FU Ori, particularly for deeper absorption lines like Fe II 5018, in \citet{Hartmann_FUOriWindConstraints_1995AJ}. As we show in Section \ref{sec:wind_slow}, the cores of the line profiles are typically centered at $-10$ to $-30$ km s$^{-1}$, while the blue wing of the profiles terminate near the blue-shifted wing of the disk model line profiles (e.g., the H$\alpha$ line in Figure \ref{fig:HIRESSpec}). This is mainly the behavior that was successfully modeled for deep atomic absorption \citep{Calvet_FUOriModel_1993,Hartmann_FUOriWindConstraints_1995AJ} and the Na I D doublet \citep{Milliner_FUOri_NaDWindModels_2019MNRAS} in FU Ori.

This low-velocity disk wind may also be traced by other absorption lines, as discussed in Section \ref{sec:narrow_blue} and shown in the CCFs in Figure \ref{fig:CCFs_HIRES}. The low EP, neutral atomic lines in both the HIRES and NIRSPEC $Y$ and $J$ band spectra all show a growing absorption component that is blue-shifted between $-5$ km s$^{-1}$ and $-15$ km s$^{-1}$, consistent with the lower range of velocities probed by the wind lines. Although this feature has not been seen in other FU Ori objects (except perhaps V960 Mon), it appears in lines that have been previously identified as wind-tracing lines at high mass outflow rates in FU Ori, V1057 Cyg, and Z CMa \citep{welty_FUOriV1057CygDiskModelAndWinds_1992ApJ}.

The structure of the absorption is also reminiscent of that identified in the bluer low EP neutral species lines in the spectrum of V1057 Cyg, dubbed the "shell features" by \citet{herbig_high-resolution_2003}. The features are attributed to condensations in the wind absorbing against the host star. They are not consistently seen in other FU Ori objects, implying V1057 Cyg may be a special case. Potentially, as the HBC 722 system ages it will reveal more narrow absorption features as the wind evolves.

The growth of the disk wind absorption in HBC 722 is also seen in the CO (2-0) band. Between the 2014 IGRINS spectrum and the 2023 NIRSPEC spectrum, the feature doubled in depth, ultimately dominating the absorption profiles of the band. Generally, the wind seems to be traced by the higher-temperature features of the CO (2-0) band, similar to the temperatures traced by the HIRES TiO 7088/7126 \AA\ absorption (see Section \ref{sec:COAbs}).


Altogether, the slow disk wind absorption component seen in the wind lines is consistently also seen absorbing against the disk spectrum across the entire visible/NIR. The growing absorption may be due to an increasing mass reservoir accumulating in the disk wind. As the outburst progresses and propagates outward and more material is lifted into the wind, the wind can be seen in increasingly weak absorption lines, like the many very weak Ti I lines that appear in the 2021 HIRES epoch. 

In the \citet{Zhu_outburst_FUOri_2020MNRAS} MHD simulation, lower portion of the wind acceleration region, where the highest layers of the disk atmosphere begin to show vertical motion, the density is still comparable to lower in the disk atmosphere, with an optical depth around $\tau = 0.1$. The model predicts that in this region the material will still rotate at roughly the local $v_\mathrm{Kep}$, though the $v_r$ is predicted to be around $10-20$ km s$^{-1}$ and $v_z \sim 50$ km s$^{-1}$. 

Considering the high inclination of HBC 722, the observed absorption line velocities would be around $-10$ km s$^{-1}$ to $-20$ km s$^{-1}$, depending on the strength of the line. This is exactly what we see for the low-velocity absorption in the wind lines, the narrow, growing atomic and visible range TiO absorption features, and the deep CO ($\delta \nu = 2$) bands. \citet{Zhu_outburst_FUOri_2020MNRAS} find that for the FU Ori system, the wind has an $\dot{M}_\mathrm{out} \sim 0.1 \ \dot{M}_\mathrm{acc}$, which would mean a $\dot{M}_\mathrm{out} \sim 10^{-6}$ to $10^{-5} \ M_\odot$ yr$^{-1}$ for HBC 722.

\subsection{Emission from the Warm Disk Wind} \label{sec:hotWind}
There are only 5 lines that show emission in any of the epochs of HIRES spectra: the Ca II K line, H$\alpha$ line and the Ca II IRT. Both H$\alpha$ and the Ca II lines are heavily wind absorbed blueward of 0 km s$^{-1}$, as discussed above, but both blue- and red-shifted emission components can be seen in almost all epochs. In both sets of lines, the line peak-to-continuum ratio is inversely correlated with the $R$ band continuum flux level, indicating the disk continuum may be out-shining most of the emission.

The H$\alpha$ emission is strongest during the first two epochs in the dip, reaching a peak-to-continuum ratio of 2.0, whereas during the plateau, due to the increased continuum brightness, the ratio is only 1.25. The blue side shows a clear emission component similar to that seen in the red side, though it is weaker in the outburst and dip epochs due to wind absorption. The blue emission strengthens to equal the red emission peak as the wind absorption weakens \citep[as in the V960 Mon system,][]{park_high-resolution_2020, Carvalho_V960MonSpectra_2023ApJ}. 

Similarly, the Ca II IRT reaches a maximum peak-to-continuum ratio of 1.35 during the dip which decreases to 1.1 during the plateau. As discussed in Section \ref{sec:wind_dip}, the ratios of the individual emission line peaks of the Ca II IRT show that the emission source is optically thick. 

Many of these points are consistent with the observations of the wind line profiles in the V960 Mon spectra \citep{Carvalho_V960MonSpectra_2023ApJ}, as well as those of FU Ori (though FU Ori does not show Ca II emission). We therefore propose that the emission in these systems is due to a warm wind, similar to those seen in high accretion rate Classical T Tauri Star (CTTS) systems \citep[e.g. the strong accreter RW Aur,][]{Alencar_RWAurWind_2005A&A}. The emission seen in lines like H$\alpha$ in these systems has been demonstrated to arise not only from the magnetospheric accretion but also in large part from hot disk and stellar winds \citep{Kurosawa_HydHelWindLineProfiles_2011MNRAS, Petrov_V1331_FacingTheWind_2014MNRAS}. 

A warm, line-emitting wind has been modeled in the past for FU Ori objects \citep{croswell_FUOriHotSphericalWind_1987ApJ}, using a spherically symmetric geometry and a plane-parallel geometry. Their H$\alpha$ line profile models for a wind with $\dot{M}_w = 10^{-6} - 10^{-5} \ M_\odot$ yr $^{-1}$ are a good match to the H$\alpha$ lines we observe in HBC 722 and V960 Mon. However, their predicted Na I D line profiles have emission components, whereas we do not see emission in our data. Their plane-parallel geometry $\dot{M}_w = 10^{-6} \ M_\odot$ yr $^{-1}$ model line profiles, on the other hand, somewhat under-predict the H$\alpha$ emission but are more similar to the Na I D line. 

To better understand the impact of different wind geometries and temperature profiles, we also compare our results models of CTTS disk winds. One example is the strong difference between the profiles generated by more collimated winds, potentially originating from the star or the innermost region of the disk, and those with wider opening angles being launched from the disk surface. In the model line profiles presented by \citet{Kurosawa_HydHelWindLineProfiles_2011MNRAS}, the more collimated wind models are consistent with the H$\alpha$ profiles observed in HBC 722 and V960 Mon shortly after their outburst peaks, though wider opening angle models better match the profiles of these systems at later times. We also see a similar progression in the H$\alpha$ lines of V1057 Cyg over time \citep{Szabo_V1057cyg_2021ApJ}. Though the wind mass loss rates modeled by \citet{Kurosawa_HydHelWindLineProfiles_2011MNRAS} are much lower than those we expect in FU Ori systems, the comparison may indicate an initially more collimated wind that becomes dominated by a slower, wider opening-angle component over time. 

Another instructive set of CTTS wind models comes from \citet{Lima_HalphaLineEmissionWinds_2010A&A}. They compute several H$\alpha$ profiles for a \citet{BlandfordPayneWind_1982MNRAS} hot disk wind under different model conditions, include the wind mass outflow rate, $\dot{M}_w$, and the maximum temperature attained in the wind $T_\mathrm{MAX,w}$. Several of their profiles are consistent with those we see in the later epochs of the HBC 722 and V960 Mon spectra, especially the emission on the red side of the profile. One of their findings is that strong blue-shifted absorption appears at particular combinations of $\dot{M}_w$ and $T_\mathrm{MAX,w}$ ($10^{-9} \ M_\odot$ yr$^{-1}$ and 9000 K, for example) and that the required $T_\mathrm{MAX,w}$ to see absorption decreases with $\dot{M}_w$. They state that for systems with very high mass loss rates ($> 10^{-8} \ M_\odot$ yr$^{-1}$), the critical $T_\mathrm{MAX,w}$ could be as low as 6000 K. In FU Ori systems, where mass loss rates may be as high as $> 10^{-7}$ to $ 10^{-6} \ M_\odot$ yr$^{-1}$, this may allow wind temperatures to be even cooler and still show the blue-shifted absorption. 

An important detail about the \citet{Lima_HalphaLineEmissionWinds_2010A&A} disk wind models is that their winds are slower (by a factor of 10-100), denser (by $\sim$1.5 dex) and cooler (by $\sim$2000 K) near the wind launching points on the disk surface. This would allow the deep, fast, blue-shifted H$\alpha$ absorption to appear in a relatively warmer part of the wind, perhaps around 5000-6000 K, while the denser, cooler region of the wind, nearer the launching point, would be traced by the low-velocity atomic and molecular absorption we highlighted in Sections \ref{sec:narrow_blue}, \ref{sec:TiOAbs}, and \ref{sec:COAbs}. In this way, the model may also be consistent with the predictions made by the cold wind models in \citet{Calvet_FUOriModel_1993} and \citet{Milliner_FUOri_NaDWindModels_2019MNRAS}.

Therefore, we propose that, as is the case in V960 Mon \citep{Carvalho_V960MonSpectra_2023ApJ}, the emission components of H$\alpha$ and the Ca II IRT originate in a boundary layer between the accretion disk and the central star. The variability in the emission is then caused by the boundary layer being absorbed by the disk atmosphere and wind. When the accretion rate is lower, the disk atmosphere is dimmer and the boundary layer appears brighter by contrast. 


\subsection{Comparison with V960 Mon} \label{sec:V960MonComp}
In contrast to the V960 Mon system we studied in \citet{Carvalho_V960MonPhotometry_2023ApJ}, we find here that the $R_\mathrm{inner}$ of HBC 722 is constant in time. The relatively colorless evolution of the system does not match the predictions made by the $R_\mathrm{inner} \propto \dot{M}^{-2/7}$ model found to match the exponential fade of V960 Mon. The constant $R_\mathrm{inner}$ may place an upper limit on the magnetic field of the HBC 722 central star or at least its ability to provide pressure support against the ram pressure of the accretion flow. For the system parameters during the low accretion rate state in the dip, using the magnetospheric truncation radius equation \citep{Hartmann_review_2016ARA&A}, we estimate an upper bound for the stellar magnetic field of $B_* < 1.3 (1/\xi)^{1/4}$ kG, where $\xi$ is an order unity correction factor. 

Another possibility is that the high inclination of the system prevents us from directly observing the variation in the innermost annulus of the disk. If the disk atmosphere is sufficiently flared, it may absorb the emission from $r < 3.65 \ R_\odot$. The HIRES spectra with coverage blueward of 4000 \AA\ are not as well-matched by the disk model and have many absorption lines that show wind excess relative to the model line profiles. This may indicate that the emission from the $r < 3.65 \ R_\odot$ annuli of the disk is obscured by self-absorption from the disk and wind absorption from the collimated, high-velocity outflow.  

The second strong contrast between the V960 Mon disk and the HBC 722 disk is the expansion of $R_\mathrm{outer}$ in HBC 722, which was not found for V960 Mon. In V960 Mon, the 3-5 $\mu$m emission is matched by an active disk component with $R_\mathrm{outer} \sim 25 \ R_\odot$ and a passive disk beyond that in both the 2015 and 2016 epochs. The HBC 722 outburst and dip spectra also require a 2-component active + passive disk model, though with different $R_\mathrm{outer}$ values between the epochs. The plateau spectrum, however, requires only an active disk with the large $R_\mathrm{outer} = 100 \ R_\odot$. The difference in the $R_\mathrm{outer}$ evolution in the two systems hints at potentially distinct outburst mechanisms in the two systems. 

The evolution of the V960 Mon system post-outburst hints at an outside-in instability, originating further out in the disk then propagating inward. This would be consistent with something like gravitational instability in the disk, which has been observed at large scales in millimeter and NIR imaging \citep{Weber_V960MonSpirals_2023ApJ}. The outward propagation of the instability in HBC 722 indicates the trigger is likelier to have originated closer to the central star. This would be more consistent with a thermal-instability-driven outburst, though how the instability can extend to larger $R_\mathrm{outer}$ is not clearly understood \citep{Nayakshin_ThermalInstability_2024MNRAS}. 

Despite the differences in the evolution and geometry of the two systems, there are several elements in common between them. An especially important one is the emergence of the narrow absorption component that strengthens over time, which in HBC 722 we can attribute to a low velocity disk wind. In V960 Mon, the essentially 0 km s$^{-1}$ central velocity of the component makes this scenario less clear, but not impossible. The fact that the absorption grows as HBC 722 brightens/remains bright, whereas in V960 Mon it grows as the target fades may be evidence of a difference between the sources of the absorption. Detailed wind models of these systems are necessary to test the low-velocity high-line-opacity disk wind picture. 

The two systems also show very similar wind line evolution, where they initially show higher velocity blue-shifted absorption that over time weakens and becomes dominated by always-present low velocity absorption component and red-shifted emission (in the case of the H$\alpha$ and Ca II lines). As discussed in Section \ref{sec:hotWind}, this is also seen in V1057 Cyg and may be due to evolution in the velocity structure of the wind over years and decades following the outburst that may be in common between the three systems.  

\subsection{Comparison with Previous Literature}
Our best-fit models tend to produce lower $T_\mathrm{max}$ and higher $\dot{M}$ than the previously published $T_\mathrm{max} > 7000$ K and $\dot{M} \sim 10^{-5} \ M_\odot$ yr$^{-1}$ \citep{kospal_hbc722_2016A&A, Rodriguez_model_2022, Liu_fuorParameterSpace_2022ApJ}. This is due to two main differences in our approach to the model fitting: we assume a much smaller mass for the central object (0.2 $M_\odot$, compared with the previously typical 0.6 $M_\odot$) and we use the line broadening from the high resolution spectra to directly constrain $R_\mathrm{inner}$ and $i$ parameters in the SED fit. 

The smaller mass, which we demonstrated to be more consistent with the progenitor spectrum than the previously assumed $M_* =$ 0.6 $M_\odot$ \citep{miller_evidence_2011, kospal_hbc722_2016A&A} implies a central star radius of $R_* = 2.8$ $R_\odot$ (see Appendix \ref{app:progenitor}); much greater than the $R_* = 1.51 \ R_\odot$ derived from matching isochrones. While our best-fit $i = 79^{+2}_{-2}$ deg agrees with the $73^{+6}_{-15}$ deg reported by \citet{kospal_hbc722_2016A&A}, our constraint on $v_\mathrm{max} \sin \ i$ also required a large $R_\mathrm{inner} = 3.65^{+0.2}_{-0.3} \ R_\odot$ for the system, which is expected from the large progenitor radius though greater than the previous to 2.0 $R_\odot$ \citep{Rodriguez_model_2022, Liu_fuorParameterSpace_2022ApJ} values derived from SED fitting alone. The inclusion of the rotational broadening from the high resolution spectra as a constraint on the radius is critical to breaking the degeneracy in SED fitting between $M_*\dot{M}$ and $R_\mathrm{inner}$ and helps to provide a more reliable estimate of $R_\mathrm{inner}$. 

Despite the differences in our derived physical parameters of the disk, our best-fit $A_V = 2.3$ mag (0.7-1.0 mag smaller than previous results) and adopted distance to the source \citep[$\sim 200$ pc farther away,][]{connelley_near-infrared_2018} counteract one another to give a similar $L_\mathrm{acc}$ estimate. The effect of our smaller $M_*$, edge-on disk orientation, and similar $L_\mathrm{bol}$ conspire to require a much larger $\dot{M}$ than before. The large $R_\mathrm{inner}$ for the system drives the $T_\mathrm{max}$ to be much lower, since $T_\mathrm{max} \propto L_\mathrm{acc}/R_\mathrm{inner}^2$. We find our $T_\mathrm{max} \sim 5700-6000$ K is a better match to high resolution spectra at outburst and in the plateau than higher $T_\mathrm{max}$ models. 

The increasing $R_\mathrm{outer}$ over time that we discuss in Section \ref{sec:thinDisk} was also identified by \citet{kospal_hbc722_2016A&A}. Using $3-5$ $\mu$m photometry from WISE and Spitzer in their SED fits, they report an initial $R_\mathrm{outer} \sim 19 \ R_\odot$ that expands to $60 \ R_\odot$ between JD 2455500 to JD 2456300, which are similar to the values we report and imply an expansion rate only $1.5 \times$ faster than the one we find.

\section{Summary and Conclusions} \label{sec:summary}
We have used several years of photometry and spectroscopy of the HBC 722 system to constrain the behavior of the components in the system. This has also enabled us to better understand the nature of each of the components in this complex accretion disk . 

The HBC 722 system is composed of 3 identifiable components: a viscous, active accretion disk, a low-velocity disk wind, and a high velocity collimated outflow. From our SED + spectroscopic fitting we are able to constrain the disk parameters at outburst and find that the disk begins with an $R_\mathrm{outer} = 25 \ R_\odot$. Varying the $\dot{M}$, $R_\mathrm{outer}$ and $T_\mathrm{min}$ in the model to match the other epochs, we see that over time the active region expands from $R_\mathrm{outer} = 25 \ R_\odot$ in 2010 to $R_\mathrm{outer} = 35 \ R_\odot$ in 2011 to $R_\mathrm{outer} = 100 \ R_\odot$ in 2015, and that $T_\mathrm{min}$ of the disk decreased over that same time.

We also see strong variability in the two outflow components. The high velocity outflow is much stronger during the dip than during the outburst or plateau, contrary to the expectation that the outflow velocity should correlate with the accretion rate. This may indicate a change in the geometry of the flow, such that during the dip the outflow is less collimated and the higher velocity material points more in the direction of the observer. 

The low-velocity wind does not show much kinematic change but grows significantly in absorption strength over time. The absorption from this component even appears stronger than the disk absorption in some lines (e.g., Li I 6707) during the most recent HIRES epoch. 

The high and low velocity wind components may also be the source of the emission in H$\alpha$ and the Ca II lines if they are warm winds, with temperatures around 3000-5000 K. The line profiles are generally consistent with the profiles of hot wind models for CTTS systems. The exact wind temperature may not need to be as hot as that expected for CTTS winds, however, but could cool enough to be consistent with the atomic and molecular wind absorption we observe \citep{Lima_HalphaLineEmissionWinds_2010A&A}. 

Our spectroscopic analysis of the progenitor of the system reveals it to be a low-mass but highly inflated CTTS, with $M_* = 0.2 \ M_\odot$, $R_* = 2.8 \ R_\odot$, and $\dot{M} = 7.8 \times 10^{-8} \ M_\odot$ yr $^{-1}$. The accretion rate is typical of Class I YSOs and among the highest for Class II YSOs in this mass range.

Comparisons between the spectral evolution of HBC 722 and V960 Mon hint at potentially different outburst mechanisms operating in the two systems. It may be that the different mechanisms can be classified according to their post-outburst photometric evolution. In that case, the outburst of HBC 722 is most similar to those of V1515 Cyg and FU Ori, whereas the V960 Mon outburst is more similar to that of V1057 Cyg, though with different fading timescales. 

We strongly encourage more high and intermediate resolution follow-up of other FU Ori systems, especially in the short time following their outbursts. Sampling the post-outburst evolution (and ideally pre-outburst when possible) is critical to constraining the physics of the outburst mechanism.

\section{Acknowledgements}
The authors thank Mike Connelley for access to the 2015 SpeX SXD and LXD spectra published in \cite{connelley_near-infrared_2018}. We also acknowledge
Luke Bouma for the program add-on to obtain the 2021 Keck/HIRES spectrum. We thank the anonymous referee for their detailed comments and suggestions that improved the final manuscript.

This work has made use of the VALD database, operated at Uppsala University, the Institute of Astronomy RAS in Moscow, and the University of Vienna.

\facilities{Keck: I (HIRES), Keck: II (NIRSPEC), Hale (TripleSpec), IRTF (SpeX), Shane (Kast), Gemini:Gillett (IGRINS), AAVSO}

\bibliography{references}{}

\begin{thebibliography}{}
\expandafter\ifx\csname natexlab\endcsname\relax\def\natexlab#1{#1}\fi
\providecommand{\url}[1]{\href{#1}{#1}}
\providecommand{\dodoi}[1]{doi:~\href{http://doi.org/#1}{\nolinkurl{#1}}}
\providecommand{\doeprint}[1]{\href{http://ascl.net/#1}{\nolinkurl{http://ascl.net/#1}}}
\providecommand{\doarXiv}[1]{\href{https://arxiv.org/abs/#1}{\nolinkurl{https://arxiv.org/abs/#1}}}

\bibitem[{{Alencar} {et~al.}(2005){Alencar}, {Basri}, {Hartmann}, \& {Calvet}}]{Alencar_RWAurWind_2005A&A}
{Alencar}, S.~H.~P., {Basri}, G., {Hartmann}, L., \& {Calvet}, N. 2005, \aap, 440, 595, \dodoi{10.1051/0004-6361:20053315}

\bibitem[{{Baraffe} {et~al.}(2015){Baraffe}, {Homeier}, {Allard}, \& {Chabrier}}]{Baraffe_isochrones_2015A&A}
{Baraffe}, I., {Homeier}, D., {Allard}, F., \& {Chabrier}, G. 2015, \aap, 577, A42, \dodoi{10.1051/0004-6361/201425481}

\bibitem[{{Blandford} \& {Payne}(1982)}]{BlandfordPayneWind_1982MNRAS}
{Blandford}, R.~D., \& {Payne}, D.~G. 1982, \mnras, 199, 883, \dodoi{10.1093/mnras/199.4.883}

\bibitem[{{Calvet} {et~al.}(1993){Calvet}, {Hartmann}, \& {Kenyon}}]{Calvet_FUOriModel_1993}
{Calvet}, N., {Hartmann}, L., \& {Kenyon}, S.~J. 1993, \apj, 402, 623, \dodoi{10.1086/172164}

\bibitem[{{Carvalho} \& {Doppmann}(2024)}]{CarvalhoDoppmann_NIRSPEC_2024ascl}
{Carvalho}, A., \& {Doppmann}, G. 2024, {PypeIt-NIRSPEC: A PypeIt Module for Reducing Keck/NIRSPEC High Resolution Spectra}, Astrophysics Source Code Library, submitted

\bibitem[{{Carvalho} {et~al.}(2023{\natexlab{a}}){Carvalho}, {Hillenbrand}, \& {Seebeck}}]{Carvalho_V960MonSpectra_2023ApJ}
{Carvalho}, A., {Hillenbrand}, L., \& {Seebeck}, J. 2023{\natexlab{a}}, \apj, 958, 140, \dodoi{10.3847/1538-4357/acff59}

\bibitem[{{Carvalho} {et~al.}(2023{\natexlab{b}}){Carvalho}, {Hillenbrand}, {Hambsch}, {Dvorak}, {Sitko}, {Russell}, {Hammond}, {Connelley}, {Ashley}, \& {Hankins}}]{Carvalho_V960MonPhotometry_2023ApJ}
{Carvalho}, A.~S., {Hillenbrand}, L.~A., {Hambsch}, F.-J., {et~al.} 2023{\natexlab{b}}, \apj, 953, 86, \dodoi{10.3847/1538-4357/ace2cb}

\bibitem[{{Cieza} {et~al.}(2018){Cieza}, {Ru{\'\i}z-Rodr{\'\i}guez}, {Perez}, {Casassus}, {Williams}, {Zurlo}, {Principe}, {Hales}, {Prieto}, {Tobin}, {Zhu}, \& {Marino}}]{cieza_v883Ori_2018MNRAS}
{Cieza}, L.~A., {Ru{\'\i}z-Rodr{\'\i}guez}, D., {Perez}, S., {et~al.} 2018, \mnras, 474, 4347, \dodoi{10.1093/mnras/stx3059}

\bibitem[{Connelley \& Reipurth(2018)}]{connelley_near-infrared_2018}
Connelley, M.~S., \& Reipurth, B. 2018, \apj, 861, 145, \dodoi{10.3847/1538-4357/aaba7b}

\bibitem[{{Croswell} {et~al.}(1987){Croswell}, {Hartmann}, \& {Avrett}}]{croswell_FUOriHotSphericalWind_1987ApJ}
{Croswell}, K., {Hartmann}, L., \& {Avrett}, E.~H. 1987, \apj, 312, 227, \dodoi{10.1086/164865}

\bibitem[{{Cushing} {et~al.}(2004){Cushing}, {Vacca}, \& {Rayner}}]{Cushing_spextool_2004PASP}
{Cushing}, M.~C., {Vacca}, W.~D., \& {Rayner}, J.~T. 2004, \pasp, 116, 362, \dodoi{10.1086/382907}

\bibitem[{{Dupree} {et~al.}(2005){Dupree}, {Brickhouse}, {Smith}, \& {Strader}}]{Dupree_HE10830HotWind_2005ApJ}
{Dupree}, A.~K., {Brickhouse}, N.~S., {Smith}, G.~H., \& {Strader}, J. 2005, \apjl, 625, L131, \dodoi{10.1086/431323}

\bibitem[{Eilers \& Boelens(2005)}]{eilers2005baseline}
Eilers, P.~H., \& Boelens, H.~F. 2005, Leiden University Medical Centre Report, 1, 5

\bibitem[{{Fang} {et~al.}(2020){Fang}, {Hillenbrand}, {Kim}, {Findeisen}, {Herczeg}, {Carpenter}, {Rebull}, \& {Wang}}]{Fang_NorthAmerica_2020ApJ}
{Fang}, M., {Hillenbrand}, L.~A., {Kim}, J.~S., {et~al.} 2020, \apj, 904, 146, \dodoi{10.3847/1538-4357/abba84}

\bibitem[{{Fiorellino} {et~al.}(2023){Fiorellino}, {Tychoniec}, {Cruz-S{\'a}enz de Miera}, {Antoniucci}, {K{\'o}sp{\'a}l}, {Manara}, {Nisini}, \& {Rosotti}}]{Fiorellino_ClassIAccretionRates_2023ApJ}
{Fiorellino}, E., {Tychoniec}, {\L}., {Cruz-S{\'a}enz de Miera}, F., {et~al.} 2023, \apj, 944, 135, \dodoi{10.3847/1538-4357/aca320}

\bibitem[{Foreman-Mackey(2016)}]{corner_FM_2016}
Foreman-Mackey, D. 2016, The Journal of Open Source Software, 1, 24, \dodoi{10.21105/joss.00024}

\bibitem[{{Foreman-Mackey} {et~al.}(2013){Foreman-Mackey}, {Hogg}, {Lang}, \& {Goodman}}]{FM_emcee_2013PASP}
{Foreman-Mackey}, D., {Hogg}, D.~W., {Lang}, D., \& {Goodman}, J. 2013, \pasp, 125, 306, \dodoi{10.1086/670067}

\bibitem[{{Hamann} \& {Persson}(1992{\natexlab{a}})}]{HamannPersonn_EmissionLineStudiesI_TTSs_1992ApJS}
{Hamann}, F., \& {Persson}, S.~E. 1992{\natexlab{a}}, \apjs, 82, 247, \dodoi{10.1086/191715}

\bibitem[{{Hamann} \& {Persson}(1992{\natexlab{b}})}]{HamannPersson_EmissionLineStudies_II_HerbigStars_1992ApJS}
---. 1992{\natexlab{b}}, \apjs, 82, 285, \dodoi{10.1086/191716}

\bibitem[{{Hartmann} \& {Calvet}(1995)}]{Hartmann_FUOriWindConstraints_1995AJ}
{Hartmann}, L., \& {Calvet}, N. 1995, \aj, 109, 1846, \dodoi{10.1086/117411}

\bibitem[{{Hartmann} {et~al.}(2016){Hartmann}, {Herczeg}, \& {Calvet}}]{Hartmann_review_2016ARA&A}
{Hartmann}, L., {Herczeg}, G., \& {Calvet}, N. 2016, \araa, 54, 135, \dodoi{10.1146/annurev-astro-081915-023347}

\bibitem[{Hartmann \& Kenyon(1996)}]{hartmann_fu_1996}
Hartmann, L., \& Kenyon, S.~J. 1996, \araa, 34, 207, \dodoi{10.1146/annurev.astro.34.1.207}

\bibitem[{{Herbig}(1966)}]{Herbig_FUOri_interpretation_1966VA}
{Herbig}, G.~H. 1966, Vistas in Astronomy, 8, 109, \dodoi{10.1016/0083-6656(66)90025-0}

\bibitem[{Herbig {et~al.}(2003)Herbig, Petrov, \& Duemmler}]{herbig_high-resolution_2003}
Herbig, G.~H., Petrov, P.~P., \& Duemmler, R. 2003, \apj, 595, 384, \dodoi{10.1086/377194}

\bibitem[{{Herczeg} \& {Hillenbrand}(2008)}]{HerczegHillenbrand_UVExcessAccretion_2008ApJ}
{Herczeg}, G.~J., \& {Hillenbrand}, L.~A. 2008, \apj, 681, 594, \dodoi{10.1086/586728}

\bibitem[{{Herter} {et~al.}(2008){Herter}, {Henderson}, {Wilson}, {Matthews}, {Rahmer}, {Bonati}, {Muirhead}, {Adams}, {Lloyd}, {Skrutskie}, {Moon}, {Parshley}, {Nelson}, {Martinache}, \& {Gull}}]{Herter_TripleSpecInstrument_2008SPIE}
{Herter}, T.~L., {Henderson}, C.~P., {Wilson}, J.~C., {et~al.} 2008, in Society of Photo-Optical Instrumentation Engineers (SPIE) Conference Series, Vol. 7014, Ground-based and Airborne Instrumentation for Astronomy II, ed. I.~S. {McLean} \& M.~M. {Casali}, 70140X, \dodoi{10.1117/12.789660}

\bibitem[{{Hillenbrand} {et~al.}(2023){Hillenbrand}, {Carvalho}, {van Roestel}, \& {De}}]{Hillenbrand_RNO54_letter_2023ApJ}
{Hillenbrand}, L.~A., {Carvalho}, A., {van Roestel}, J., \& {De}, K. 2023, \apjl, 958, L27, \dodoi{10.3847/2041-8213/ad0be0}

\bibitem[{Hillenbrand {et~al.}(2015)Hillenbrand, Reipurth, \& Connelley}]{hillenbrand_optical_2015}
Hillenbrand, L.~A., Reipurth, B., \& Connelley, M.~S. 2015, The Astronomer's Telegram, 8331.
\newblock \url{http://adsabs.harvard.edu/abs/2015ATel.8331....1H}

\bibitem[{{Howard} {et~al.}(2010){Howard}, {Johnson}, {Marcy}, {Fischer}, {Wright}, {Bernat}, {Henry}, {Peek}, {Isaacson}, {Apps}, {Endl}, {Cochran}, {Valenti}, {Anderson}, \& {Piskunov}}]{Howard_CPSI_2010ApJ}
{Howard}, A.~W., {Johnson}, J.~A., {Marcy}, G.~W., {et~al.} 2010, \apj, 721, 1467, \dodoi{10.1088/0004-637X/721/2/1467}

\bibitem[{{Husser} {et~al.}(2013){Husser}, {Wende-von Berg}, {Dreizler}, {Homeier}, {Reiners}, {Barman}, \& {Hauschildt}}]{Husser_Phoenix_2013A&A}
{Husser}, T.~O., {Wende-von Berg}, S., {Dreizler}, S., {et~al.} 2013, \aap, 553, A6, \dodoi{10.1051/0004-6361/201219058}

\bibitem[{{Kenyon} {et~al.}(1988){Kenyon}, {Hartmann}, \& {Hewett}}]{Kenyon_FUOri_disks_1988ApJ}
{Kenyon}, S.~J., {Hartmann}, L., \& {Hewett}, R. 1988, \apj, 325, 231, \dodoi{10.1086/165999}

\bibitem[{{Korotin} {et~al.}(2020){Korotin}, {Andrievsky}, {Caffau}, {Bonifacio}, \& {Oliva}}]{Korotin_SrII_NLTE_2020MNRAS}
{Korotin}, S.~A., {Andrievsky}, S.~M., {Caffau}, E., {Bonifacio}, P., \& {Oliva}, E. 2020, \mnras, 496, 2462, \dodoi{10.1093/mnras/staa1707}

\bibitem[{{K{\'o}sp{\'a}l} {et~al.}(2016){K{\'o}sp{\'a}l}, {{\'A}brah{\'a}m}, {Acosta-Pulido}, {Dunham}, {Garc{\'\i}a-{\'A}lvarez}, {Hogerheijde}, {Kun}, {Mo{\'o}r}, {Farkas}, {Hajdu}, {Hodos{\'a}n}, {Kov{\'a}cs}, {Kriskovics}, {Marton}, {Moln{\'a}r}, {P{\'a}l}, {S{\'a}rneczky}, {S{\'o}dor}, {Szak{\'a}ts}, {Szalai}, {Szegedi-Elek}, {Szing}, {T{\'o}th}, {Vida}, \& {Vink{\'o}}}]{kospal_hbc722_2016A&A}
{K{\'o}sp{\'a}l}, {\'A}., {{\'A}brah{\'a}m}, P., {Acosta-Pulido}, J.~A., {et~al.} 2016, \aap, 596, A52, \dodoi{10.1051/0004-6361/201528061}

\bibitem[{{Kuhn} \& {Hillenbrand}(2020)}]{Kuhn_NorthAmericaDistance_2020RNAAS}
{Kuhn}, M.~A., \& {Hillenbrand}, L.~A. 2020, Research Notes of the American Astronomical Society, 4, 224, \dodoi{10.3847/2515-5172/abd18a}

\bibitem[{{Kurosawa} {et~al.}(2011){Kurosawa}, {Romanova}, \& {Harries}}]{Kurosawa_HydHelWindLineProfiles_2011MNRAS}
{Kurosawa}, R., {Romanova}, M.~M., \& {Harries}, T.~J. 2011, \mnras, 416, 2623, \dodoi{10.1111/j.1365-2966.2011.19216.x}

\bibitem[{{Lee} {et~al.}(2015){Lee}, {Park}, {Green}, {Cochran}, {Kang}, {Lee}, \& {Sung}}]{Lee_HBC722_2015ApJ}
{Lee}, J.-E., {Park}, S., {Green}, J.~D., {et~al.} 2015, \apj, 807, 84, \dodoi{10.1088/0004-637X/807/1/84}

\bibitem[{{Lee} {et~al.}(2017){Lee}, {Gullikson}, \& {Kaplan}}]{Lee_IGRINS_Pipeline_2017zndo}
{Lee}, J.-J., {Gullikson}, K., \& {Kaplan}, K. 2017, {igrins/plp 2.2.0},  Zenodo, \dodoi{10.5281/zenodo.845059}

\bibitem[{{Lima} {et~al.}(2010){Lima}, {Alencar}, {Calvet}, {Hartmann}, \& {Muzerolle}}]{Lima_HalphaLineEmissionWinds_2010A&A}
{Lima}, G.~H.~R.~A., {Alencar}, S.~H.~P., {Calvet}, N., {Hartmann}, L., \& {Muzerolle}, J. 2010, \aap, 522, A104, \dodoi{10.1051/0004-6361/201014490}

\bibitem[{{Liu} {et~al.}(2022){Liu}, {Herczeg}, {Johnstone}, {Contreras-Pe{\~n}a}, {Lee}, {Yang}, {Zhou}, {Yoon}, {Lee}, {Kunitomo}, \& {Jose}}]{Liu_fuorParameterSpace_2022ApJ}
{Liu}, H., {Herczeg}, G.~J., {Johnstone}, D., {et~al.} 2022, \apj, 936, 152, \dodoi{10.3847/1538-4357/ac84d2}

\bibitem[{{L{\'o}pez} {et~al.}(2020){L{\'o}pez}, {Hoffman}, {Doppmann}, {Fitzgerald}, {Johnson}, {Kassis}, {Lanclos}, {Lyke}, {Martin}, {McLean}, {Sohn}, \& {Weiss}}]{Lopez_NIRSPECUpgradePerformance_2020SPIE11447E}
{L{\'o}pez}, R.~A., {Hoffman}, E.~B., {Doppmann}, G., {et~al.} 2020, in Society of Photo-Optical Instrumentation Engineers (SPIE) Conference Series, Vol. 11447, Ground-based and Airborne Instrumentation for Astronomy VIII, ed. C.~J. {Evans}, J.~J. {Bryant}, \& K.~{Motohara}, 114476B, \dodoi{10.1117/12.2563075}

\bibitem[{{Maehara} {et~al.}(2014){Maehara}, {Kojima}, \& {Fujii}}]{Maehara_2014ATel.6770}
{Maehara}, H., {Kojima}, T., \& {Fujii}, M. 2014, The Astronomer's Telegram, 6770, 1

\bibitem[{{Manara} {et~al.}(2023){Manara}, {Ansdell}, {Rosotti}, {Hughes}, {Armitage}, {Lodato}, \& {Williams}}]{Manara_PPVIIChapter_2023ASPC}
{Manara}, C.~F., {Ansdell}, M., {Rosotti}, G.~P., {et~al.} 2023, in Astronomical Society of the Pacific Conference Series, Vol. 534, Protostars and Planets VII, ed. S.~{Inutsuka}, Y.~{Aikawa}, T.~{Muto}, K.~{Tomida}, \& M.~{Tamura}, 539, \dodoi{10.48550/arXiv.2203.09930}

\bibitem[{{Martin} {et~al.}(2018){Martin}, {Fitzgerald}, {McLean}, {Doppmann}, {Kassis}, {Aliado}, {Canfield}, {Johnson}, {Kress}, {Lanclos}, {Magnone}, {Sohn}, {Wang}, \& {Weiss}}]{Martin_NIRSPECupgrade_2018SPIE10702E}
{Martin}, E.~C., {Fitzgerald}, M.~P., {McLean}, I.~S., {et~al.} 2018, in Society of Photo-Optical Instrumentation Engineers (SPIE) Conference Series, Vol. 10702, Ground-based and Airborne Instrumentation for Astronomy VII, ed. C.~J. {Evans}, L.~{Simard}, \& H.~{Takami}, 107020A, \dodoi{10.1117/12.2312266}

\bibitem[{{McLean} {et~al.}(1998){McLean}, {Becklin}, {Bendiksen}, {Brims}, {Canfield}, {Figer}, {Graham}, {Hare}, {Lacayanga}, {Larkin}, {Larson}, {Levenson}, {Magnone}, {Teplitz}, \& {Wong}}]{McLean_nirspecDesign_1998SPIE}
{McLean}, I.~S., {Becklin}, E.~E., {Bendiksen}, O., {et~al.} 1998, in Society of Photo-Optical Instrumentation Engineers (SPIE) Conference Series, Vol. 3354, Infrared Astronomical Instrumentation, ed. A.~M. {Fowler}, 566--578, \dodoi{10.1117/12.317283}

\bibitem[{Miller {et~al.}(2011)Miller, Hillenbrand, Covey, Poznanski, Silverman, Kleiser, Rojas-Ayala, Muirhead, Cenko, Bloom, Kasliwal, Filippenko, Law, Ofek, Dekany, Rahmer, Hale, Smith, Quimby, Nugent, Jacobsen, Zolkower, Velur, Walters, Henning, Bui, McKenna, Kulkarni, Klein, Kandrashoff, \& Morton}]{miller_evidence_2011}
Miller, A.~A., Hillenbrand, L.~A., Covey, K.~R., {et~al.} 2011, \apj, 730, 80, \dodoi{10.1088/0004-637X/730/2/80}

\bibitem[{{Miller} \& {Stone}(1993)}]{MillerStone_KastSpec_1993}
{Miller}, J., \& {Stone}, R. 1993, Lick Observatory Technical Report, 66

\bibitem[{{Milliner} {et~al.}(2019){Milliner}, {Matthews}, {Long}, \& {Hartmann}}]{Milliner_FUOri_NaDWindModels_2019MNRAS}
{Milliner}, K., {Matthews}, J.~H., {Long}, K.~S., \& {Hartmann}, L. 2019, \mnras, 483, 1663, \dodoi{10.1093/mnras/sty3197}

\bibitem[{{Nayakshin} {et~al.}(2024){Nayakshin}, {Cruz S{\'a}enz de Miera}, {K{\'o}sp{\'a}l}, {{\'C}alovi{\'c}}, {Eisl{\"o}ffel}, \& {Lin}}]{Nayakshin_ThermalInstability_2024MNRAS}
{Nayakshin}, S., {Cruz S{\'a}enz de Miera}, F., {K{\'o}sp{\'a}l}, {\'A}., {et~al.} 2024, \mnras, 530, 1749, \dodoi{10.1093/mnras/stae877}

\bibitem[{{Nguyen} {et~al.}(2022){Nguyen}, {Costa}, {Girardi}, {Volpato}, {Bressan}, {Chen}, {Marigo}, {Fu}, \& {Goudfrooij}}]{Nguyen_parsec_2022A&A}
{Nguyen}, C.~T., {Costa}, G., {Girardi}, L., {et~al.} 2022, \aap, 665, A126, \dodoi{10.1051/0004-6361/202244166}

\bibitem[{Oller-Moreno {et~al.}(2014)Oller-Moreno, Pardo, Jim{\'e}nez-Soto, Samitier, \& Marco}]{oller2014adaptive}
Oller-Moreno, S., Pardo, A., Jim{\'e}nez-Soto, J.~M., Samitier, J., \& Marco, S. 2014, in 2014 IEEE 11th International Multi-Conference on Systems, Signals \& Devices (SSD14), IEEE, 1--5

\bibitem[{{Park} {et~al.}(2014){Park}, {Jaffe}, {Yuk}, {Chun}, {Pak}, {Kim}, {Pavel}, {Lee}, {Oh}, {Jeong}, {Sim}, {Lee}, {Nguyen Le}, {Strubhar}, {Gully-Santiago}, {Oh}, {Cha}, {Moon}, {Park}, {Brooks}, {Ko}, {Han}, {Nah}, {Hill}, {Lee}, {Barnes}, {Yu}, {Kaplan}, {Mace}, {Kim}, {Lee}, {Hwang}, \& {Park}}]{Park_IGRINS_2014SPIE}
{Park}, C., {Jaffe}, D.~T., {Yuk}, I.-S., {et~al.} 2014, in Society of Photo-Optical Instrumentation Engineers (SPIE) Conference Series, Vol. 9147, Ground-based and Airborne Instrumentation for Astronomy V, ed. S.~K. {Ramsay}, I.~S. {McLean}, \& H.~{Takami}, 91471D, \dodoi{10.1117/12.2056431}

\bibitem[{Park {et~al.}(2020)Park, Lee, Pyo, Jaffe, Mace, Sung, Lee, Kang, Oh, Yoon, Yoon, \& Green}]{park_high-resolution_2020}
Park, S., Lee, J.-E., Pyo, T.-S., {et~al.} 2020, \apj, 900, 36, \dodoi{10.3847/1538-4357/aba532}

\bibitem[{{Petrov} \& {Herbig}(2008)}]{PetrovHerbig_FUOriLineStructure_2008AJ}
{Petrov}, P.~P., \& {Herbig}, G.~H. 2008, \aj, 136, 676, \dodoi{10.1088/0004-6256/136/2/676}

\bibitem[{{Petrov} {et~al.}(2014){Petrov}, {Kurosawa}, {Romanova}, {Gameiro}, {Fernandez}, {Babina}, \& {Artemenko}}]{Petrov_V1331_FacingTheWind_2014MNRAS}
{Petrov}, P.~P., {Kurosawa}, R., {Romanova}, M.~M., {et~al.} 2014, \mnras, 442, 3643, \dodoi{10.1093/mnras/stu1131}

\bibitem[{{Prochaska} {et~al.}(2020){Prochaska}, {Hennawi}, {Westfall}, {Cooke}, {Wang}, {Hsyu}, {Davies}, {Farina}, \& {Pelliccia}}]{prochaska_pypeit_2020JOSS}
{Prochaska}, J., {Hennawi}, J., {Westfall}, K., {et~al.} 2020, The Journal of Open Source Software, 5, 2308, \dodoi{10.21105/joss.02308}

\bibitem[{{Rayner} {et~al.}(2009){Rayner}, {Cushing}, \& {Vacca}}]{Rayner_IRTF_2009}
{Rayner}, J.~T., {Cushing}, M.~C., \& {Vacca}, W.~D. 2009, \apjs, 185, 289, \dodoi{10.1088/0067-0049/185/2/289}

\bibitem[{{Rayner} {et~al.}(2003){Rayner}, {Toomey}, {Onaka}, {Denault}, {Stahlberger}, {Vacca}, {Cushing}, \& {Wang}}]{Rayner_spex_2003PASP}
{Rayner}, J.~T., {Toomey}, D.~W., {Onaka}, P.~M., {et~al.} 2003, \pasp, 115, 362, \dodoi{10.1086/367745}

\bibitem[{{Rodriguez} \& {Hillenbrand}(2022)}]{Rodriguez_model_2022}
{Rodriguez}, A.~C., \& {Hillenbrand}, L.~A. 2022, \apj, 927, 144, \dodoi{10.3847/1538-4357/ac496b}

\bibitem[{{Ryabchikova} {et~al.}(2015){Ryabchikova}, {Piskunov}, {Kurucz}, {Stempels}, {Heiter}, {Pakhomov}, \& {Barklem}}]{vald_reference_2015PhyS}
{Ryabchikova}, T., {Piskunov}, N., {Kurucz}, R.~L., {et~al.} 2015, \physscr, 90, 054005, \dodoi{10.1088/0031-8949/90/5/054005}

\bibitem[{{Semkov} {et~al.}(2021){Semkov}, {Ibryamov}, \& {Peneva}}]{Semkov_HBC722Photometry_2021Symm}
{Semkov}, E., {Ibryamov}, S., \& {Peneva}, S. 2021, Symmetry, 13, 2433, \dodoi{10.3390/sym13122433}

\bibitem[{{Semkov} \& {Peneva}(2010)}]{Semkov_HBC722Detection_2010ATel.2801}
{Semkov}, E., \& {Peneva}, S. 2010, The Astronomer's Telegram, 2801, 1

\bibitem[{Semkov {et~al.}(2013)Semkov, Peneva, Munari, Dennefeld, Mito, Dimitrov, Ibryamov, \& Stoyanov}]{semkov_photometric_2013}
Semkov, E.~H., Peneva, S.~P., Munari, U., {et~al.} 2013, Astronomy \& Astrophysics, 556, A60, \dodoi{10.1051/0004-6361/201321732}

\bibitem[{{Shakura} \& {Sunyaev}(1973)}]{Shakura_sunyaev_alpha_1973A&A}
{Shakura}, N.~I., \& {Sunyaev}, R.~A. 1973, \aap, 24, 337

\bibitem[{{Szab{\'o}} {et~al.}(2021){Szab{\'o}}, {K{\'o}sp{\'a}l}, {{\'A}brah{\'a}m}, {Park}, {Siwak}, {Green}, {Mo{\'o}r}, {P{\'a}l}, {Acosta-Pulido}, {Lee}, {Cseh}, {Cs{\"o}rnyei}, {Hanyecz}, {K{\"o}nyves-T{\'o}th}, {Krezinger}, {Kriskovics}, {Ordasi}, {S{\'a}rneczky}, {Seli}, {Szak{\'a}ts}, {Szing}, \& {Vida}}]{Szabo_V1057cyg_2021ApJ}
{Szab{\'o}}, Z.~M., {K{\'o}sp{\'a}l}, {\'A}., {{\'A}brah{\'a}m}, P., {et~al.} 2021, \apj, 917, 80, \dodoi{10.3847/1538-4357/ac04b3}

\bibitem[{{Vacca} {et~al.}(2003){Vacca}, {Cushing}, \& {Rayner}}]{Vacca_telluricMethod_2003PASP}
{Vacca}, W.~D., {Cushing}, M.~C., \& {Rayner}, J.~T. 2003, \pasp, 115, 389, \dodoi{10.1086/346193}

\bibitem[{{Vogt} {et~al.}(1994){Vogt}, {Allen}, {Bigelow}, {Bresee}, {Brown}, {Cantrall}, {Conrad}, {Couture}, {Delaney}, {Epps}, {Hilyard}, {Hilyard}, {Horn}, {Jern}, {Kanto}, {Keane}, {Kibrick}, {Lewis}, {Osborne}, {Pardeilhan}, {Pfister}, {Ricketts}, {Robinson}, {Stover}, {Tucker}, {Ward}, \& {Wei}}]{Vogt1994}
{Vogt}, S.~S., {Allen}, S.~L., {Bigelow}, B.~C., {et~al.} 1994, in Society of Photo-Optical Instrumentation Engineers (SPIE) Conference Series, Vol. 2198, Instrumentation in Astronomy VIII, ed. D.~L. {Crawford} \& E.~R. {Craine}, 362, \dodoi{10.1117/12.176725}

\bibitem[{{Weber} {et~al.}(2023){Weber}, {P{\'e}rez}, {Zurlo}, {Miley}, {Hales}, {Cieza}, {Principe}, {C{\'a}rcamo}, {Garufi}, {K{\'o}sp{\'a}l}, {Takami}, {Kastner}, {Zhu}, \& {Williams}}]{Weber_V960MonSpirals_2023ApJ}
{Weber}, P., {P{\'e}rez}, S., {Zurlo}, A., {et~al.} 2023, \apjl, 952, L17, \dodoi{10.3847/2041-8213/ace186}

\bibitem[{{Welty} {et~al.}(1992){Welty}, {Strom}, {Edwards}, {Kenyon}, \& {Hartmann}}]{welty_FUOriV1057CygDiskModelAndWinds_1992ApJ}
{Welty}, A.~D., {Strom}, S.~E., {Edwards}, S., {Kenyon}, S.~J., \& {Hartmann}, L.~W. 1992, \apj, 397, 260, \dodoi{10.1086/171785}

\bibitem[{{Zhu} {et~al.}(2009){Zhu}, {Espaillat}, {Hinkle}, {Hernandez}, {Hartmann}, \& {Calvet}}]{Zhu_FUOriDifferentialRotation_2009ApJ}
{Zhu}, Z., {Espaillat}, C., {Hinkle}, K., {et~al.} 2009, \apjl, 694, L64, \dodoi{10.1088/0004-637X/694/1/L64}

\bibitem[{{Zhu} {et~al.}(2020){Zhu}, {Jiang}, \& {Stone}}]{Zhu_outburst_FUOri_2020MNRAS}
{Zhu}, Z., {Jiang}, Y.-F., \& {Stone}, J.~M. 2020, \mnras, 495, 3494, \dodoi{10.1093/mnras/staa952}

\end{thebibliography}
\bibliographystyle{aasjournal}

\appendix 

\restartappendixnumbering

\section{Progenitor Spectral Type} \label{app:progenitor}
\begin{figure}[!h]
    \centering
    \includegraphics[width = 0.54\linewidth]{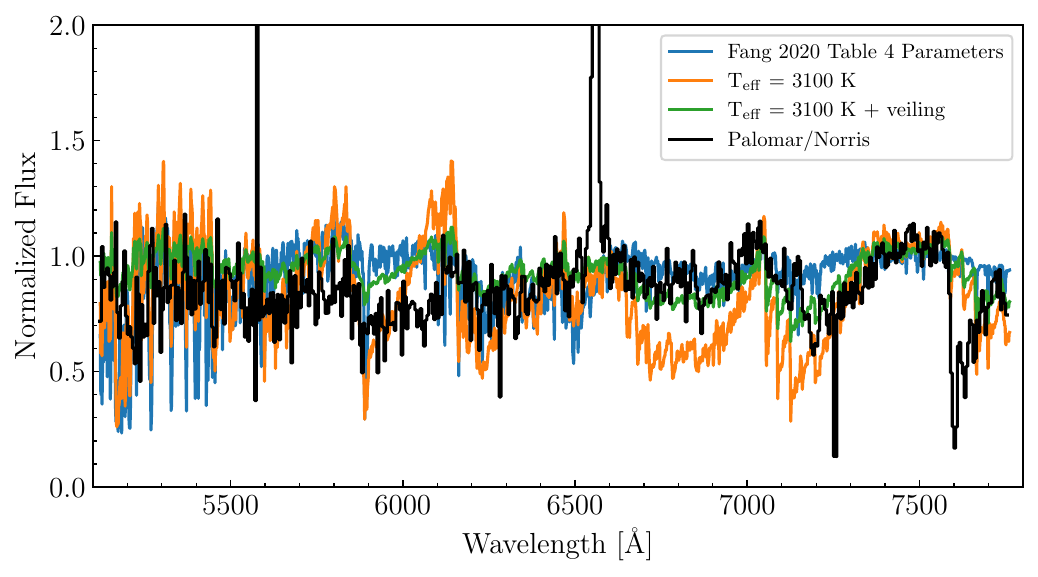}
    \includegraphics[width = 0.44\linewidth]{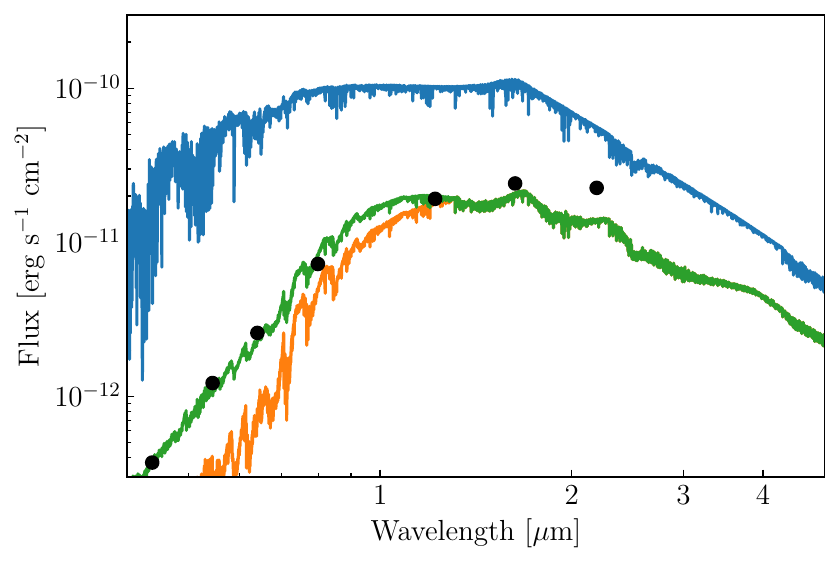}
    \caption{The progenitor data compared with models using the \citet{Fang_NorthAmerica_2020ApJ} reported best-fit parameters and our adopted parameters. \textbf{Left:} The 1998 Norris spectrum, binned to $R = 1000$, showing deep TiO absorption more consistent with the veiled late-type (M4-M5, green) model spectrum than the unveiled K7 model (blue). \textbf{Right:} The progenitor SED (black dots), using photometry from \citet{miller_evidence_2011}, compared with the different models for the progenitor spectrum (color lines). The model with our adopted $T_\mathrm{eff} = 3100$ K, $R_* = 2.8 \ R_\odot$, $r_{7465} = 0.7$ and $A_V = 2.5$ mag, shown in green, is a good fit to both the SED and the Norris spectrum.  }
    \label{fig:norris}
\end{figure}

To better constrain the physical parameters of our disk model, in particular $M_*$, we made use of the pre-outburst spectrum of the HBC 722 progenitor published in \citet{Fang_NorthAmerica_2020ApJ}. The spectrum, spanning 5100-7800 \AA, was obtained in 1998 with the Norris multi-object spectrograph on the Palomar 5-m Hale telescope at an estimated resolution $R \sim 2000$. \citet{Fang_NorthAmerica_2020ApJ} fit both empirical and model template spectra to the Norris spectrum to determine the spectral type, veiling, and $A_V$. In their first spectral type calculation, they report a spectral type of M4.4 ($T_\mathrm{eff} \sim 3000$ K) for the progenitor. However, the second step of their analysis included a post-outburst spectrum and their final spectral type of K6.6 adopted for the source, as reported in their Table 4, is not consistent with the pre-outburst Norris spectrum (see Figure \ref{fig:norris}). As a result, their final adopted $L_*$ and $A_V$ for the system are unreliable because they may have been derived by fitting the outburst spectrum, rather than the progenitor spectrum.

To establish reliable $T_\mathrm{eff}$, $R_*$, $A_V$, and veiling values for the HBC 722 progenitor, we reanalyze the Norris spectrum and the pre-outburst SED \citep[using the photometry from][]{miller_evidence_2011}. We first use a $T_\mathrm{eff} = 3100$ K BT-Settl model spectrum simultaneously produce models of both the SED and medium-resolution spectrum. This consistent with an M4.3 spectral type according to \citet{Fang_NorthAmerica_2020ApJ} (and close to their reported M4.4 from their initial spectral type fit), which we do find to be a good match for deep TiO absorption in the progenitor spectrum. 

We then vary the angular size of the progenitor to match the $H$ band flux and find $R_* = 2.8 \ R_\odot$, assuming our adopted distance of 745 pc, indicating the star was highly inflated relative to typical CTTSs \citep{Baraffe_isochrones_2015A&A, Nguyen_parsec_2022A&A}. The final step is to vary the veiling, $r_{7465}$, and the $A_V$, until both the SED and Norris spectra are well-matched. Our best-fit values for the progenitor are then $T_\mathrm{eff} = 3100$ K, a veiling of $r_{7465}=0.7$ , $A_V = 2.5$ mag, and a $R_* = 2.8 \ R_\odot$. The $T_\mathrm{eff}$ and $R_*$ correspond to $M_* \sim 0.2 \ M_\odot$ in the $100$ kyr PARSEC isochrone \citep{Nguyen_parsec_2022A&A}. The best-fit models (with and without veiling) are shown in Figure \ref{fig:norris} alongside a model that adopts the \citet{Fang_NorthAmerica_2020ApJ} parameters for the system. 

We can also follow the procedure in \citet{HerczegHillenbrand_UVExcessAccretion_2008ApJ} to estimate the mass accretion rate of the progenitor based on the H$\alpha$ line luminosity, $L_\mathrm{H\alpha}$. We first use the $A_V$-corrected continuum flux from the best-fit SED model, $F_c = 2.388 \times 10^{-15}$ erg s$^{-1}$ cm$^{-2}$ $\mathrm{\AA}^{-1}$, to estimate the flux of the H$\alpha$ line. We then integrate the line from 6545 \AA\ to 6580 \AA\ to get $F_\mathrm{H\alpha} = 2.43 \times 10^{-13}$ erg s$^{-1}$ cm$^{-2}$ or , $L_\mathrm{H\alpha} = 4.19 \times 10^{-3} \ L_\odot$, using the distance to the target, 745 pc. We convert this to an accretion luminosity using the prescription adopted in \citet{HerczegHillenbrand_UVExcessAccretion_2008ApJ}, $\log L_\mathrm{acc} = 1.2 \log H_\alpha + 2$, yielding $L_\mathrm{acc}$ = 0.14 $L_\odot$ and $L_\mathrm{acc}/L_\mathrm{bol} = 0.22$. Using Equation 1 from \citet{HerczegHillenbrand_UVExcessAccretion_2008ApJ}, $\dot{M} \sim 1.25 \ L_\mathrm{acc} R_* / (G M_*)$, we estimate $\dot{M} = 7.8 \times 10^{-8} \ M_\odot \ \mathrm{yr}^{-1}$. This would make the HBC 722 progenitor one of the most rapidly accreting stars with $M_* \sim 0.2 \ M_\odot$ \citep{Manara_PPVIIChapter_2023ASPC}, although it is consistent with measured accretion rates of Class I protostars \citep{Fiorellino_ClassIAccretionRates_2023ApJ}. 

\restartappendixnumbering

\section{Line Profiles of \ion{O}{1} Wind Lines} \label{app:KandOAbsorption}

\begin{figure*}[!b]
    \centering
    \includegraphics[width = 0.32\linewidth]{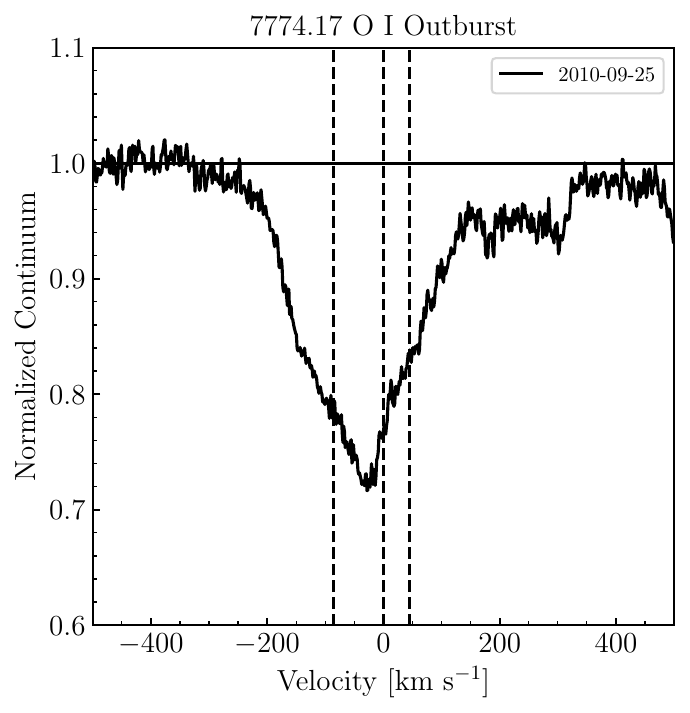}
    \includegraphics[width = 0.32\linewidth]{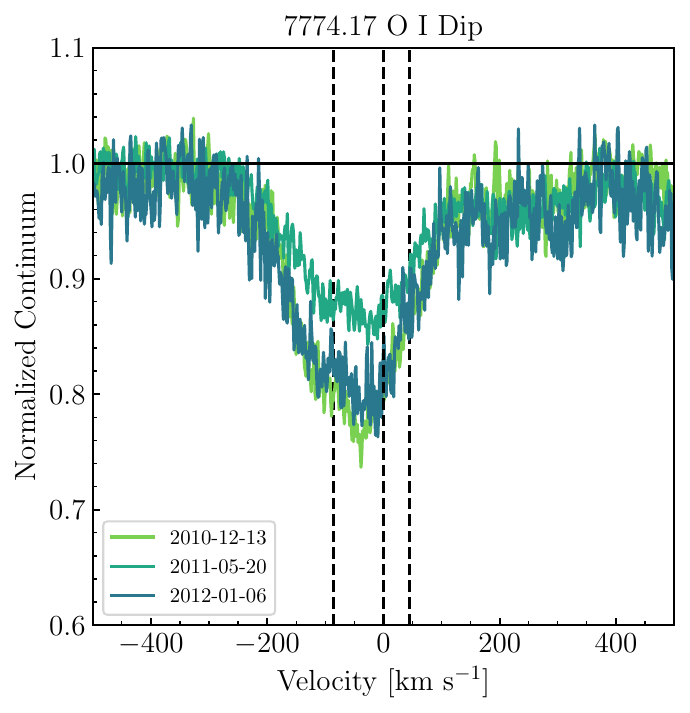}
    \includegraphics[width = 0.32\linewidth]{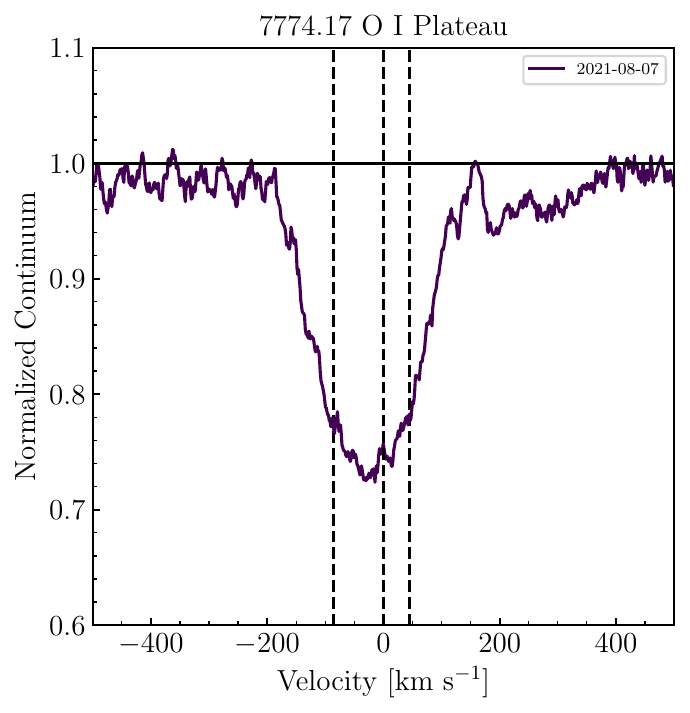}
    \newline
    \hfill
    \includegraphics[width = 0.32\linewidth]{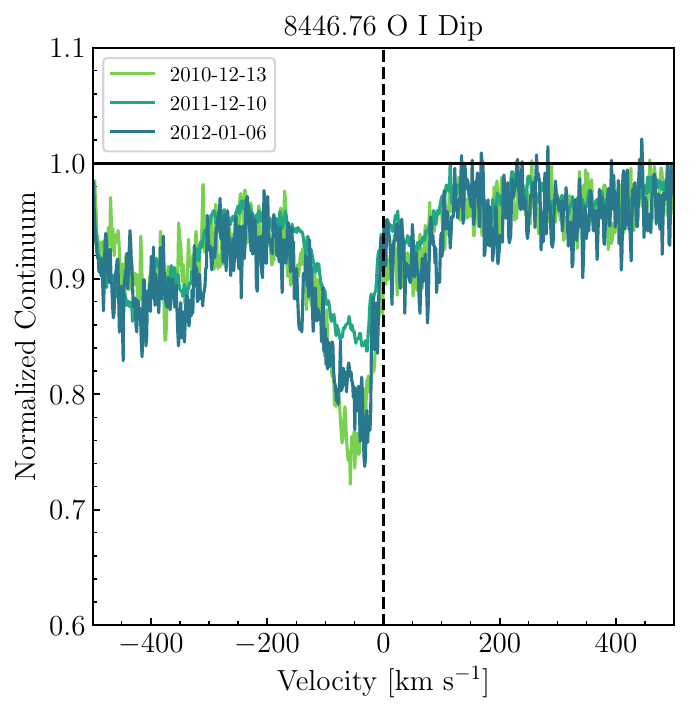}
    \includegraphics[width = 0.32\linewidth]{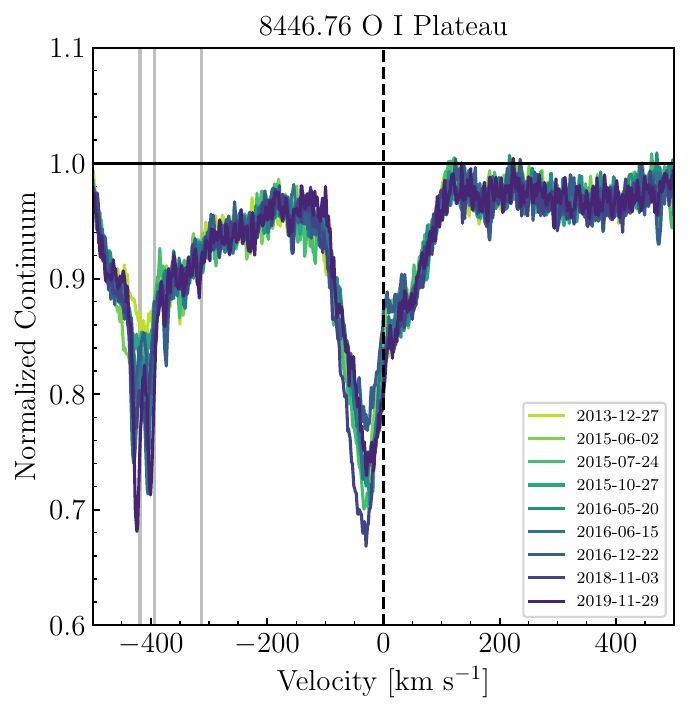}
    \caption{Line profiles of red/NIR O features known to trace winds, shown in the HIRES spectra for the outburst epoch (left column), dip epochs (middle column) and plateau epochs (right column). \textbf{Upper Row:} The line profiles of the O I 7773 \AA\ triplet, which is typically only seen in higher temperature atmospheres. The absorption at $+ 250$ km s$^{-1}$ is disk atmospheric absorption from the Fe I 7780 feature. \textbf{Lower Row:} The O I 8446 \AA\ triplet, which appeared only in our dip and plateau epochs. The grey lines at $-390$ and $-420$ km s$^{-1}$ mark the locations of the Ti I 8434.961 and 8435.653 \AA\ lines, both of which have EP $= 0.8$ eV. The broader absorption marked at $-310$ km s$^{-1}$ is mostly due to the Paschen line at 8837 \AA.}
    \label{fig:OandKWindLines}
\end{figure*}

In Section \ref{sec:windLines}, we identify the main absorption components of several lines typically understood to trace outflows in young stars. Here, we include 2 lines that were omitted from that discussion: the O I 7773 \AA\ triplet (at 7771, 7774, and 7775 \AA) and the O I 8446 triplet (at 8446.25, 8446.36, and 8446.76 \AA), both shown in Figure \ref{fig:OandKWindLines}. We omit the \ion{O}{1} triplets because the three components blended together make it difficult to identify the distinct velocity components we see in the other wind lines. 

The O I 7773 triplet, which has an EP$=9.1$ eV, is typically only present in higher temperature atmospheres or high $T_\mathrm{max}$ FU Ori objects \citep{Hillenbrand_RNO54_letter_2023ApJ}. Here, it appears almost entirely in excess of the disk model, which predicts a maximum depth of $\sim 5 \%$ for the line. This indicates the triplet must arise entirely from some hotter region in the disk, potentially near the launch point of the collimated wind (i.e., near $R_\mathrm{inner}$). This may be corroborated by the blue-shifted absorption that can be seen at $-30$ km s$^{-1}$ relative to the 8774 line. The triplet absorption is also shallower in the dip than in the outburst or plateau, which is the opposite of what we see in the other wind lines. If the triplet is sensitive here to the hottest region in the disk, the weaker absorption can be considered further evidence that the disk cools during the dip. 

The behavior of the O I 8446 triplet, which also has an EP$=9.1$ eV, differs from the O I 7773 lines. This triplet clearly shows blue-shifted wind absorption that extends to $-100$ km s$^{-1}$ and red-shifted disk profile absorption, like we see in the Fe II 5018 line profiles. The lower velocity component of O I 8446 is at $-50$ km s$^{-1}$, faster than we see for the other wind features, and like O I 7773, it is shallower in the dip than in the plateau. Again this may indicate that these high EP O I triplets trace the hottest absorbing material in the system, close to the central star.

\restartappendixnumbering

\section{The Temperature Sensitivity and Time Evolution of the CO (3-0) Band} \label{app:COBandTemps}
In this appendix we explore the potential temperature ranges for the disk and wind absorption components of the CO bands in the NIRSPEC and IGRINS spectra. We expand the CCF analysis used in Section \ref{sec:COAbs} for the CO (2-0) band in the NIRSPEC spectrum to the CO (2-0) band in the IGRINS spectrum and the CO (3-0) bands of both spectra. For the CO (3-0) and (3-1) bands, the wavelength ranges over which we compute the CCFs are $15540-15640$ and $23205-23305$ \AA. We also investigate the relative sensitivity of the CO (3-0), (3-1), and (2-0) bands to the low-velocity wind absorption.

\begin{figure}[!b]
    \includegraphics[width = 0.98\linewidth]{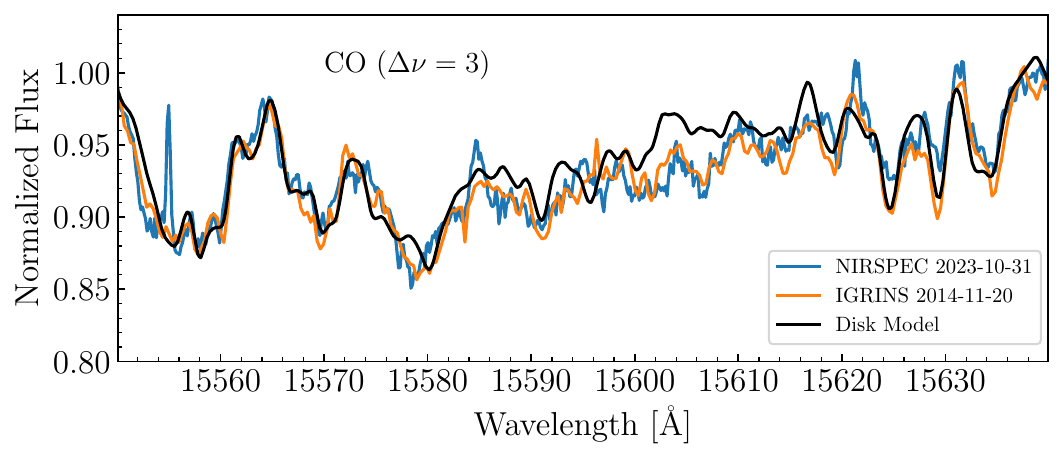}        
    \includegraphics[width = 0.32\linewidth]{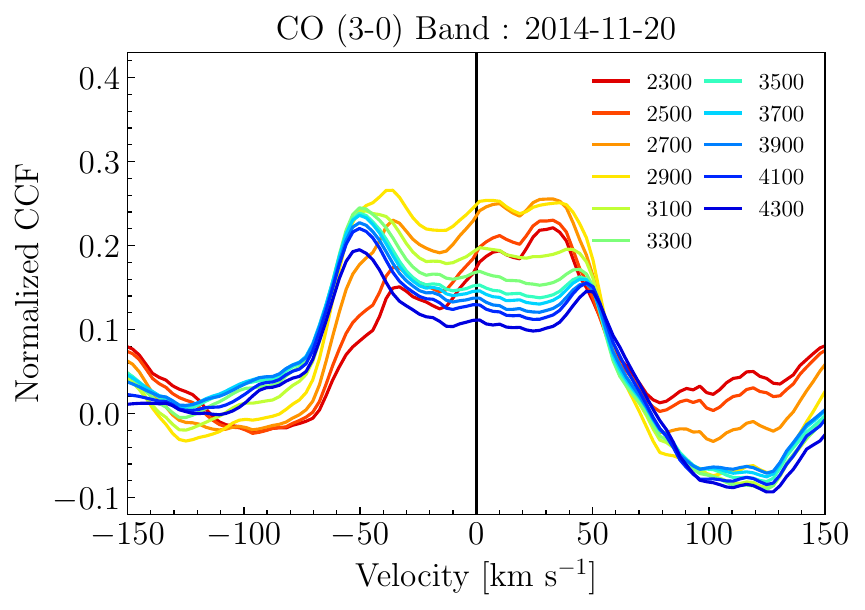} 
    \includegraphics[width = 0.32\linewidth]{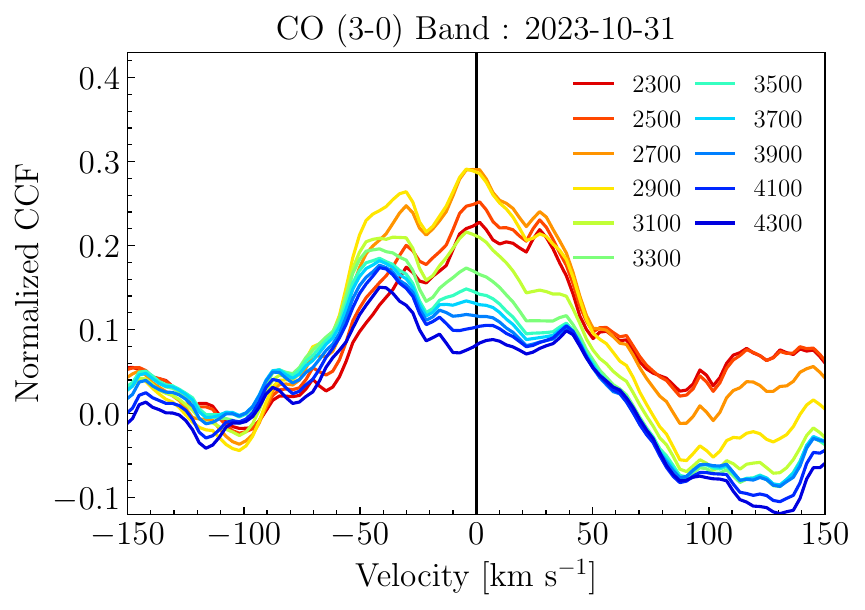}
    \includegraphics[width = 0.32\linewidth]{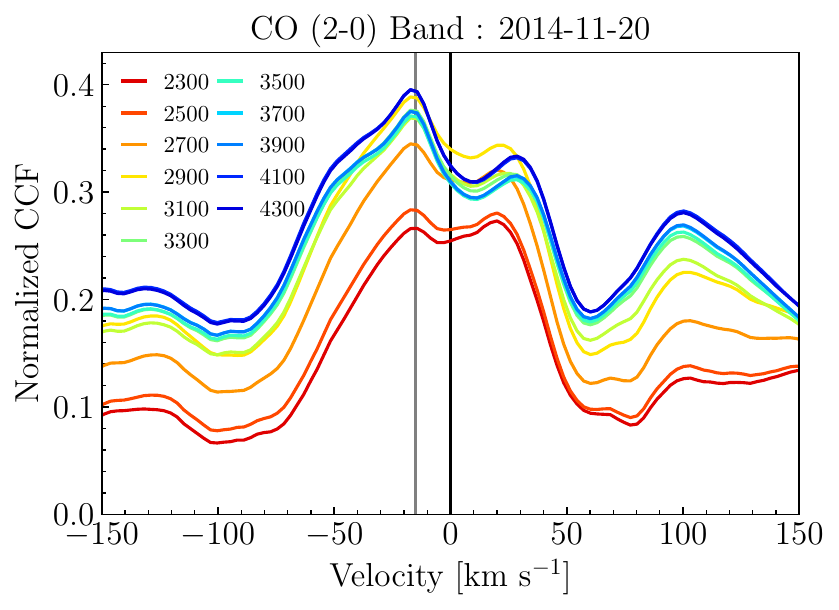}
    
    \caption{\textbf{Top Panel:} The CO (3-0) band from the Keck/NIRSPEC 30 Oct 2023 spectrum (blue), the HJST/IGRINS 20 Nov 2014 spectrum (orange) compared with the high resolution disk model spectrum (black). \textbf{Bottom Panels:} CCFs between the HBC 722 spectra and PHOENIX model atmospheres with different $T_\mathrm{eff}$. \textbf{Left:} The CCFs for the CO (3-0) band in the IGRINS spectrum. \textbf{Middle:} The CCFs for the CO (3-0) band in the NIRSPEC spectrum. \textbf{Right:} CCFs computed with (2-0) band for the IGRINS spectrum. Here, the disk absorption is seen at lower temperatures than in the NIRSPEC epoch, but at higher temperatures, the profile still shows strong narrow blue-shifted absorption at -15 km s$^{-1}$.}    
    \label{fig:CCFs_temps}
\end{figure}

Considering first the CO (3-0) bands (top panel of Figure \ref{fig:CCFs_temps}) we see that there is little evolution between the IGRINS and NIRSPEC epochs and that the disk model generally matches the data closely. Accordingly, CO (3-0) band CCFs (middle row, Figure \ref{fig:CCFs_temps}) show very clear disk profiles in both the NIRSPEC and IGRINS spectra. This remains the case almost regardless of the $T_\mathrm{eff}$ of the model used to compute the CCF. 

One trend that appears is that higher $T_\mathrm{eff}$ models produce CCFs with slightly broader disk profiles, whereas the lower $T_\mathrm{eff}$ models show narrower profiles. This demonstrates two important properties of the accretion disk model: that a large region of the disk representing several $T_\mathrm{eff}$ values can contribute to any particular spectral feature and that the Keplerian broadening of the hotter (closer in) regions is greater than that of the cooler (further out) regions. The only significant difference between the two epochs in this band is that the cooler temperature CCFs in the NIRSPEC spectrum are narrower than those of the IGRINS spectrum. This may trace the coolest, slowest-moving material at large radii away from the star and may be evidence of $R_\mathrm{outer}$ continuing to expand between 2014 and 2023.

The CCFs computed for the CO (2-0) band in the IGRINS spectrum (bottom right panel, Figure \ref{fig:CCFs_temps}) highlight the difference between the band in the two spectral epochs, as discussed in Section \ref{sec:COAbs}. In the IGRINS spectrum, the disk profiles seen at lower temperatures are much more clear than in the NIRSPEC spectrum. Although we also see the wind absorption component growing in prominence from the cooler model CCFs to the hotter ones, the feature does not dominate the profiles like we see in the NIRSPEC spectrum. 

The IGRINS CO (2-0) band notably shows similar behavior to the CO (3-0) band: the CCFs computed with hotter models have broader disk profiles than those computed with the cooler models. The CO (2-0) disk profiles are also generally narrower than the CO (3-0) profiles, which was also identified by \citet{Lee_HBC722_2015ApJ}. We generally see the trend in the NIRSPEC spectra for the cooler temperature CCFs. This is consistent with the idea that the $K$ band continuum is dominated by emission from greater (and therefore slower-orbiting) radii than $H$ band as predicted by the expected $T(r)$ and $v(r)$ profiles of viscous accretion disks in Keplerian rotation (demonstrated in Appendix \ref{app:RadLambda}).

\restartappendixnumbering

\section{Continuum Emission Radius Relation} \label{app:RadLambda}
\begin{figure}[!b]
    \centering
    \includegraphics[width=0.50\linewidth]{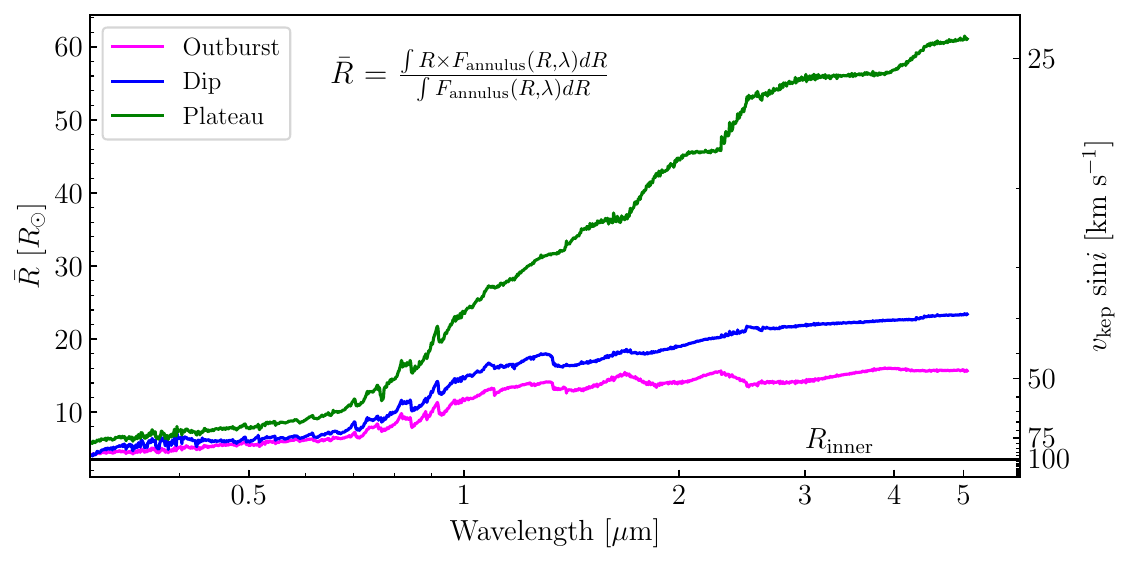}
    \includegraphics[width=0.47\linewidth]{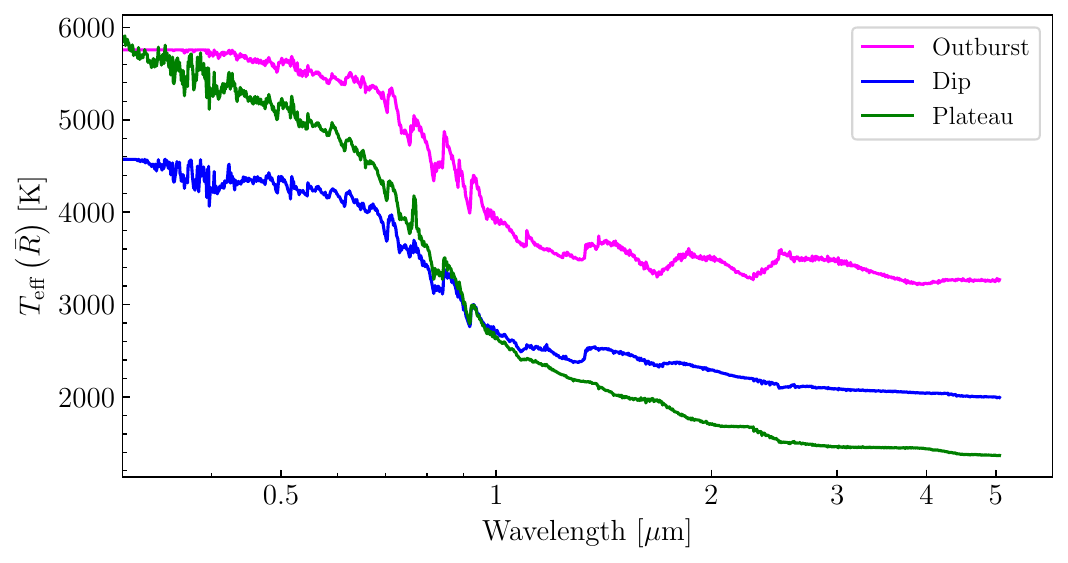}
    \caption{The flux-weighted mean radius from which the continuum for each wavelength bin arises. As the $R_\mathrm{outer}$ value for the disk model increases, the more the annuli at larger radii dominate the continuum NIR emission. At outburst, the $K$ band continuum arises mostly from the $R \sim 12 \ R_\odot$ region of the disk, whereas in the plateau the $K$ band continuum probes out to $R \sim 45 \ R_\odot$. However, the radii probed by the visible range (0.3 - 0.8 $\mu$m) are relatively unchanged. }
    \label{fig:WaveRad}
\end{figure}

With our disk model, we can compute the flux-weighted mean radius $\overline{R}$ and corresponding effective temperature $T_\mathrm{eff}(\overline{R})$ that dominates the continuum emission for each wavelength bin in the spectrum, shown in Figure \ref{fig:WaveRad}. From the $\overline{R}$ ``spectrum", we can also compute the expected $v_\mathrm{kep}(\overline{R}) \sin \ i$ for each wavelength bin. As is discussed in detail in \citet{Carvalho_V960MonSpectra_2023ApJ}, the expected velocity in a given wavelength bin is not directly probed by the atomic absorption in that region of the spectrum, as the disk region probed by a given absorption line is also dependent on the atomic properties of that line. However, we do generally see that the $v_\mathrm{kep}(\lambda)$ relation is relatively flat in the visible, as see in the data (Section \ref{sec:HIRESModelsComp}). We also see that the velocity decrease from the visible range ($\sim 80$ km s$^{-1}$) to 1.55 $\mu$m ($\sim 50$ km s$^{-1}$) is similar to the decrease in line broadening from the HIRES spectra to the CO (3-0) band.

In Figure \ref{fig:WaveRad} we plot the $\overline{R}$ (left panel) and $T_\mathrm{eff}(\overline{R})$ (right panel) spectra for the outburst, dip, and plateau models. As the $R_\mathrm{outer}$ increases, the $\overline{R}$ probed in the visible remains constant, while the in the NIR, the contribution from larger annuli becomes large. In the visible range, the outburst and dip models probe temperatures near the $T_\mathrm{max}$ of each model, whereas the plateau epoch does not quite appear to reach its maximum temperature redward of 0.3 $\mu$m. In the $H$ and $K$ bands, the dip and outburst spectra probe much cooler temperatures than the outburst spectrum, which explains the deepening molecular absorption in those bands and supports the need to include cooler atmosphere models in the disk model grid. In the plateau model, the CO (2-0) band at 2.29 $\mu$m probes very low disk temperatures ($<$ 2000 K), as predicted from the CCF analysis described in Appendix \ref{app:COBandTemps}. 

\end{document}